\documentclass[twoside,10pt]{article}
\usepackage{amsmath}
\usepackage{amssymb}
\usepackage{graphicx}

\renewcommand{\theequation}{\thesection.\arabic{equation}}
\renewcommand{\section}{\secdef\sct\sect}
\newcommand{\sct}[2][default]{\refstepcounter{section}
\vspace{0.5cm} \setcounter{equation}{0}
\centerline{ \scshape \arabic{section}.\ #1} \vspace{0.3cm}}
\newcommand{\sect}[1]{
\vspace{0.5cm} \centerline{\large\scshape #1} \vspace{0.3cm}}
\renewcommand{\subsection}{\secdef \subsct\sbsect}
\newcommand{\subsct}[2][default]{\refstepcounter{subsection}
\nopagebreak \vspace{0.5\baselineskip} {\flushleft\bf
\arabic{section}.\arabic{subsection}~\bf #1  } \nopagebreak}
\newcommand{\sbsect}[1]{\vspace{0.1cm}\noindent
{\bf #1}\vspace{0.1cm}}
\renewcommand{\subsubsection}{\secdef \subsubsect\sbsbsect}
\newcommand{\subsubsect}[2][default]{\refstepcounter{subsubsection} \nopagebreak
\vspace{0.1\baselineskip} \nopagebreak {\flushleft
\sffamily\slshape
\arabic{section}.\arabic{subsection}.\arabic{subsubsection}
\ \sffamily #1\/.}\ }
\newcommand{\sbsbsect}[1]{\vspace{0.1cm}\noindent
{\bf #1}\ }
\newtheorem{theorem}{Theorem}[section]
\newtheorem{lemma}[theorem]{Lemma}
\newtheorem{corollary}[theorem]{Corollary}

\newtheorem{remark}[theorem]{Remark}

\newtheorem{notation}[theorem]{Notation}
\begin{document}

\title{Microscopic Foundations of the Mei{\ss }ner Effect -- Thermodynamic
Aspects}
\author{J.-B. Bru\thanks{%
jb.bru@ikerbasque.org} \; and W. de Siqueira Pedra\thanks{%
wpedra@if.usp.br}}

\maketitle

\begin{abstract}
We analyze the Mei{\ss }ner effect from first principles of quantum
mechanics. We show in particular the existence of superconducting states
minimizing the magnetic free--energy of BCS--like models and carrying
surface currents which annihilate the total magnetic induction inside the
bulk in the thermodynamic limit. This study is a step towards a complete
explanation of the Mei{\ss }ner effect from microscopic models. It remains
indeed to prove that those states are dynamically stable, i.e.,
quasi--stationary at low temperatures. Note that our analysis shows that the
Mei{\ss }ner effect is not necessarily related to an effective magnetic
susceptibility equal to $-1$.\\[1.3ex]
\end{abstract}

% \altaffiliation[Also at ]{home.}  %  optional

\section{Introduction}

The so--called Mei{\ss }ner (or Mei{\ss }ner--Ochsenfeld) effect was
discovered in 1933 by the physicists W. Mei{\ss }ner and R. Ochsenfeld,
twenty--two years after the discovery of mercury superconductivity in 1911.
This represented an important experimental breakthrough and demonstrated,
among other things, that superconductors cannot be seen as perfect classical
conductors. This effect is well--described by phenomenological theories like
the celebrated London equations. We observe however that its microscopic
origin is far from being fully understood almost eighty years later. In
other words, there is no rigorous microscopic foundation of the Mei{\ss }ner
effect starting from first principles of quantum mechanics only.

In our papers \cite{BruPedra1,BruPedraAniko} we have recently showed, from a
microscopic theory, a weak version of the Mei{\ss }ner effect defined by the
absence of magnetization in presence of superconductivity, provided the
(space--homogeneous) external magnetic induction does not reach a critical
value. \cite[Section VI.B]{BruPedra-homog} extends these results to
space--inhomogeneous magnetic inductions. Nevertheless, the (full) Mei{\ss }%
ner effect also includes the existence of currents, concentrated near the
surface of the bulk, which annihilate the total magnetic induction inside
the superconductor. This phenomenon has not been analyzed in \cite%
{BruPedra1,BruPedraAniko,BruPedra-homog}. Such a study is the main subject
of the present paper.

We base our microscopic theory on the strong--coupling BCS--Hubbard model
with a \emph{self--generated} magnetic induction, which is driven by a
space--inhomogeneous external magnetic induction. Indeed, the
strong--coupling BCS--Hubbard model at fixed magnetic induction shows
qualitatively the same density dependency of the critical temperature
observed in high--$T_{c}$\emph{\ }superconductors \cite%
{BruPedra1,BruPedraAniko}. Depending on the choice of parameters, properties
of conventional superconductors are also qualitatively well--described by
such a model. Moreover, adding a sufficiently small hopping term to the
strong--coupling BCS--Hubbard model we obtain a more realistic model which
has essentially the same correlation functions, by Grassmann integration and
Brydges--Kennedy tree expansion methods together with determinant bounds
(see \cite{Pedra-Salmhofer} and reference therein). We outline the proof of
the Mei\ss ner effect for models with hopping terms in\ Section \ref{Section
Universality}. All these assertions result from the method described in \cite%
{BruPedra2}, which gives access to domains of the phase diagram usually
difficult to reach via other standard mathematical tools.

The analysis of the (full) Mei{\ss }ner effect from first principles of
quantum mechanics is highly non--trivial and in the present paper we provide
results concerning the free--energy, taking into account contributions of
the magnetic energy due to currents. Note indeed that Gibbs states of the
model under consideration do not manifest\ currents, at least for
space--homogeneous external magnetic inductions. By adding a magnetic term
to the usual free--energy density (similar to \cite[Eq. (2.11)]%
{sewellmeisner}), our results show that the minimizers of this new magnetic
free--energy can create surface\ currents which annihilate the total
magnetic induction inside the bulk, in the thermodynamic limit. The
corresponding Euler--Lagrange equations for these minimizers seem to
indicate that an effective magnetic susceptibility equal to $-1$ is not the
mechanism behind the Mei{\ss }ner effect.

Note that such magnetic free--energy minimizing states should be, in some
sense, stable with respect to dynamics to be named equilibrium states. Their
existence is only a necessary condition to have the Mei{\ss }ner effect.
Indeed, also for high temperatures, we can have minimizers of the magnetic
free--energy density suppressing the magnetic induction within the bulk.
This comes from the fact that the finite volume system can produce any
current density by creating local superconducting patches within a
negligible volume. Therefore, we conjecture that the quantum dynamics
rapidly destroys all currents in the non--superconducting phase. In
particular, the second step will be to show the dynamical
instability/stability of such a phenomenon in the
non--superconducting/superconducting phase. Such an analysis is not
performed here because it requires an extension of \cite{BruPedra2} to
include dynamics. We postpone it to a further paper.

Finally, note that thermodynamic studies of the Mei{\ss }ner effect have
been performed in \cite{sewellmeisner,sewell2} from an axiomatic point of
view. They are based on assumptions, not proven for some concrete
microscopic model, like the existence of equilibrium states with
off--diagonal long range order. In the same spirit, we also discuss some
model independent conditions for the existence of the Mei\ss ner effect in
Section \ref{Section Universality}.

The present paper is organized as follows. In Section \ref{meisneer sect 1}
we set up the quantum many--body problem at fixed magnetic induction and
give one important result concerning the possibility of having currents with
no energy cost in the thermodynamic limit. Section \ref{Section Magnetic
Energy} explains the Biot--Savart operator used to define magnetic
inductions from currents. The \emph{self--generated} magnetic induction is
then discussed in\ Section \ref{section Meissner effect} to obtain a proper
definition of the magnetic free--energy density. The Mei{\ss }ner effect is
finally discussed in Sections \ref{section Meissner} and \ref{Section
Universality}. Section \ref{meisneer sect 2} explains in detail all
technical proofs required to show the assertions of previous sections. Our
main result is Theorem \ref{theorem principal copy(1)}. See also Theorems %
\ref{Theorem inv trans} and \ref{Theorem inv trans copy(1)}.

\begin{notation}[Norms]
\mbox{ }\newline
For any $x=(x_{1},x_{2},x_{3})\in \mathbb{R}^{3}$, $|x|:=\sqrt{%
x_{1}^{2}+x_{2}^{2}+x_{3}^{2}}$. For any subset $\Lambda \subset \mathbb{Z}$%
, $\left\vert \Lambda \right\vert $ is by definition the cardinality of $%
\Lambda $. For any $p\in \mathbb{N}$, $\Vert -\Vert _{p}$ stands for the $%
L^{p}$--norm, whereas $\Vert -\Vert $ is the operator norm. Additionally, $%
\Vert -\Vert _{\mathrm{Tr}}$ denotes the trace norm.
\end{notation}

\section{Thermodynamic stability of currents\label{meisneer sect 1}}

The host material for superconducting electrons is assumed to be a (perfect)
cubic crystal. Other lattices could also be studied, but for simplicity we
refrain from considering them. The unit of length is chosen so that the
lattice spacing in this crystal is exactly $1$. We thus use $\mathbb{Z}^{3}$
to represent the crystal. Our microscopic theory is based on the strong
coupling BCS--Hubbard model studied in \cite{BruPedra1,BruPedraAniko}.

In absence of magnetic induction, it is defined in the box $\Lambda _{l}:=\{%
\mathbb{Z}\cap \lbrack -l,l-1]\}^{3}$ of side length $2l$ for $l\in \mathbb{N%
}$ by the Hamiltonian%
\begin{equation}
T_{l}:=-\mu \sum\limits_{x\in \Lambda _{l}}\left( n_{x,\uparrow
}+n_{x,\downarrow }\right) +2\lambda \sum_{x\in \Lambda _{l}}n_{x,\uparrow
}n_{x,\downarrow }-\frac{\gamma }{\left\vert \Lambda _{l}\right\vert }%
\sum_{x,y\in \Lambda _{l}}a_{x,\uparrow }^{\ast }a_{x,\downarrow }^{\ast
}a_{y,\downarrow }a_{y,\uparrow }  \label{Hamiltonian BCS-Hubbard0}
\end{equation}%
with real parameters $\mu ,\lambda \in \mathbb{R}$ and $\gamma \in \mathbb{R}%
^{+}$ (i.e., $\gamma >0$). The operator $a_{x,\mathrm{s}}^{\ast }$ resp. $%
a_{x,\mathrm{s}}$ creates resp. annihilates a fermion with spin $\mathrm{s}%
\in \{\uparrow ,\downarrow \}$ at lattice position $x\in \mathbb{Z}^{3}$
whereas $n_{x,\mathrm{s}}:=a_{x,\mathrm{s}}^{\ast }a_{x,\mathrm{s}}$ is the
particle number operator at position $x$ and spin $\mathrm{s}$.

The first term of the right hand side of (\ref{Hamiltonian BCS-Hubbard0})
represents the strong coupling limit of the kinetic energy, with $\mu $
being the chemical potential of the system. Note that this \textquotedblleft
strong coupling limit\textquotedblright\ is also called \textquotedblleft
atomic limit\textquotedblright\ in the context of the Hubbard model. See,
e.g., \cite{atomiclimit1,atomiclimit2}. For further discussions, we also
strongly recommend Section \ref{Section Universality} (cf. \textbf{2.}).

The second term in the right hand side of (\ref{Hamiltonian BCS-Hubbard0})
represents the (screened) Coulomb repulsion as in the celebrated Hubbard
model. So, the parameter $\lambda $ should be taken as a positive number but
our results are also valid for any real Hubbard coupling $\lambda \in
\mathbb{R}$.

The last term is the BCS interaction written in the $x$--space since%
\begin{equation*}
\frac{\gamma }{\left\vert \Lambda _{l}\right\vert }\sum_{x,y\in \Lambda
_{l}}a_{x,\uparrow }^{\ast }a_{x,\downarrow }^{\ast }a_{y,\downarrow
}a_{y,\uparrow }=\frac{\gamma }{\left\vert \Lambda _{l}\right\vert }%
\sum_{k,q\in \Lambda _{l}^{\ast }}\tilde{a}_{k,\uparrow }^{\ast }\tilde{a}%
_{-k,\downarrow }^{\ast }\tilde{a}_{q,\downarrow }\tilde{a}_{-q,\uparrow }
\end{equation*}%
with $\Lambda _{l}^{\ast }$ being the reciprocal lattice of quasi--momenta
and where $\tilde{a}_{q,\mathrm{s}}$ is the corresponding annihilation
operator for $\mathrm{s}\in \{\uparrow ,\downarrow \}$ and $q\in \Lambda
_{l}^{\ast }$. Observe that the thermodynamics of the model for $\gamma =0$
can easily be computed. Therefore, we restrict the analysis to the case $%
\gamma \in \mathbb{R}^{+}$. Note also that the BCS interaction can imply a
superconducting phase. The mediator implying this effective interaction does
not matter here, i.e., it could be a phonon, as in conventional type I
superconductors, or anything else.

We now fix a magnetic induction $\mathrm{B}\in L^{2}(\mathfrak{C};\mathbb{R}%
^{3})$, where $\mathfrak{C}:=[-1/2,1/2]^{3}$. The strong coupling
BCS--Hubbard model with space--inhomogeneous magnetic induction is then
defined by $H_{l}:=T_{l}+\mathcal{M}_{l}$ with
\begin{eqnarray}
\mathcal{M}_{l} &:=&-\vartheta \sum_{x\in \Lambda _{l}}\left( a_{x,\uparrow
}^{\ast }a_{x,\downarrow }+a_{x,\downarrow }^{\ast }a_{x,\uparrow }\right)
\int_{\mathfrak{C}}\mathrm{b}_{1}\left( \frac{x+y}{2l}\right) \mathrm{d}^{3}y
\notag \\
&&+i\vartheta \sum_{x\in \Lambda _{l}}\left( a_{x,\uparrow }^{\ast
}a_{x,\downarrow }-a_{x,\downarrow }^{\ast }a_{x,\uparrow }\right) \int_{%
\mathfrak{C}}\mathrm{b}_{2}\left( \frac{x+y}{2l}\right) \mathrm{d}^{3}y
\notag \\
&&-\vartheta \sum_{x\in \Lambda _{l}}\left( n_{x,\uparrow }-n_{x,\downarrow
}\right) \int_{\mathfrak{C}}\mathrm{b}_{3}\left( \frac{x+y}{2l}\right)
\mathrm{d}^{3}y  \label{Hamiltonian BCS-Hubbard-inhom0}
\end{eqnarray}%
for any fixed parameter $\vartheta \in \mathbb{R}^{+}$ ($\vartheta >0$) and%
\begin{equation*}
\mathrm{B}\left( \mathfrak{t}\right) \equiv \left( \mathrm{b}_{1}\left(
\mathfrak{t}\right) ,\mathrm{b}_{2}\left( \mathfrak{t}\right) ,\mathrm{b}%
_{3}\left( \mathfrak{t}\right) \right) \in \mathbb{R}^{3}
\end{equation*}%
for $\mathfrak{t}\in \mathfrak{C}$ almost everywhere (a.e.). Indeed, the
terms of $\mathcal{M}_{l}$ correspond to the interaction between spins and
the total magnetic induction $\mathrm{B}((x+y)/(2l))$ within a unit cell $%
\mathfrak{C}$ around $x\in \Lambda _{l}$.

Note that, for continuous fields $\mathrm{B}\in C^{0}(\mathfrak{C};\mathbb{R}%
^{3})$ and in the thermodynamic limit $l\rightarrow \infty $,%
\begin{equation}
\int_{\mathfrak{C}}\mathrm{B}\left( \frac{x+y}{2l}\right) \mathrm{d}%
^{3}y=\left\vert \Lambda _{l}\right\vert \int_{(2l)^{-1}\mathfrak{C}}\mathrm{%
B}\left( \frac{x}{2l}+\mathfrak{t}\right) \mathrm{d}^{3}\mathfrak{t}=\mathrm{%
B}\left( \frac{x}{2l}\right) +o(1)\ .  \label{mean value}
\end{equation}%
If $\mathrm{B}$ is continuous, then we can equivalently take either (a) the
integral of $\mathrm{B}((x+y)/(2l))$ (with respect to $y$) in the unit cell $%
\mathfrak{C}$ or (b) the value $\mathrm{B}(x/(2l))$ in the definition of $%
\mathcal{M}_{l}$. In fact, the thermodynamic limit of both systems (a)--(b)
are identical for continuous magnetic inductions $\mathrm{B}\in C^{0}(%
\mathfrak{C};\mathbb{R}^{3})$ (cf. Section \ref{meisneer sect 2}), but an
extension of our results to all $\mathrm{B}\in L^{2}(\mathfrak{C};\mathbb{R}%
^{3})$ leads us to consider the definition (a) in (\ref{Hamiltonian
BCS-Hubbard-inhom0}) and not (b). Additionally, (a) is also more natural if
one considers $\mathrm{B}$ as an effective field coming from a quantum
magnetic induction. See, e.g., \cite{BruPedra2}.

The scaling factor $(2l)^{-1}$ used in (\ref{Hamiltonian BCS-Hubbard-inhom0}%
) means that the space fluctuations of the inhomogeneous magnetic induction
involve a macroscopic number of lattice sites. This obviously does not
prevent the space scale of these fluctuations from being extremely small as
compared to the side--length $2l$ of the box $\Lambda _{l}$. Similarly, we
could also model mesoscopic fluctuations meaning that -- in the
thermodynamic limit -- the space scale of inhomogeneities is infinitesimal
with respect to the box side--length $2l$ whereas the lattice spacing is
infinitesimal with respect to the space scale of inhomogeneities. See, e.g.,
\cite[Section V]{BruPedra-homog}. Microscopic fluctuations can also be
handled provided they are periodic, see \cite[Section III]{BruPedra-homog}.
Both situations (or any combination of them with the macroscopic one) are
however omitted to simplify discussions and proofs.

We observe that $T_{l}$, $\mathcal{M}_{l}$ and $H_{l}=T_{l}+\mathcal{M}_{l}$
belong to the CAR $C^{\ast }$--algebra $\mathcal{U}_{\Lambda _{l}}$ with
identity $\mathbf{1}$ and generators $\{a_{x,\mathrm{s}}\}_{x\in \Lambda
_{l},\mathrm{s}\in \{\uparrow ,\downarrow \}}$ satisfying the canonical
anti--commutation relations (CAR):%
\begin{equation}
\left\{
\begin{array}{lll}
a_{x,\mathrm{s}}a_{x^{\prime },\mathrm{s}^{\prime }}+a_{x^{\prime },\mathrm{s%
}^{\prime }}a_{x,\mathrm{s}} & = & 0\ , \\[0.15cm]
a_{x,\mathrm{s}}a_{x^{\prime },\mathrm{s}^{\prime }}^{\ast }+a_{x^{\prime },%
\mathrm{s}^{\prime }}^{\ast }a_{x,\mathrm{s}} & = & \delta _{x,x^{\prime
}}\delta _{\mathrm{s},\mathrm{s}^{\prime }}\mathbf{1}\ .%
\end{array}%
\right.  \label{CAR}
\end{equation}%
$\mathcal{U}_{\Lambda _{l}}$ is isomorphic to the $C^{\ast }$--algebra $%
\mathcal{L}(\bigwedge \mathcal{H}_{\Lambda _{l}})$ of all linear operators
on the fermion Fock space $\bigwedge \mathcal{H}_{\Lambda _{l}}$, where%
\begin{equation}
\mathcal{H}_{\Lambda _{l}}:=\bigoplus\limits_{x\in \Lambda _{l}}\mathcal{H}%
_{x}\ .  \label{hilbert space}
\end{equation}%
Here, for every $x\in \mathbb{Z}^{3}$, $\mathcal{H}_{x}$ is a copy of some
fixed two dimensional Hilbert space $\mathcal{H}$ with orthonormal basis $%
\{|\uparrow \rangle ,|\downarrow \rangle \}$. States on the $C^{\ast }$%
--algebra $\mathcal{U}_{\Lambda _{l}}$ are linear functionals $\rho \in
\mathcal{U}_{\Lambda _{l}}^{\ast }$ which are positive, i.e., for all $A\in
\mathcal{U}_{\Lambda _{l}}$, $\rho (A^{\ast }A)\geq 0$, and normalized,
i.e., $\rho (\mathbf{1})=1$. We denote by $E_{\Lambda _{l}}\subset \mathcal{U%
}_{\Lambda _{l}}^{\ast }$ the set of all states on $\mathcal{U}_{\Lambda
_{l}}$ for any $l\in \mathbb{N}$.

It is well--known that the physics of the system at thermodynamical
equilibrium is given by the Gibbs state $\mathfrak{g}_{l}\in E_{\Lambda
_{l}} $ defined by
\begin{equation}
\mathfrak{g}_{l}\left( A\right) :=\mathrm{Trace}_{\wedge \mathcal{H}%
_{\Lambda _{l}}}(\mathrm{d}_{\mathfrak{g}_{l}}A)\ ,\quad A\in \mathcal{U}%
_{\Lambda _{l}}\ ,  \label{Gibbs0}
\end{equation}%
with density matrix%
\begin{equation}
\mathrm{d}_{\mathfrak{g}_{l}}:=\frac{\mathrm{e}^{-\beta H_{l}}}{\mathrm{Trace%
}_{\wedge \mathcal{H}_{\Lambda _{l}}}(\mathrm{e}^{-\beta H_{l}})}
\label{Gibbs}
\end{equation}%
for any inverse temperature $\beta \in \mathbb{R}^{+}$ and $l\in \mathbb{N}$%
. Indeed, given any state $\rho \in E_{\Lambda _{l}}$ on $\mathcal{U}%
_{\Lambda _{l}}$, the energy observable $H_{l}=H_{l}^{\ast }\in \mathcal{U}%
_{\Lambda _{l}}$ fixes the finite volume free--energy density
\begin{equation}
f_{l}\left( \mathrm{B},\rho \right) :=\left\vert \Lambda _{l}\right\vert
^{-1}\left\{ \rho (H_{l})-\beta ^{-1}S_{l}(\rho )\right\}
\label{free energy density}
\end{equation}%
at fixed magnetic induction $\mathrm{B}\in L^{2}(\mathfrak{C};\mathbb{R}%
^{3}) $ and inverse temperature $\beta \in \mathbb{R}^{+}$ for any $l\in
\mathbb{N} $. If $\mathrm{B}\in L^{2}\equiv L^{2}(\mathbb{R}^{3};\mathbb{R}%
^{3})$, then we set $f_{l}(\mathrm{B},\rho )\equiv f_{l}(\mathrm{B}|_{%
\mathfrak{C}},\rho ) $. The first term in $f_{l}$ is the mean energy per
unit of volume of the physical system found in the state $\rho \in
E_{\Lambda _{l}}$, whereas $S_{l}$ is the von Neumann entropy defined, for
all $\rho \in E_{\Lambda _{l}} $, by
\begin{equation}
S_{l}(\rho ):=\mathrm{Trace}_{\wedge \mathcal{H}_{\Lambda _{l}}}\,\left(
\eta (\mathrm{d}_{\rho })\right) \geq 0\ .  \label{neuman entropy}
\end{equation}%
Here, $\eta (t):=-t\log (t)$ for $t\in \mathbb{R}^{+}$, $\eta (0):=0$, and $%
\mathrm{d}_{\rho }$ is the density matrix of $\rho \in E_{\Lambda _{l}}$.
The state of a system in thermal equilibrium and at fixed mean energy per
unit of volume maximizes the entropy, by the second law of thermodynamics.
Therefore, it minimizes the free--energy density functional $\rho \mapsto
f_{l}(\mathrm{B},\rho )$. Such well--known arguments lead to the study of
the variational problem $\inf f_{l}\left( \mathrm{B},E_{\Lambda _{l}}\right)
$. The value of this variational problem is directly related to the
so--called pressure $p_{l}\left( \mathrm{B}\right) $ as%
\begin{equation}
p_{l}\left( \mathrm{B}\right) :=(\beta \left\vert \Lambda _{l}\right\vert
)^{-1}\ln \mathrm{Trace}_{\wedge \mathcal{H}_{\Lambda _{l}}}\left( \mathrm{e}%
^{-\beta H_{l}}\right) =-\underset{\rho \in E_{\Lambda _{l}}}{\inf }%
f_{l}\left( \mathrm{B},\rho \right)  \label{pressure0}
\end{equation}%
for any magnetic induction $\mathrm{B}\in L^{2}(\mathfrak{C};\mathbb{R}^{3})$%
. (If $\mathrm{B}\in L^{2}$, then $p_{l}(\mathrm{B})\equiv p_{l}(\mathrm{B}%
|_{\mathfrak{C}})$.) For any $\beta \in \mathbb{R}^{+}$ and $l\in \mathbb{N}$%
, the unique solution of this variational problem is precisely the Gibbs
state $\mathfrak{g}_{l}\in E_{\Lambda _{l}}$ (\ref{Gibbs0})--(\ref{Gibbs}).
This fact is named in the literature the passivity of Gibbs states and is a
consequence of\ Jensen's inequality.

Our microscopic approach to the Mei{\ss }ner effect requires a definition of
(charged) currents. Indeed, we would like to study the existence of currents
near the surface of the bulk of the model. To this end, we note that, for
all $x\in \Lambda _{l}$,%
\begin{equation*}
\frac{\mathrm{d}}{\mathrm{d}t}\left\{ \mathrm{e}^{itH_{l}}\left(
n_{x,\uparrow }+n_{x,\downarrow }\right) \mathrm{e}^{-itH_{l}}\right\} =%
\mathrm{e}^{itH_{l}}i\left[ H_{l},n_{x,\uparrow }+n_{x,\downarrow }\right]
\mathrm{e}^{-itH_{l}}
\end{equation*}%
and%
\begin{equation}
i\left[ H_{l},n_{x,\uparrow }+n_{x,\downarrow }\right] =\sum_{y\in \Lambda
_{l}}\frac{4\gamma }{\left\vert \Lambda _{l}\right\vert }\mathop{\rm Im}%
\left( a_{y,\uparrow }^{\ast }a_{y,\downarrow }^{\ast }a_{x,\downarrow
}a_{x,\uparrow }\right) \ .  \notag
\end{equation}%
The quantum observable describing the (charged) current from $x$ to $y$ is
thus defined by%
\begin{equation}
\mathrm{I}_{l}^{x,y}:=\frac{4\gamma }{\left\vert \Lambda _{l}\right\vert }%
\mathop{\rm Im}\left( a_{x,\uparrow }^{\ast }a_{x,\downarrow }^{\ast
}a_{y,\downarrow }a_{y,\uparrow }\right)  \label{curents observable}
\end{equation}%
for any $x,y\in \Lambda _{l}$. These current observables naturally give rise
to a magnetic induction functional which we define below by using the
Biot--Savart law.

Indeed, given any state $\rho \in E_{\Lambda _{l}}$, we interpret the real
number $\rho \left( \mathrm{I}_{l}^{x,y}\right) $ as the current passing
from $x$ to $y$. We use this observation to define a current density induced
by the system in the state $\rho $. One expects that the full current $\rho
\left( \mathrm{I}_{l}^{x,y}\right) $ between $x$ and $y$ is smoothly
distributed in some region of size $\left\vert x-y\right\vert $ around $%
(x+y)/2$. The current profile is fixed by an arbitrary smooth, compactly
supported, spherical symmetric and non--negative function $\xi \in
C_{0}^{\infty }\equiv C_{0}^{\infty }(\mathbb{R}^{3};\mathbb{R}^{3})$ such
that $\xi \left( 0\right) >0$,%
\begin{equation}
\int_{\mathbb{R}^{3}}\xi \left( \mathfrak{t}\right) \mathrm{d}^{3}\mathfrak{t%
}=1\quad \mathrm{and}\quad \int_{\mathbb{R}^{2}}\xi \left( 0,\mathfrak{t}%
_{2},\mathfrak{t}_{3}\right) \mathrm{d}\mathfrak{t}_{2}\mathrm{d}\mathfrak{t}%
_{3}=1\ .  \label{carlos1}
\end{equation}%
For any $l\in \mathbb{N}$, the current density induced by the system in the
state $\rho \in E_{\Lambda _{l}}$ at $x\in \mathbb{R}^{3}$ is defined by
\begin{equation}
\rho \mapsto j_{\rho }\left( x\right) :=\sum\limits_{y,z\in \Lambda _{l},\
y\neq z}\frac{z-y}{\left\vert z-y\right\vert ^{3}}\ \xi \left( \frac{x-\frac{%
y+z}{2}}{\left\vert z-y\right\vert }\right) \rho \left( \mathrm{I}%
_{l}^{y,z}\right) \ .  \label{carlos_currents}
\end{equation}%
It defines a map $j\ $from $E_{\Lambda _{l}}$ to the real vector space $%
C_{0}^{\infty }$ of compactly supported smooth fields. This map is named
here the current density functional of the box $\Lambda _{l}$. Observe that
the second condition of (\ref{carlos1}) ensures that the flow of the field
\begin{equation*}
\frac{z-y}{\left\vert z-y\right\vert ^{3}}\ \xi \left( \frac{x-\frac{y+z}{2}%
}{\left\vert z-y\right\vert }\right) \rho \left( \mathrm{I}_{l}^{y,z}\right)
\end{equation*}%
through the hyperplane perpendicular to $z-y$ at $(y+z)/2$ equals the full
current $\rho \left( \mathrm{I}_{l}^{y,z}\right) $ passing from $y$ to $z$.

As we are interested in magnetic effects induced by the quantum system, for
any state $\rho \in E_{\Lambda _{l}}$, we shall consider the smooth magnetic
induction $B_{\rho }\in C^{\infty }\equiv C^{\infty }(\mathbb{R}^{3};\mathbb{%
R}^{3})$ created by the current density $j_{\rho }$ together with some fixed
external magnetic induction $\mathrm{B}_{\mathrm{ext}}$. Its definition uses
the Biot--Savart operator $\mathcal{S}$ (Section \ref{Section Magnetic
Energy}) and requires further explanations given in Section \ref{section
Meissner effect}. We only note at this point that the smooth magnetic
induction $B_{\rho }$ has to be rescaled in order to be compared to $\mathrm{%
B}$, see (\ref{Hamiltonian BCS-Hubbard-inhom0}). We thus define the rescaled
magnetic induction $B_{\rho }^{(l)}\in C^{\infty }$ by
\begin{equation}
B_{\rho }^{(l)}\left( \mathfrak{t}\right) :=B_{\rho }\left( 2l\mathfrak{t}%
\right) \ ,\quad \mathfrak{t}\in \mathbb{R}^{3}\ ,
\label{magnetic induction0}
\end{equation}%
for all $l\in \mathbb{N}$. Keeping in mind the Biot--Savart law (cf. (\ref%
{def.B.rho})), we similarly need to define a rescaled current density $%
j_{\rho }^{(l)}\in C_{0}^{\infty }$ from the current density $j_{\rho }$ as
follows:%
\begin{equation}
j_{\rho }^{(l)}(\mathfrak{t}):=2l\ j_{\rho }(2l\mathfrak{t})\ ,\quad
\mathfrak{t}\in \mathbb{R}^{3}\ ,  \label{rescaled current density}
\end{equation}%
for all $l\in \mathbb{N}$ and $\rho \in E_{\Lambda _{l}}$. Here, the support
\begin{equation}
\mathrm{supp}(j_{\rho }^{(l)}):=\{\mathfrak{t}\in \mathbb{R}^{3}\ :\ j_{\rho
}^{(l)}(\mathfrak{t})\neq 0\}  \label{support}
\end{equation}%
of $j_{\rho }^{(l)}$ is contained in a sufficiently large box $%
[-L,L]^{3}\supset \mathfrak{C}$ which depends on the size of the support of
the function $\xi $ used in the definition of $j_{\rho }$ but not on the
length $l\in \mathbb{N}$. However, because of the prefactor $|\Lambda
_{l}|^{-1}$ in the definition (\ref{curents observable}) of current
observables, $j_{\rho }^{(l)}$ is strongly concentrated inside $\mathfrak{C}$%
, as $l\rightarrow \infty $.

The system also shows magnetization due to spinning charged particles,
electrons in our case. The magnetization observables are seen as coordinates
of an observable vector $\mathrm{M}^{x}:=(\mathrm{m}_{1}^{x},\mathrm{m}%
_{2}^{x},\mathrm{m}_{3}^{x})$ where, for all $x\in \mathbb{Z}^{3}$,
\begin{eqnarray}
\mathrm{m}_{1}^{x} &:=&\vartheta \left( a_{x,\uparrow }^{\ast
}a_{x,\downarrow }+a_{x,\downarrow }^{\ast }a_{x,\uparrow }\right) \ ,
\notag \\
\mathrm{m}_{2}^{x} &:=&i\vartheta (a_{x,\downarrow }^{\ast }a_{x,\uparrow
}-a_{x,\uparrow }^{\ast }a_{x,\downarrow })\ ,  \label{magne1} \\
\mathrm{m}_{3}^{x} &:=&\vartheta \left( n_{x,\uparrow }-n_{x,\downarrow
}\right) \ .  \notag
\end{eqnarray}%
For any strictly positive fixed parameter $\epsilon \in \mathbb{R}^{+}$, let
\begin{equation}
\xi _{\epsilon }(\mathfrak{t}):=\epsilon ^{-3}\ \xi \left( \epsilon ^{-1}%
\mathfrak{t}\right) \ ,\quad \mathfrak{t}\in \mathbb{R}^{3}\ .
\label{xi eps m}
\end{equation}%
Then, for any $l\in \mathbb{N}$ and $\rho \in E_{\Lambda _{l}}$, we define
the coarse--grained magnetization density at $x\in \mathbb{R}^{3}$ by
\begin{equation}
\rho \mapsto m_{\rho }\left( x\right) :=\frac{1}{\left\vert \Lambda
_{l}\right\vert }\sum\limits_{\substack{ y\in \Lambda _{l}  \\ \left(
2l\right) ^{-1}y+\mathrm{supp}\left( \Xi _{\epsilon }\right) \subset
\mathfrak{C}}}\Xi _{\epsilon }\left( \frac{x-y}{2l}\right) \rho \left(
\mathrm{M}^{y}\right) \ ,  \label{magnetization spin}
\end{equation}%
where%
\begin{equation*}
\Xi _{\epsilon }\left( \mathfrak{t}\right) :=\int_{\mathfrak{C}}\xi
_{\epsilon }\left( \mathfrak{t}-\frac{z}{2l}\right) \mathrm{d}z\ ,\quad
\mathfrak{t}\in \mathbb{R}^{3}\ ,
\end{equation*}%
for any $\epsilon \in \mathbb{R}^{+}$, whereas%
\begin{equation}
\rho \left( \mathrm{M}^{x}\right) :=%
%TCIMACRO{\TeXButton{\Big(}{\Big(}}%
%BeginExpansion
\Big(%
%EndExpansion
\rho (\mathrm{m}_{1}^{x})\ ,\ \rho (\mathrm{m}_{2}^{x}),\ \rho (\mathrm{m}%
_{3}^{x})%
%TCIMACRO{\TeXButton{\Big)}{\Big)}}%
%BeginExpansion
\Big)%
%EndExpansion
\in \mathbb{R}^{3}\ .  \label{magne2}
\end{equation}%
It is again a map from $E_{\Lambda _{l}}$ to $C_{0}^{\infty }$. Similar to
the rescaled magnetic induction $B_{\rho }^{(l)}\in C^{\infty }$, the
rescaled magnetization density is defined by
\begin{equation}
m_{\rho }^{(l)}\left( \mathfrak{t}\right) :=m_{\rho }\left( 2l\mathfrak{t}%
\right) \ ,\quad \mathfrak{t}\in \mathbb{R}^{3}\ ,
\label{rescaled magnetization density}
\end{equation}%
for any state $\rho \in E_{\Lambda _{l}}$.

The use of $\Xi _{\epsilon }$ in the definition of the magnetization density
is technically convenient but not essential for our analysis. It is only a
specific choice of a function with integral equal to $1$ that implements the
coarse--graining of the magnetization. The first condition of (\ref{carlos1}%
) ensures indeed that the full magnetization produced by one lattice site $%
x\in \Lambda _{l}$ equals $\rho \left( \mathrm{M}^{x}\right) $.

The restriction
\begin{equation*}
\left( 2l\right) ^{-1}y+\mathrm{supp}\left( \Xi _{\epsilon }\right) \subset
\mathfrak{C}
\end{equation*}%
in the definition of $m_{\rho }$ guarantees that $\mathrm{supp}(m_{\rho
}^{(l)})\subset \mathfrak{C}$. This is also technically convenient. We add
that the scaling factor $(2l)^{-1}$ in (\ref{magnetization spin}) means that
$m_{\rho }^{(l)}\in C_{0}^{\infty }$ is a \emph{macroscopic} magnetization,
used in Section \ref{section Meissner effect} to define the (also
macroscopic) self--generated magnetic induction $B_{\rho }^{(l)}$ via the
Maxwell equations in matter.

\begin{remark}[Coarse--graining of the magnetization]
\label{small eps}\mbox{ }\newline
The coarse--grained magnetization density $m_{\rho }^{(l)}$ is defined for
all $\epsilon \in \mathbb{R}^{+}$. However, we are only interested in the
case where the space--scale of the coarse--graining of $m_{\rho }^{(l)}$ is
very small as compared to the side length of the unit box $\mathfrak{C}$.
This corresponds to take $\epsilon <<\epsilon _{\xi }:=1/(2R_{\xi })$ with
\begin{equation}
R_{\xi }:=\sup \left\{ \left\vert x\right\vert :\xi \left( x\right) \neq
0\right\} \in \mathbb{R}^{+}  \label{radius1}
\end{equation}%
being the radius of the support of the function $\xi \in C_{0}^{\infty }$.
\end{remark}

\begin{remark}[Smooth from discrete]
\label{remark Smooth from discrete}\mbox{ }\newline
The quantum many--body problem considered here uses discrete space
coordinates. On the other hand, the Maxwell equations require differentiable
fields. We have thus defined smooth magnetization and current densities $%
m_{\rho },j_{\rho }$ on $\mathbb{R}^{3}$. The latter is done without
introducing any arbitrariness in our thermodynamic results since we take $%
\epsilon \rightarrow 0^{+}$ after the thermodynamic limit $l\rightarrow
\infty $. The thermodynamics then becomes independent of the choice of $\xi
\in C_{0}^{\infty }$.
\end{remark}

We give now one of our main (technical) result about the creation of any
smooth current density without energy costs:

\begin{theorem}[Thermodynamic stability of currents]
\label{coolthml1 copy(2)}\mbox{ }\newline
For every $\mathrm{B}\in L^{2}(\mathfrak{C};\mathbb{R}^{3})$ and any smooth
current density $\mathrm{j}\in C^{\infty }(\mathfrak{C};\mathbb{R}^{3})$,
there are states $\rho _{l}\in E_{\Lambda _{l}}$ for $l\in \mathbb{N}$
satisfying%
\begin{equation*}
\underset{l\rightarrow \infty }{\lim }%
%TCIMACRO{\TeXButton{\big |}{\big |}}%
%BeginExpansion
\big |%
%EndExpansion
f_{l}\left( \mathrm{B},\rho _{l}\right) -\underset{\rho \in E_{\Lambda _{l}}}%
{\inf }f_{l}\left( \mathrm{B},\rho \right)
%TCIMACRO{\TeXButton{\big |}{\big |}}%
%BeginExpansion
\big |%
%EndExpansion
=0
\end{equation*}%
(cf. (\ref{free energy density})--(\ref{neuman entropy})) as well as%
\begin{equation*}
\underset{l\rightarrow \infty }{\lim }\ \underset{\mathfrak{t}\in \mathbb{R}%
^{3}}{\sup }\
%TCIMACRO{\TeXButton{\big |}{\big |}}%
%BeginExpansion
\big |%
%EndExpansion
j_{\rho _{l}}^{(l)}(\mathfrak{t})-\mathrm{j}(\mathfrak{t})%
%TCIMACRO{\TeXButton{\big |}{\big |}}%
%BeginExpansion
\big |%
%EndExpansion
=0\ .
\end{equation*}
\end{theorem}

\noindent \textit{Proof. }The proof is a direct consequence of Lemmata \ref%
{lemma free energy} and \ref{lemma1 current} with, for instance, $\eta
^{\bot }=0.8$ and $\eta =0.95$. Indeed, for any $\mathrm{j}_{1}\in
C_{0}^{\infty }(\mathbb{R}^{3};\mathbb{R})$, we construct in Lemma \ref%
{lemma1 current} a sequence $\{\rho _{l}\}_{l\in \mathbb{N}}$ of
approximating minimizers which creates in the thermodynamic limit a current
density $(\mathrm{j}_{1}(\mathfrak{t}),0,0)$ at rescaled (macroscopic)
position $\mathfrak{t}\in \mathfrak{C}$. In the same way, one constructs a
sequence of approximating minimizers which creates in the thermodynamic
limit a current density $(0,\mathrm{j}_{2}(\mathfrak{t}),0)$ or $(0,0,%
\mathrm{j}_{3}(\mathfrak{t}))$ at $\mathfrak{t}\in \mathfrak{C}$. Using the
convexity of the free--energy density and the affinity of the current
density functional $\rho \mapsto j_{\rho }$ together with a convex
combination of such three approximating minimizers, one proves the
assertion. \hfill $\Box $

\begin{remark}[Dynamical stability of currents]
\label{dynamics}\mbox{ }\newline
The proof of Theorem \ref{coolthml1 copy(2)} is based on the existence of
mesoscopic superconducting domains in the (macroscopic) bulk, see (\ref%
{local superconducting})--(\ref{carlos mini}). Therefore, we expect that the
quantum dynamics rapidly destroys all currents $j_{\rho _{l}}$ in the
non--superconducting phase, even if this dynamics does not change the
free--energy density.
\end{remark}

Theorem \ref{coolthml1 copy(2)} shows that the macroscopic system can create
any smooth current density $\mathrm{j}$ by paying an infinitesimal energy
price in the thermodynamic limit. Indeed, $\{\rho _{l}\}_{l\in \mathbb{N}}$
is a sequence\ of approximating minimizers of the free--energy density
functional $\rho \mapsto f_{l}(\mathrm{B},\rho )$ in the thermodynamic limit
$l\rightarrow \infty $. Therefore, one could use this phenomenon to create\
currents near the surface of the bulk which annihilate the magnetic
induction inside the box $\Lambda _{l}$. This fact suggests the existence of
a Mei{\ss }ner effect for the model under consideration.

Note however that the true minimizer of the free--energy density, i.e., the
Gibbs state (\ref{Gibbs0})--(\ref{Gibbs}), does\emph{\ not} manifest any
current, at least for space--homogeneous magnetic inductions $\mathrm{B}$ on
$\mathfrak{C}$. This can be seen by using the symmetry properties of the
model $H_{l}$. Therefore, one should also take into account the energy
carried by the total magnetic induction, by adding a magnetic term to the
free--energy density. Minimizers of this new magnetic free--energy density
functional do carry currents, in general.

Such a magnetic term is introduced in Section \ref{section Meissner effect},
after the definition of the Biot--Savart operator $\mathcal{S}$ given in the
next section.

\section{The Biot--Savart Operator\label{Section Magnetic Energy}}

The energy contained in the static configuration $\mathrm{B}$ of the total
magnetic induction is given as usual by%
\begin{equation}
E_{\mathrm{mag}}(\mathrm{B}):=\frac{1}{2}\left\Vert \mathrm{B}\right\Vert
_{2}^{2}:=\frac{1}{2}\int_{\mathbb{R}^{3}}\left\vert \mathrm{B}(\mathfrak{t}%
)\right\vert ^{2}\mathrm{d}^{3}\mathfrak{t}\ .  \label{magnetic energy}
\end{equation}%
See, e.g., \cite[Chap. 5, 6]{Vassallo} for an interesting derivation of the
magnetic energy. As a consequence, we only consider magnetic inductions $%
\mathrm{B}$ which belong to the (real) Hilbert space $L^{2}\equiv L^{2}(%
\mathbb{R}^{3};\mathbb{R}^{3})$ with $L^{2}$--norm $\Vert -\Vert _{2}$ and
scalar product defined by
\begin{equation}
\left\langle \mathrm{B}_{1},\mathrm{B}_{2}\right\rangle _{2}:=\int_{\mathbb{R%
}^{3}}\mathrm{B}_{1}\left( \mathfrak{t}\right) \cdot \mathrm{B}_{2}\left(
\mathfrak{t}\right) \mathrm{d}^{3}\mathfrak{t}\ ,\quad \mathrm{B}_{1},%
\mathrm{B}_{2}\in L^{2}\ .  \label{magnetic energybis}
\end{equation}%
A dense set of $L^{2}$\ is of course given by the real vector space $%
C_{0}^{\infty }\equiv C_{0}^{\infty }(\mathbb{R}^{3};\mathbb{R}^{3})$ of
compactly supported smooth fields.

Units have here been chosen so that the magnetic permeability of free space
equals $1$, keeping in mind that the unit of length is already fixed to have
a lattice spacing also equal to $1$. Hence, a static magnetic induction $%
\mathrm{B}$ and a current density $j$ satisfy in our units the Maxwell
equation $\nabla \times \mathrm{B}=j$.

Now therefore, the natural Hilbert space $\mathfrak{H}$ of current densities
is defined as follows:
\begin{equation*}
\left\langle j_{1},j_{2}\right\rangle _{\mathfrak{H}}:=\frac{1}{4\pi }\int_{%
\mathbb{R}^{3}}\int_{\mathbb{R}^{3}}\frac{j_{1}\left( \mathfrak{t}\right)
\cdot j_{2}\left( \mathfrak{s}\right) }{|\mathfrak{t}-\mathfrak{s}|}\mathrm{d%
}^{3}\mathfrak{t}\ \mathrm{d}^{3}\mathfrak{s},\ \ j_{1},j_{2}\in
C_{0}^{\infty },
\end{equation*}%
defines a (energy) scalar product in the real vector space $C_{0}^{\infty }$
of compactly supported smooth current densities. This is easily seen by
using the Fourier transform $F$. In particular, the (magnetic energy) norm
\begin{equation}
\left\Vert j\right\Vert _{\mathfrak{H}}:=\left\langle j,j\right\rangle _{%
\mathfrak{H}}^{1/2}\ ,\quad j\in C_{0}^{\infty }\ ,
\label{magnetic energy norm}
\end{equation}%
clearly satisfies the parallelogram identity and we define the Hilbert space
$\mathfrak{H}\equiv (\mathfrak{H},\left\langle -,-\right\rangle _{\mathfrak{H%
}})$ to be the completion of $(C_{0}^{\infty },\left\langle -,-\right\rangle
_{\mathfrak{H}})$. The divergence--free subspaces of respectively $\mathfrak{%
H}$ and $L^{2}$ are isomorphic as Hilbert spaces. One natural isomorphism is
given by the \emph{Biot--Savart operator }$\mathcal{S}$ defined below.

Observe that the Fourier transform $F$ defines a unitary map from $\mathfrak{%
H}$ to $L^{2}(\mathbb{R}^{3},\left\vert k\right\vert ^{-2}\mathrm{d}^{3}k;%
\mathbb{R}^{3})$. Since
\begin{equation*}
L^{2}(\mathbb{R}^{3},|k|^{-2}\mathrm{d}^{3}k;\mathbb{R}^{3})\hookrightarrow
L^{2}(\mathbb{R}^{3},(|k|^{2}+1)^{-1}\mathrm{d}^{3}k;\mathbb{R}^{3})\ ,
\end{equation*}%
$\mathfrak{H}$ can be seen as a subspace of distributions in $W^{-1,2}(%
\mathbb{R}^{3};\mathbb{R}^{3})$, where $W^{-1,2}(\mathbb{R}^{3};\mathbb{R}%
^{3})$ is the dual of the Sobolev space $W^{1,2}(\mathbb{R}^{3};\mathbb{R}%
^{3})$. The energy interpretation of the expression defining the norm $\Vert
-\Vert _{\mathfrak{H}}$ above is well--known (see (\ref{magnetic energy2})
below). Nevertheless, remark that, at least to our knowledge, the space $%
\mathfrak{H}$ does not seem to have been previously used in a similar
context.

Note further that, for any arbitrary smooth compactly supported field $\Psi
\in C_{0}^{\infty }$, there is a unique (Helmholtz) decomposition $\Psi
=\Psi ^{\parallel }+\Psi ^{\perp }$ with $\Psi ^{\parallel },\Psi ^{\perp
}\in C^{\infty }$ (not necessarily compactly supported), $\nabla \times \Psi
^{\parallel }=0$, $\nabla \cdot \Psi ^{\perp }=0$, and $\Psi ^{\parallel }(%
\mathfrak{t}),\Psi ^{\perp }(\mathfrak{t})\rightarrow 0$, as $|\mathfrak{t}%
|\rightarrow \infty $. Moreover, $\Psi ^{\perp }$ is the curl of some smooth
field whereas $\Psi ^{\parallel }$ is the gradient of a smooth function.
This well--known result is the Helmholtz (decomposition) theorem. See, e.g.,
\cite[Section 9.2, Theorem 3]{Agricola}. The fields $\Psi ^{\parallel }$ and
$\Psi ^{\perp }$ are sometimes called the longitudinal and transverse
components of $\Psi $, respectively.

Indeed, for any $j\in C_{0}^{\infty }$, $j^{\parallel }=P^{\parallel }j$ and
$j^{\bot }=P^{\bot }j$, where $P^{\parallel },P^{\bot }$ are the orthogonal
projections respectively defined in Fourier space by%
\begin{eqnarray}
F[P^{\parallel }j](k) &:=&\frac{k}{\left\vert k\right\vert ^{2}}\ k\cdot F%
\left[ j\right] (k)\ ,  \label{P parallele} \\
F[P^{\bot }j](k) &:=&F\left[ j\right] (k)-\frac{k}{\left\vert k\right\vert
^{2}}\ k\cdot F\left[ j\right] (k)\ ,  \label{P purp}
\end{eqnarray}%
for $k\in \mathbb{R}^{3}$ and all current densities $j$ in the dense subset $%
C_{0}^{\infty }\subset \mathfrak{H}$. Recall that $F$ stands for the Fourier
transform. Straightforward\ computations show that $P^{\parallel }j,P^{\bot
}j\in C^{\infty }$ are smooth functions satisfying%
\begin{equation*}
\nabla \times \lbrack P^{\parallel }j]=0\ ,\text{ \ \ }\nabla \cdot \lbrack
P^{\bot }j]=0\ ,
\end{equation*}%
and $[P^{\parallel }j](\mathfrak{t})$, $[P^{\bot }j](\mathfrak{t}%
)\rightarrow 0$, as $|\mathfrak{t}|\rightarrow \infty $.

We denote again by $P^{\parallel }$ and $P^{\bot }$ the unique orthogonal
projections with ranges being respectively the closures of $P^{\parallel
}C_{0}^{\infty }$ and $P^{\bot }C_{0}^{\infty }$ in $\mathfrak{H}$. The sets
$P^{\parallel }\mathfrak{H}$ and $P^{\bot }\mathfrak{H}$ are clearly
orthogonal. In fact, these projections can still be explicitly defined a.e.
by (\ref{P parallele})--(\ref{P purp}) for any $j\in \mathfrak{H}$. The same
construction can be carried out in $L^{2}$ and so, $P^{\parallel }$ and $%
P^{\bot }$ are also seen as mutually orthogonal projections acting on $L^{2}$%
. In fact,%
\begin{equation}
\mathfrak{H}=P^{\parallel }\mathfrak{H}\oplus P^{\bot }\mathfrak{H}\ ,\quad
L^{2}=P^{\parallel }L^{2}\oplus P^{\bot }L^{2}\ .
\label{projection important}
\end{equation}%
In other words, $P^{\bot }=1-P^{\parallel }$ and $P^{\bot }P^{\parallel
}=P^{\parallel }P^{\bot }=0$ as operators acting either on $\mathfrak{H}$ or
$L^{2}$.

We define now the\emph{\ restricted }Biot--Savart operator $\mathcal{S}_{0}$
on the dense set $C_{0}^{\infty }\subset \mathfrak{H}$ of smooth, compactly
supported current densities $j\in C_{0}^{\infty }$ by
\begin{equation}
\mathcal{S}_{0}(j)\left( \mathfrak{t}\right) :=\frac{1}{4\pi }\int_{\mathbb{R%
}^{3}}\frac{\left( \nabla \times j\right) \left( \mathfrak{s}\right) }{%
\left\vert \mathfrak{t}-\mathfrak{s}\right\vert }\mathrm{d}^{3}\mathfrak{s}\
,\quad \mathfrak{t}\in \mathbb{R}^{3}\ .  \label{biosavat2}
\end{equation}%
In Fourier space, for any $j\in C_{0}^{\infty }$,
\begin{equation}
F[\mathcal{S}_{0}\left( j\right) ](k)=\frac{ik}{\left\vert k\right\vert ^{2}}%
\times F[j](k)\ ,\quad k\in \mathbb{R}^{3}\ .  \label{fourier biot}
\end{equation}%
Using the elementary equality
\begin{equation}
\Theta \times (\Psi \times \Phi )=(\Theta \cdot \Phi )\Psi -(\Theta \cdot
\Psi )\Phi  \label{inequality idiote}
\end{equation}%
together with (\ref{P purp}), we remark that%
\begin{equation}
ik\times F[\mathcal{S}_{0}\left( j\right) ](k)=F[P^{\bot }j](k)\ ,\quad k\in
\mathbb{R}^{3}\ .  \label{Maxwell equation restricted0}
\end{equation}%
In other words, $\mathcal{S}_{0}(j)$ and $j$ satisfy the (generalized)
Maxwell equation
\begin{equation}
\nabla \times \mathcal{S}_{0}(j)=j^{\perp }\ ,\qquad j\in C_{0}^{\infty }\ .
\label{Maxwell equation restricted}
\end{equation}%
Additionally, we infer from (\ref{fourier biot}) that
\begin{equation}
\nabla \cdot \mathcal{S}_{0}(j)=0\ ,\qquad j\in C_{0}^{\infty }\ .
\label{biot savart divergence free}
\end{equation}

The restricted Biot--Savart operator $\mathcal{S}_{0}$ also maps the dense
set $C_{0}^{\infty }\subset \mathfrak{H}$ of current densities to the space $%
L^{2}$ of magnetic inductions. Indeed, for any $j_{1},j_{2}\in C_{0}^{\infty
}$,%
\begin{equation*}
\left\langle \mathcal{S}_{0}(j_{1}),\mathcal{S}_{0}(j_{2})\right\rangle
_{2}=\int_{\mathbb{R}^{3}}\mathcal{S}_{0}(j_{1})(\mathfrak{t})\cdot \lbrack
\nabla \times \mathcal{A}\left( j_{2}\right) (\mathfrak{t})]\mathrm{d}^{3}%
\mathfrak{t}
\end{equation*}%
with the vector potential
\begin{equation}
\mathcal{A}\left( j_{2}\right) (\mathfrak{t}):=\frac{1}{4\pi }\int_{\mathbb{R%
}^{3}}\frac{j_{2}\left( \mathfrak{s}\right) }{\left\vert \mathfrak{t}-%
\mathfrak{s}\right\vert }\mathrm{d}^{3}\mathfrak{s}\ ,\quad \mathfrak{t}\in
\mathbb{R}^{3}\ .  \label{vector potential1}
\end{equation}%
By using the well--known identity%
\begin{equation}
\nabla \cdot (\Psi \times \Phi )=\Phi \cdot \nabla \times \Psi -\Psi \cdot
\nabla \times \Phi  \label{standard identity}
\end{equation}%
for smooth fields $\Psi ,\Phi \in C^{\infty }$, the Maxwell equation (\ref%
{Maxwell equation restricted}), the Gauss Theorem, and decay of $\mathcal{S}%
_{0}(j_{1})(\mathfrak{t})\times \mathcal{A}\left( j_{2}\right) (\mathfrak{t}%
) $ as $|\mathfrak{t}|\rightarrow \infty $, one gets%
\begin{eqnarray*}
\left\langle \mathcal{S}_{0}(j_{1}),\mathcal{S}_{0}(j_{2})\right\rangle _{2}
&=&\int_{\mathbb{R}^{3}}[\nabla \times \mathcal{S}_{0}(j_{1})(\mathfrak{t}%
)]\cdot \mathcal{A}\left( j_{2}\right) (\mathfrak{t})\mathrm{d}^{3}\mathfrak{%
t} \\
&=&\int_{\mathbb{R}^{3}}j_{1}^{\perp }\left( \mathfrak{t}\right) \cdot
\mathcal{A}\left( j_{2}\right) (\mathfrak{t})\mathrm{d}^{3}\mathfrak{t}%
=\left\langle j_{1}^{\perp },j_{2}\right\rangle _{\mathfrak{H}}\ .
\end{eqnarray*}%
The above computation is standard. Since $P^{\parallel }$, $P^{\bot }$ are
mutually orthogonal projections acting on $\mathfrak{H}$, we infer from the
last equality that
\begin{equation}
\left\langle \mathcal{S}_{0}(j_{1}),\mathcal{S}_{0}(j_{2})\right\rangle
_{2}=\left\langle j_{1}^{\perp },j_{2}^{\perp }\right\rangle _{\mathfrak{H}%
},\quad j_{1},j_{2}\in C_{0}^{\infty }\ .  \label{isometry}
\end{equation}%
Therefore, we can extend $\mathcal{S}_{0}$ to a bounded operator $\mathcal{S}
$ acting on $\mathfrak{H}$, named the \emph{Biot--Savart operator}. By (\ref%
{isometry}), the operator $\mathcal{S}$ restricted to $P^{\bot }\mathfrak{H}$
is an isometry:%
\begin{equation}
\left\langle \mathcal{S}(j_{1}),\mathcal{S}(j_{2})\right\rangle
_{2}=\left\langle j_{1}^{\perp },j_{2}^{\perp }\right\rangle _{\mathfrak{H}%
}\ ,\quad j_{1},j_{2}\in \mathfrak{H}\ .  \label{magnetic energy2bis}
\end{equation}%
In particular,%
\begin{equation}
E_{\mathrm{mag}}(\mathcal{S}(j)):=\frac{1}{2}\left\Vert \mathcal{S}%
(j)\right\Vert _{2}^{2}=\frac{1}{2}\left\Vert j^{\bot }\right\Vert _{%
\mathfrak{H}}^{2}\ ,\quad j\in \mathfrak{H}\ .  \label{magnetic energy2}
\end{equation}%
Clearly, $\ker \mathcal{S}=P^{\parallel }\mathfrak{H}$, i.e., $\mathcal{S}%
(j)=\mathcal{S}(j^{\bot })$.

Note that Equality (\ref{fourier biot}) can be extended to all $j\in
\mathfrak{H}$ because the Fourier transform $F$ is a unitary map from $%
\mathfrak{H}$ to $L^{2}(\mathbb{R}^{3},\left\vert k\right\vert ^{-2}\mathrm{d%
}^{3}k;\mathbb{R}^{3})$. In other words, for all $j\in \mathfrak{H}$,
\begin{equation}
F[\mathcal{S}\left( j\right) ](k)=\frac{ik}{\left\vert k\right\vert ^{2}}%
\times F[j](k)\ ,\quad k\in \mathbb{R}^{3}\ \mathrm{(a.e.)}\ .
\label{bio savart fourier}
\end{equation}%
Since (\ref{P purp}) holds on the whole space $\mathfrak{H}$, we can also
extend (\ref{Maxwell equation restricted}) to get
\begin{equation*}
ik\times F[\mathcal{S}\left( j\right) ](k)=F[P^{\bot }j](k)\ ,\quad k\in
\mathbb{R}^{3}\ \mathrm{(a.e.)}\ ,
\end{equation*}%
for all $j\in \mathfrak{H}$, where
\begin{equation*}
F[P^{\bot }j]\in L^{2}(\mathbb{R}^{3},\left\vert k\right\vert ^{-2}\mathrm{d}%
^{3}k;\mathbb{R}^{3})\ .
\end{equation*}%
Consequently, the curl $\nabla \times $, which is seen here as an operator
from the Hilbert space $P^{\bot }L^{2}$ to $P^{\bot }\mathfrak{H}$, defined
in Fourier space by $ik\times $ is the left inverse of $\mathcal{S}$ on the
subspace $P^{\bot }\mathfrak{H}$ of divergence--free currents. In
particular, $\mathcal{S}(j)$ and $j$ satisfy the (generalized) Maxwell
equation
\begin{equation}
\nabla \times \mathcal{S}(j)=j^{\bot }\ ,\qquad j\in \mathfrak{H}\ .
\label{weak curl BS}
\end{equation}

Analogously, one shows that $\nabla \times :P^{\bot }L^{2}\rightarrow
P^{\bot }\mathfrak{H}$ is the right inverse of $\mathcal{S}|_{P^{\bot
}L^{2}} $:
\begin{equation*}
\mathcal{S}(\nabla \times \mathrm{B})=\mathrm{B}\ ,\qquad \mathrm{B}\in
P^{\bot }L^{2}\ .
\end{equation*}%
In particular, by (\ref{magnetic energy2bis}), $\mathcal{S}:P^{\bot }%
\mathfrak{H}\rightarrow P^{\bot }L^{2}$ is an isomorphism of Hilbert spaces.

Similar to (\ref{biot savart divergence free}), one can use (\ref{bio savart
fourier}) to get also that
\begin{equation*}
\nabla \cdot \mathcal{S}(j)=0\ ,\qquad j\in \mathfrak{H}\ ,
\end{equation*}%
i.e., $\mathcal{S}(\mathfrak{H})\subseteq P^{\bot }L^{2}$. Here, $\nabla
\cdot $ is defined in Fourier space by $ik\cdot $ .

For further details on the Biot--Savart operator\emph{\ }we recommend \cite%
{Biot--Savart1} where the latter is studied on the euclidean space $\mathbb{R%
}^{3}$. The results of \cite{Biot--Savart1} are also extended to the
three--dimensional sphere in \cite{Biot--Savart2}. Note, however, that in
\cite{Biot--Savart1,Biot--Savart2} the magnetic induction is restricted to
bounded domains and the energy norm $\Vert -\Vert _{\mathfrak{H}}$ is not
used. For example, in \cite{Biot--Savart1}, the Biot--Savart operator is
seen as a map from the Hilbert space $L^{2}(\Lambda ;\mathbb{R}^{3})$ to $%
L^{2}(\Lambda ;\mathbb{R}^{3})$ with $\Lambda \subset \mathbb{R}^{3}$ and $%
|\Lambda |<\infty $.

\begin{remark}[Vector potentials]
\label{remark projection magn copy(1)}\mbox{ }\newline
To any $j\in C_{0}^{\infty }$ we associate a vector potential $\mathcal{A}%
\left( j\right) \in C^{\infty }$ as defined by (\ref{vector potential1}).
The definition of vector potentials $\mathcal{A}\left( j\right) $ for all $%
j\in \mathfrak{H}$ is given in Section \ref{Vector potentials for all
current densities}. In this case, $\mathcal{A}\left( j\right) $ is not
anymore a function but a distribution, in general.
\end{remark}

\section{Magnetic Free--Energy Density\label{section Meissner effect}}

We fix a smooth external magnetic induction $\mathrm{B}_{\mathrm{ext}}\in
C^{\infty }\equiv C^{\infty }(\mathbb{R}^{3};\mathbb{R}^{3})$ that fulfills
the Maxwell equation $\nabla \cdot \mathrm{B}_{\mathrm{ext}}=0$ and has a
finite magnetic energy:
\begin{equation*}
E_{\mathrm{mag}}(\mathrm{B}_{\mathrm{ext}}):=\frac{1}{2}\left\Vert \mathrm{B}%
_{\mathrm{ext}}\right\Vert _{2}^{2}:=\frac{1}{2}\int_{\mathbb{R}%
^{3}}\left\vert \mathrm{B}_{\mathrm{ext}}\left( \mathfrak{t}\right)
\right\vert ^{2}\mathrm{d}^{3}\mathfrak{t}<\infty \ .
\end{equation*}%
This field results from some fixed divergence--free current density $\mathrm{%
j}_{\mathrm{ext}}$, outside the electron (quantum) system.

More precisely, $\mathrm{B}_{\mathrm{ext}}=\mathcal{S}_{0}(\mathrm{j}_{%
\mathrm{ext}})$ for $\mathrm{j}_{\mathrm{ext}}\in C_{0}^{\infty }\cap
P^{\bot }\mathfrak{H}$ with compact support
\begin{equation}
\mathrm{supp}(\mathrm{j}_{\mathrm{ext}}):=\{\mathfrak{t}\in \mathbb{R}^{3}\
:\ \mathrm{j}_{\mathrm{ext}}(\mathfrak{t})\neq 0\}\subset \mathbb{R}%
\backslash \mathfrak{C}\ .  \label{support condition}
\end{equation}%
Recall that $\mathcal{S}_{0}$ is the restricted Biot--Savart operator
defined by (\ref{biosavat2}), whereas $\mathfrak{C}:=[-1/2,1/2]^{3}$. Note
that the assumption $\mathrm{B}_{\mathrm{ext}}\in C^{\infty }$ is clearly
not necessary since the Biot--Savart operator $\mathcal{S}$ is defined for
all $j\in \mathfrak{H}$. This stronger assumption is only used to simplify
arguments using the usual Maxwell equations.

The external magnetic induction $\mathrm{B}_{\mathrm{ext}}$ can produce a
magnetization density $m_{\rho }^{(l)}$ in the system. The latter can
happen, for instance, if one assumes that $\mathrm{B}=\mathrm{B}_{\mathrm{ext%
}}\neq 0$ in (\ref{Hamiltonian BCS-Hubbard-inhom0}) and $\rho =\mathfrak{g}%
_{l}$ is the corresponding Gibbs state. However, there is no reason for the
magnetic induction $(\mathrm{B}_{\mathrm{ext}}+m_{\rho }^{(l)})$ to satisfy
Gauss's law for magnetism $\nabla \cdot (\mathrm{B}_{\mathrm{ext}}+m_{\rho
}^{(l)})=0$ when $m_{\rho }^{(l)}\neq 0$. On the other hand, in
magnetostatics the total magnetic induction $B_{\rho }^{(l)}$ of the system
must always satisfy $\nabla \cdot B_{\rho }^{(l)}=0$. Observe moreover that
spins of electrons interact with the total magnetic induction within the
material and not only with the external magnetic induction $\mathrm{B}_{%
\mathrm{ext}}$. In other words, one must take $\mathrm{B}=B_{\rho }^{(l)}$
in (\ref{Hamiltonian BCS-Hubbard-inhom0}) and not $\mathrm{B}=\mathrm{B}_{%
\mathrm{ext}}$. Therefore, it is crucial to properly define a total magnetic
induction $B_{\rho }^{(l)}$ of the system satisfying $\nabla \cdot B_{\rho
}^{(l)}=0$ for any state $\rho \in E_{\Lambda _{l}}$.

It is well--known that the magnetization density $m_{\rho }^{(l)}\in
C_{0}^{\infty }$ creates an effective current density, named bound current
density, defined by%
\begin{equation}
\mathfrak{j}_{m_{\rho }}^{(l)}:=\nabla \times m_{\rho }^{(l)}\ .
\label{bound currents}
\end{equation}%
(Recall that units have here been chosen so that the magnetic permeability
of free space equals $1$.)

\begin{remark}[Transverse projection of $m_{\protect\rho }^{(l)}$]
\label{remark projection magn}\mbox{ }\newline
As $\mathfrak{j}_{m_{\rho }}^{(l)}\in C_{0}^{\infty }$, by (\ref{Maxwell
equation restricted}) and \cite[Section 9.2, Theorem 2]{Agricola}, $\mathcal{%
S}_{0}(\mathfrak{j}_{m_{\rho }}^{(l)})$ is the unique smooth
divergence--free field satisfying $\nabla \times \mathcal{S}_{0}(\mathfrak{j}%
_{m_{\rho }}^{(l)})=\mathfrak{j}_{m_{\rho }}^{(l)}$. By definition of the
bound current density $\mathfrak{j}_{m_{\rho }}^{(l)}$, $\mathcal{S}_{0}(%
\mathfrak{j}_{m_{\rho }}^{(l)})=(m_{\rho }^{(l)})^{\perp }$ must be the
transverse component $(m_{\rho }^{(l)})^{\perp }$ of $m_{\rho }^{(l)}\in
C_{0}^{\infty }$. In particular, the longitudinal component $(m_{\rho
}^{(l)})^{\parallel }$ has no effect on the total magnetic induction.
\end{remark}

Using Maxwell--Ampere's law, one then gets that the total magnetic induction
$B_{\rho }^{(l)}$ must satisfy the equation
\begin{equation}
\nabla \times B_{\rho }^{(l)}=J_{\rho }^{(l)}+\frac{\partial D_{\rho }}{%
\partial t}  \label{Maxwell}
\end{equation}%
on $\mathbb{R}^{3}$, where%
\begin{equation}
J_{\rho }^{(l)}:=j_{\rho }^{(l)}+\mathfrak{j}_{m_{\rho }}^{(l)}+\mathrm{j}_{%
\mathrm{ext}}\in C_{0}^{\infty }  \label{total current density}
\end{equation}%
is the total current density and $j_{\rho }^{(l)}$ is the rescaled internal
(free) current density (\ref{rescaled current density}), whereas $D_{\rho }$
is the electric induction produced by the system in the state $\rho \in
E_{\Lambda _{l}}$.

If, at fixed time, $(J_{\rho }^{(l)}+\partial D_{\rho }/\partial t)\in
C_{0}^{\infty }$ then there is a unique $B_{\rho }^{(l)}\in C^{\infty }$
satisfying the Maxwell equations $\nabla \cdot B_{\rho }^{(l)}=0$ and (\ref%
{Maxwell}) with $B_{\rho }^{(l)}(\mathfrak{t})\rightarrow 0$, as $|\mathfrak{%
t}|\rightarrow \infty $. This unique magnetic induction is given, for any $%
\rho \in E_{\Lambda _{l}}$ and $l\in \mathbb{N}$, by%
\begin{equation}
B_{\rho }^{(l)}:=\mathcal{S}_{0}(J_{\rho }^{(l)}+\partial D_{\rho }/\partial
t)\ .  \label{Maxwell1}
\end{equation}%
For more details, we again recommend \cite[Section 9.2, Theorem 2]{Agricola}.

Note that $\nabla \cdot \mathfrak{j}_{m_{\rho }}^{(l)}=0$ because $\mathfrak{%
j}_{m_{\rho }}^{(l)}$ is the curl of $m_{\rho }^{(l)}$, whereas $\nabla
\cdot \mathrm{j}_{\mathrm{ext}}=0$, by assumption. In other words, $(%
\mathfrak{j}_{m_{\rho }}^{(l)})=(\mathfrak{j}_{m_{\rho }}^{(l)})^{\perp }$
and $\mathrm{j}_{\mathrm{ext}}=\mathrm{j}_{\mathrm{ext}}^{\perp }$. See the
discussion in Section \ref{Section Magnetic Energy} about the Helmholtz
theorem. The current density $j_{\rho }^{(l)}\in C_{0}^{\infty }$ (\ref%
{rescaled current density}) has generally a non--trivial decomposition: the
rescaled longitudinal component $(j_{\rho }^{(l)})^{\parallel }$ could be
non--zero, i.e., $\nabla \cdot j_{\rho }^{(l)}\neq 0$. Because of (\ref%
{Maxwell}), one must have
\begin{equation}
0=\left( \frac{\partial D_{\rho }}{\partial t}+J_{\rho }\right) ^{\parallel
}=\left( \frac{\partial D_{\rho }}{\partial t}\right) ^{\parallel }+\left(
j_{\rho }^{(l)}\right) ^{\parallel }\ .  \label{ptt equation}
\end{equation}%
In particular, only the divergence--free part $j_{\rho }^{\perp }$ of the
current density functional $j_{\rho }\in C_{0}^{\infty }$ (\ref%
{carlos_currents}) is relevant with respect to the total magnetic induction $%
B_{\rho }^{(l)}$.

As we are interested in the system at equilibrium, it is natural to consider
the stationary case. In particular, the electric induction $D_{\rho }$
should be constant in time, i.e., we shall assume that $\partial D_{\rho
}/\partial t=0$. As a consequence, $(j_{\rho }^{(l)})^{\parallel }$ should
vanish, to be consistent with (\ref{ptt equation}). Because of the discrete
nature of the system under consideration, $(j_{\rho }^{(l)})^{\parallel }$
is generally not exactly zero at fixed $l\in \mathbb{N}$, but the energy
norm $\Vert (j_{\rho }^{(l)})^{\parallel }\Vert _{\mathfrak{H}}$ is taken
below as arbitrarily small at large $l\in \mathbb{N}$, see (\ref{new set of
states}).

In any case, the total magnetic induction functional is thus defined, for
any $\rho \in E_{\Lambda _{l}}$ and $l\in \mathbb{N}$, by
\begin{equation}
B_{\rho }^{(l)}:=\mathcal{S}_{0}(j_{\rho }^{(l)}+\mathfrak{j}_{m_{\rho
}}^{(l)}+\mathrm{j}_{\mathrm{ext}})\in C^{\infty }\ ,  \label{def.B.rho}
\end{equation}%
see (\ref{Maxwell1}) with $\partial D_{\rho }/\partial t=0$. By definition
of the operator $\mathcal{S}_{0}$, (\ref{def.B.rho}) corresponds to the
Biot--Savart law which gives the total magnetic induction $B_{\rho }^{(l)}$
of the system from the total current density $J_{\rho }^{(l)}$. By
construction, it is an affine map from $E_{\Lambda _{l}}$ to $C_{0}^{\infty
} $ satisfying $\nabla \cdot B_{\rho }^{(l)}=0$, cf. (\ref{biot savart
divergence free}). Using the linearity of the (restricted) Biot--Savart
operator\emph{\ }$\mathcal{S}_{0}$, Remark \ref{remark projection magn}, $%
\mathcal{S}_{0}=\mathcal{S}_{0}P^{\perp }$ and $\mathrm{B}_{\mathrm{ext}}=%
\mathcal{S}_{0}(\mathrm{j}_{\mathrm{ext}})$, we observe that
\begin{equation}
B_{\rho }^{(l)}=\mathcal{S}_{0}((j_{\rho }^{(l)})^{\perp })+(m_{\rho
}^{(l)})^{\perp }+\mathrm{B}_{\mathrm{ext}}  \label{magnetic induction bis}
\end{equation}%
and, by (\ref{Maxwell equation restricted}), its curl equals
\begin{equation*}
\nabla \times B_{\rho }^{(l)}=(J_{\rho }^{(l)})^{\perp }=(j_{\rho
}^{(l)})^{\perp }+\mathfrak{j}_{m_{\rho }}^{(l)}+\mathrm{j}_{\mathrm{ext}}
\end{equation*}%
for any $\rho \in E_{\Lambda _{l}}$ and $l\in \mathbb{N}$.

Using the magnetic energy $E_{\mathrm{mag}}$ defined by (\ref{magnetic
energy}), we then finally introduce a magnetic, finite volume free--energy
density functional $\mathcal{F}_{l}^{(\epsilon )}$ defined, for any $l\in
\mathbb{N}$ and strictly positive parameter $\epsilon \in \mathbb{R}^{+}$,
by
\begin{equation}
\rho \mapsto \mathcal{F}_{l}^{(\epsilon )}(\rho ):=f_{l}(0,\rho )-\langle
B_{\rho }^{(l)},m_{\rho }^{(l)}\rangle _{2}+E_{\mathrm{mag}}(B_{\rho }^{(l)})
\label{magnetic free energy}
\end{equation}%
on the set $E_{\Lambda _{l}}$ of states. Here, $-\langle B_{\rho
}^{(l)},m_{\rho }^{(l)}\rangle _{2}$ is the magnetic interaction energy per
unit of volume, whereas $E_{\mathrm{mag}}(B_{\rho }^{(l)})$ is the magnetic
energy of $B_{\rho }$ per unit of volume. Indeed, by (\ref{magnetic energy}%
)--(\ref{magnetic energybis}),
\begin{equation}
\begin{array}{c}
E_{\mathrm{mag}}(B_{\rho })=|\Lambda _{l}|E_{\mathrm{mag}}(B_{\rho }^{(l)})\
, \\
\langle B_{\rho },m_{\rho }\rangle _{2}=|\Lambda _{l}|\langle B_{\rho
}^{(l)},m_{\rho }^{(l)}\rangle _{2}%
\end{array}
\label{interaction magnetic}
\end{equation}%
are respectively the magnetic energy of $B_{\rho }$ (\ref{magnetic
induction0}) and, up to a minus sign, the magnetic interaction energy with
the system in the state $\rho $. Note that $\mathcal{F}_{l}^{(\epsilon )}$
is defined for all $\epsilon \in \mathbb{R}^{+}$, but we are interested in
the situation where $\epsilon <<\epsilon _{\xi }$ is an arbitrarily small
parameter, see Remark \ref{small eps}.

Our choice of the magnetic interaction energy is consistent with (\ref{bound
currents}) and (\ref{magnetic induction bis}). This interaction energy can
equivalently be seen as a quantum magnetic interaction energy with the
divergence--free magnetic induction $B_{\rho }^{(l)}\ast \xi _{\epsilon }$
as
\begin{equation}
f_{l}(0,\rho )-\langle B_{\rho }^{(l)},m_{\rho }^{(l)}\rangle
_{2}=f_{l}(B_{\rho }^{(l)}\ast \xi _{\epsilon },\rho )
\label{important equality}
\end{equation}%
for any $\epsilon \in \mathbb{R}^{+}$, $l\in \mathbb{N}$ and all states $%
\rho \in E_{\Lambda _{l}}$. Recall that $\xi _{\epsilon }\in C_{0}^{\infty }$
is defined by (\ref{xi eps m}).

Note that a similar conceptual approach based on self--generated magnetic
fields has been recently used in \cite{Solovej} to study electrons in atoms.
See also \cite{Solovej2,Solovej3}.

\begin{remark}[The $E_{\protect\rho }^{(l)}\equiv 0$ assumption]
\label{dielectric}\mbox{ }\newline
Let $E_{\rho }^{(l)}$ be the rescaled static electric field induced by the
system in the state $\rho $. Considering a purely magnetic energy implicitly
corresponds to the situation where $E_{\rho }^{(l)}$ vanishes. One cannot
expect this if the electron density is non--constant in space. One
consequence of our analysis is that the electron density corresponding to
minimizers of $\mathcal{F}_{l}^{(\epsilon )}$ is space--homogeneous within
the superconducting regime, at large $l\in \mathbb{N}$ and small $\epsilon
<<\epsilon _{\xi }$ (cf. Remark \ref{small eps}). See discussions below
Theorem \ref{theorem principal copy(1)} as well as Equation (\ref{electron
density3}) with $h_{\mathfrak{t}}=0$. Thus, the a priori assumption $E_{\rho
}^{(l)}\equiv 0$ is justified within such a phase because of the electric
neutrality of matter. In fact, minimizers of $\mathcal{F}_{l}^{(\epsilon )}$
with space--homogeneous electron density (as $\epsilon \rightarrow 0^{+}$)
must also minimize the electromagnetic free--energy density functional
\begin{equation*}
\rho \mapsto \mathcal{F}_{l}^{(\epsilon )}(\rho )+\frac{\mathrm{\epsilon }%
_{0}}{2}\int_{\mathbb{R}^{3}}|E_{\rho }^{(l)}(\mathfrak{t})|^{2}\ \mathrm{d}%
^{3}\mathfrak{t}
\end{equation*}%
in the limit $\epsilon \rightarrow 0^{+}$. (Here, $\mathrm{\epsilon }_{0}$
is the relative permittivity of free space.) In the non--superconducting
phase, this assumption is generally not consistent with the structure of
minimizers of $\mathcal{F}_{l}^{(\epsilon )}$.
\end{remark}

\section{Thermodynamics of the Mei{\ss }ner Effect\label{section Meissner}}

We analyze now the thermodynamics corresponding to the magnetic free--energy
density functional $\mathcal{F}_{l}^{(\epsilon )}$ defined, for any $l\in
\mathbb{N}$ and $\epsilon \in \mathbb{R}^{+}$, by (\ref{magnetic free energy}%
) on the set $E_{\Lambda _{l}}$ of states. In finite volume, equilibrium
states $\omega _{\epsilon ,l}\in E_{\Lambda _{l}}$ should minimize this
functional.

We expect moreover their associated currents $j_{\omega _{\epsilon
,l}}^{(l)} $ to be divergence--free in the limit $l\rightarrow \infty $.
Indeed, the system should not be able, at thermodynamical equilibrium, to
transmit energy in form of electromagnetic waves, i.e., the generated
electromagnetic field should be static. The latter is consistent with (\ref%
{ptt equation}) and the stationarity assumption $\partial D_{\rho }/\partial
t=0$. As a consequence, we only consider states creating
quasi--divergence--free currents $j_{\rho }^{(l)}$, that is, states which
belong to the set
\begin{equation}
E_{\Lambda _{l}}^{\bot }:=%
%TCIMACRO{\TeXButton{\Big \{}{\Big \{}}%
%BeginExpansion
\Big \{%
%EndExpansion
\rho \in E_{\Lambda _{l}}:\ \Vert (j_{\rho }^{(l)})^{\parallel }\Vert _{%
\mathfrak{H}}\leq l^{-\varkappa }%
%TCIMACRO{\TeXButton{\Big \}}{\Big \}} }%
%BeginExpansion
\Big \}
%EndExpansion
\label{new set of states}
\end{equation}%
for some small, but strictly positive parameter $\varkappa \in \mathbb{R}%
^{+} $. (For instance, take $\varkappa <0.05$ with $\eta ^{\bot }=0.8$ and $%
\eta =0.95$, see proof of Theorem \ref{coolthml1 copy(2)}.) It means in the
thermodynamic limit $l\rightarrow \infty $ that, for all $\rho \in
E_{\Lambda _{l}}^{\bot }$, the current density $j_{\rho }^{(l)}$ is
divergence--free in the sense of the energy norm.

By (\ref{magnetic energy2bis})--(\ref{magnetic energy2}) together with (\ref%
{bound currents}) and (\ref{def.B.rho}) note that, for any $\rho \in
E_{\Lambda _{l}}$ and $l\in \mathbb{N}$,
\begin{equation*}
\mathcal{F}_{l}^{(\epsilon )}(\rho )=f_{l}(0,\rho )-\langle J_{\rho }^{(l)},%
\mathfrak{j}_{m_{\rho }}^{(l)}\rangle _{\mathfrak{H}}+\frac{1}{2}\Vert
(J_{\rho }^{(l)})^{\perp }\Vert _{\mathfrak{H}}^{2}\ ,
\end{equation*}%
where $J_{\rho }^{(l)}$ is the total current density defined by (\ref{total
current density}) with transverse component $(J_{\rho }^{(l)})^{\perp
}=P^{\bot }J_{\rho }^{(l)}$. If $\rho \notin E_{\Lambda _{l}}\backslash
E_{\Lambda _{l}}^{\bot }$ then one should consider the energy of the induced
electric field, but we refrain from considering this case in order to keep
technical aspects as simple as possible. See also Remark \ref{dielectric}.

Therefore, we shall consider the variational problem
\begin{equation}
\mathcal{P}_{l}^{(\epsilon )}:=-\inf_{\rho \in E_{\Lambda _{l}}^{\bot }}%
\mathcal{F}_{l}^{(\epsilon )}(\rho )\ ,\qquad l\in \mathbb{N}\ ,\ \epsilon
\in \mathbb{R}^{+}.  \label{magnetic pressure}
\end{equation}%
The value $\mathcal{P}_{l}^{(\epsilon )}$ is named \emph{finite volume
magnetic pressure} by analogy with (\ref{pressure0}). Recall that the
functionals $\rho \mapsto B_{\rho }^{(l)}$ and $\rho \mapsto m_{\rho }^{(l)}$
are affine. Hence, they are both continuous maps from $E_{\Lambda
_{l}}^{\bot }$ to $L^{2}$, by finite dimensionality of $E_{\Lambda
_{l}}^{\bot }$. The map $\rho \mapsto f_{l}(0,\rho )$ from $E_{\Lambda
_{l}}^{\bot }$ to $\mathbb{R}$ is also continuous. Therefore, the functional
$\mathcal{F}_{l}^{(\epsilon )}$ is continuous on $E_{\Lambda _{l}}^{\bot }$
for every $l\in \mathbb{N}$, $\epsilon \in \mathbb{R}^{+}$, and by
compactness of $E_{\Lambda _{l}}^{\bot }$ and the Weierstra{\ss } theorem,
the set%
\begin{equation}
\mathit{\Omega }_{l}^{(\epsilon )}:=%
%TCIMACRO{\TeXButton{\Big \{}{\Big \{}}%
%BeginExpansion
\Big \{%
%EndExpansion
\omega _{\epsilon ,l}\in E_{\Lambda _{l}}^{\bot }\ :\ \mathcal{F}%
_{l}^{(\epsilon )}\left( \omega _{\epsilon ,l}\right) =\inf\limits_{\rho \in
E_{\Lambda _{l}}^{\bot }}\,\mathcal{F}_{l}^{(\epsilon )}(\rho )%
%TCIMACRO{\TeXButton{\Big \}}{\Big \}} }%
%BeginExpansion
\Big \}
%EndExpansion
\label{finite volume minimizers}
\end{equation}%
of finite volume minimizers is \emph{non--empty} for any $l\in \mathbb{N}$
and $\epsilon \in \mathbb{R}^{+}$. In general, such minimizers are not Gibbs
states $\mathfrak{g}_{l}\in E_{\Lambda _{l}}$ for $\mathrm{B}=B_{\rho
}^{(l)} $, see (\ref{Gibbs0})--(\ref{Gibbs}). Moreover, $\mathcal{F}%
_{l}^{(\epsilon )}$ is a priori not a convex functional on $E_{\Lambda
_{l}}^{\bot }$, i.e., its minimizer may not be unique.

\begin{remark}[Quantum magnetic fields]
\label{small eps copy(2)}\mbox{ }\newline
Considering a quantum (electro--) magnetic field interacting with the
quantum system defined by $H_{l}$, we should obtain a convex free--energy
density functional. In this context, $\mathcal{F}_{l}^{(\epsilon )}$ may be
seen as an approximating free--energy density functional and elements of $%
\mathit{\Omega }_{l}^{(\epsilon )}$ as extreme states of the fully quantum
system, as $l\rightarrow \infty $, $\epsilon \rightarrow 0^{+}$. An analogue
situation is found in \cite{BruPedra2} where $\mathcal{F}_{l}^{(\epsilon )}$
would play the r\^{o}le of \cite[Definition 2.6]{BruPedra2}. In particular,
one shall instead consider the $\Gamma $--regularization of $\mathcal{F}%
_{l}^{(\epsilon )}$ on $E_{\Lambda _{l}}$ and the new set of minimizers
would then be the closed convex hull of $\mathit{\Omega }_{l}^{(\epsilon )}$%
, see \cite[Theorem 1.4]{BruPedraconvex}. However, in order to keep
mathematical aspects as simple as possible, we refrain from considering such
a framework.
\end{remark}

The aim of the present section is to analyze the thermodynamics of the
quantum system under consideration with a self--generated magnetic induction
in relation with the existence of the Mei{\ss }ner effect. To this end, we
first observe that the thermodynamic pressure
\begin{equation}
\mathrm{B}\mapsto p_{\infty }\left( \mathrm{B}\right) :=\underset{%
l\rightarrow \infty }{\lim }p_{l}\left( \mathrm{B}\right) <\infty
\label{thermodynamic limit of the pressure}
\end{equation}%
is a well--defined continuous map from $L^{2}$ to $\mathbb{R}$. Its main
properties are given by Theorem \ref{thm limit pressure}. Similarly, the
thermodynamic limit
\begin{equation*}
\mathcal{P}_{\infty }^{(\epsilon )}:=\underset{l\rightarrow \infty }{\lim }%
\mathcal{P}_{l}^{(\epsilon )}<\infty
\end{equation*}%
of the magnetic pressure $\mathcal{P}_{l}^{(\epsilon )}$ exists for all $%
\epsilon \in \mathbb{R}^{+}$, see Theorem \ref{coolthml1}. It is given by a
variational problem over a closed subspace $\mathcal{B}\subset P^{\bot
}L^{2} $ defined as follows: Consider the set
\begin{equation}
\mathcal{J}:=C_{0}^{\infty }(\mathfrak{C};\mathbb{R}^{3})\cap P^{\bot }%
\mathfrak{H}  \label{biosavat1}
\end{equation}%
of divergence--free smooth current densities supported in $\mathfrak{C}$.
Recall that $\mathcal{S}_{0}$ is the restricted Biot--Savart operator
defined by (\ref{biosavat2}). Then, we denote by
\begin{equation}
\mathcal{B}:=\overline{\mathcal{S}_{0}(\mathcal{J})}\subset P^{\bot }L^{2}
\label{definition B}
\end{equation}%
the closure of the set $\mathcal{S}_{0}(\mathcal{J})\subset C^{\infty }$ in
the weak topology of $L^{2}$.

We focus on the Mei{\ss }ner effect, that is, the existence of
superconducting states $\omega _{\epsilon ,l}\in \mathit{\Omega }%
_{l}^{(\epsilon )}$ with self--generated magnetic inductions $B_{\omega
_{\epsilon ,l}}^{(l)}$ which vanish inside the unit box $\mathfrak{C}$ while
being created by currents supported on the boundary $\partial \mathfrak{C}$
of $\mathfrak{C}$, in the limit $\epsilon \rightarrow 0^{+}$ after $%
l\rightarrow \infty $. Indeed, we analyze in the limit $\epsilon \rightarrow
0^{+}$ the sets $\mathbb{B}_{\epsilon }^{(\pm )}$ defined by%
\begin{equation}
\mathbb{B}_{\epsilon }^{(\pm )}:=\underset{\{\omega _{\epsilon ,l}\}_{l\in
\mathbb{N}}\subset \mathit{\Omega }^{(\epsilon )}}{\bigcup }\mathbb{B}%
_{\epsilon }^{(\pm )}\left( \{\omega _{\epsilon ,l}\}_{l\in \mathbb{N}%
}\right)  \label{set of magnetic inductions}
\end{equation}%
for any $\epsilon \in \mathbb{R}^{+}$. Here, $\mathit{\Omega }^{(\epsilon )}$
is the set of all sequences $\{\omega _{\epsilon ,l}\}_{l\in \mathbb{N}}$
with $\omega _{\epsilon ,l}\in \mathit{\Omega }_{l}^{(\epsilon )}$, and $%
\mathbb{B}_{\epsilon }^{(\pm )}\left( \{\omega _{\epsilon ,l}\}_{l\in
\mathbb{N}}\right) $ are the sets of all weak ($-$) and norm ($+$) cluster
points of $\{B_{\omega _{\epsilon ,l}}^{(l)}\}_{l\in \mathbb{N}}$.

Using Theorem \ref{thm limit pressure} (ii) and $\mathcal{P}_{\infty
}^{(\epsilon )}<\infty $, one verifies the existence of a radius $R\in
\mathbb{R}^{+}$ such that $\Vert B_{\omega _{\epsilon ,l}}^{(l)}\Vert
_{2}\leq R$ for all $\epsilon \in \mathbb{R}^{+}$, $l\in \mathbb{N}$ and $%
\omega _{\epsilon ,l}\in \mathit{\Omega }_{l}^{(\epsilon )}$. Because of the
Banach--Alaoglu theorem and the separability of $L^{2}$, the set $%
\{B_{\omega _{\epsilon ,l}}^{(l)}\}_{l\in \mathbb{N},\omega _{\epsilon
,l}\in \mathit{\Omega }_{l}^{(\epsilon )}}$ is sequentially weak--precompact
and $\mathbb{B}_{\epsilon }^{(-)}$ is not empty. Therefore, one primary aim
is to prove that elements of $\mathbb{B}_{\epsilon }^{(-)}$ can vanish
inside the unit box $\mathfrak{C}$ while being created by currents supported
on the boundary $\partial \mathfrak{C}$ of $\mathfrak{C}$, in the limit $%
\epsilon \rightarrow 0^{+}$.

As explained in Remark \ref{remark Smooth from discrete}, we take the limit $%
\epsilon \rightarrow 0^{+}$ after $l\rightarrow \infty $ to avoid any
arbitrariness. We thus define the infinite volume magnetic pressure by
\begin{equation}
\mathcal{P}_{\infty }:=\underset{\epsilon \rightarrow 0^{+}}{\lim }\mathcal{P%
}_{\infty }^{(\epsilon )}<\infty \ .  \label{perfect magnetic pressure}
\end{equation}%
This pressure exists and equals:

\begin{theorem}[Thermodynamics]
\label{theorem principal}\mbox{ }\newline
Let $\mathrm{B}_{\mathrm{ext}}=\mathcal{S}_{0}(\mathrm{j}_{\mathrm{ext}})$
with $\mathrm{j}_{\mathrm{ext}}\in C_{0}^{\infty }\cap P^{\bot }\mathfrak{H}$%
. Then : \newline
\emph{(i)} The infinite volume magnetic pressure equals
\begin{equation*}
\mathcal{P}_{\infty }=\underset{\mathrm{B}\in \mathcal{B}}{\sup }\left\{ -%
\frac{1}{2}\left\Vert \mathrm{B}+\mathrm{B}_{\mathrm{ext}}\right\Vert
_{2}^{2}+p_{\infty }\left( \mathrm{B}+\mathrm{B}_{\mathrm{ext}}\right)
\right\} \ .
\end{equation*}%
\emph{(ii)} For any $\epsilon \in \mathbb{R}^{+}$,
\begin{equation*}
\mathbb{B}_{\epsilon }:=\mathbb{B}_{\epsilon }^{(+)}=\mathbb{B}_{\epsilon
}^{(-)}\neq \emptyset \ .
\end{equation*}%
\emph{(iii)} For any family $\{\mathrm{B}_{\epsilon }\}_{\epsilon \in
\mathbb{R}^{+}}$ with $\mathrm{B}_{\epsilon }\in \mathbb{B}_{\epsilon }$,
\begin{equation*}
\underset{\epsilon \rightarrow 0^{+}}{\lim }\left\{ -\frac{1}{2}\Vert
\mathrm{B}_{\epsilon }+\mathrm{B}_{\mathrm{ext}}\Vert _{2}^{2}+p_{\infty }(%
\mathrm{B}_{\epsilon }+\mathrm{B}_{\mathrm{ext}})\right\} =\mathcal{P}%
_{\infty }\ .
\end{equation*}
\end{theorem}

\noindent \textit{Proof. }(i) is Corollary \ref{corollary de funes}.
(ii)--(iii) result from Lemma \ref{lemma var prob3 copy(12)} and Corollary %
\ref{lemma var prob3 copy(1)}.\hfill $\Box $

\begin{remark}[Existence of maximizer(s)]
\label{small eps copy(1)}\mbox{ }\newline
As explained in Section \ref{section Self--Generated}, the variational
problem $\mathcal{P}_{\infty }$ could have no maximizer. Indeed, the map
\begin{equation}
\mathrm{B}\mapsto \mathcal{G}\left( \mathrm{B}\right) :=-\frac{1}{2}%
\left\Vert \mathrm{B}+\mathrm{B}_{\mathrm{ext}}\right\Vert
_{2}^{2}+p_{\infty }\left( \mathrm{B}+\mathrm{B}_{\mathrm{ext}}\right)
\label{map+sup}
\end{equation}%
from $\mathcal{B}$\ to $\mathbb{R}$ is neither concave nor upper
semi--continuous in the weak topology. However, under certain conditions, we
show in Theorem \ref{theorem principal copy(1)} that $\mathcal{G}$ has a
unique maximizer $\mathrm{B}_{\mathrm{int}}\in \mathcal{B}$.
\end{remark}

We prove now the Mei{\ss }ner effect at large enough inverse temperatures $%
\beta >>1$ and large BCS couplings $\gamma >>1$, \ i.e., in presence of a
superconducting phase defined as follow: Consider the annihilation and
creation operators
\begin{equation*}
\mathfrak{c}_{0}:=\frac{1}{\left\vert \Lambda _{l}\right\vert ^{1/2}}%
\sum_{x\in \Lambda _{l}}a_{x,\downarrow }a_{x,\uparrow }\ \ \text{and }\
\mathfrak{c}_{0}^{\ast }:=\frac{1}{\left\vert \Lambda _{l}\right\vert ^{1/2}}%
\sum_{x\in \Lambda _{l}}a_{x,\uparrow }^{\ast }a_{x,\downarrow }^{\ast }
\end{equation*}%
of Cooper pairs within the condensate, i.e., in the zero--mode for electron
pairs. A \emph{superconducting} phase is then characterized by a strictly
positive (global) Cooper pair condensate density for all minimizers in the
thermodynamic limit, that is,
\begin{equation}
\mathbf{r}_{\beta }:=\underset{\epsilon \rightarrow 0^{+}}{\lim }\ \underset{%
l\rightarrow \infty }{\lim \inf }\left\{ \underset{\omega _{\epsilon ,l}\in
\mathit{\Omega }_{l}^{(\epsilon )}}{\inf }\omega _{\epsilon ,l}\left( \frac{%
\mathfrak{c}_{0}^{\ast }\mathfrak{c}_{0}}{\left\vert \Lambda _{l}\right\vert
}\right) \right\} >0\ .  \label{limit00}
\end{equation}%
This inequality corresponds to the existence of an off--diagonal long range
order. The domain of parameters $(\beta ,\mu ,\lambda ,\gamma ,\mathrm{B}_{%
\mathrm{ext}})$ where $\mathbf{r}_{\beta }$ is strictly positive is
non--empty. At sufficiently large inverse temperatures $\beta >>1$, the
latter holds for instance when $\mu <-\vartheta ^{2}$ and $\gamma >|\mu
-\lambda |\Gamma _{0}$ with%
\begin{equation*}
\Gamma _{0}:=\frac{4}{1-\vartheta ^{2}|\mu |^{-1}}>4\ .
\end{equation*}%
See Theorem \ref{lemma var prob3 copy(5)}. The Mei{\ss }ner effect appears
in this regime:

\begin{theorem}[Mei\ss ner effect]
\label{theorem principal copy(1)}\mbox{ }\newline
Let $\mu <-\vartheta ^{2}$, $\gamma >|\mu -\lambda |\Gamma _{0}$ and $%
\mathrm{B}_{\mathrm{ext}}=\mathcal{S}_{0}(\mathrm{j}_{\mathrm{ext}})$ with $%
\mathrm{j}_{\mathrm{ext}}\in C_{0}^{\infty }\cap P^{\bot }\mathfrak{H}$.
Then, there is $\beta _{0}\in \mathbb{R}^{+}$ such that, for all $\beta
>\beta _{0}$: \newline
\emph{(i)} For any sequence of minimizers $\omega _{\epsilon ,l}\in \mathit{%
\Omega }_{l}^{(\epsilon )}$,%
\begin{equation*}
\underset{\epsilon \rightarrow 0^{+}}{\lim }\liminf_{l\rightarrow \infty
}\omega _{\epsilon ,l}\left( \frac{\mathfrak{c}_{0}^{\ast }\mathfrak{c}_{0}}{%
\left\vert \Lambda _{l}\right\vert }\right) =\underset{\epsilon \rightarrow
0^{+}}{\lim }\limsup_{l\rightarrow \infty }\omega _{\epsilon ,l}\left( \frac{%
\mathfrak{c}_{0}^{\ast }\mathfrak{c}_{0}}{\left\vert \Lambda _{l}\right\vert
}\right) =\mathrm{r}_{\beta }(0)
\end{equation*}%
with
\begin{equation}
\mathrm{r}_{\beta }(0)\geq \Gamma _{0}^{-2}-\gamma ^{-2}\left( \mu -\lambda
\right) ^{2}>0  \label{inequality r beta}
\end{equation}%
being the unique solution of (\ref{var prob principal}) for $\mathrm{B}=0$.%
\newline
\emph{(ii)} For any sequence of minimizers $\omega _{\epsilon ,l}\in \mathit{%
\Omega }_{l}^{(\epsilon )}$,%
\begin{equation*}
\underset{\epsilon \rightarrow 0^{+}}{\lim }\ \limsup_{l\rightarrow \infty
}\Vert B_{\omega _{\epsilon ,l}}^{(l)}-\mathrm{B}_{\mathrm{int}}\Vert _{2}=0
\end{equation*}%
with $\mathrm{B}_{\mathrm{int}}\in \mathcal{B}$ being the unique maximizer
of the variational problem $\mathcal{P}_{\infty }$. See Theorem \ref{theorem
principal} (i).\newline
\emph{(iii)} The total magnetic induction vanishes inside the unit box $%
\mathfrak{C}$: $\mathrm{B}_{\mathrm{int}}+\mathrm{B}_{\mathrm{ext}}=0$ a.e.
in $\mathfrak{C}$. \newline
\emph{(iv)} If (\ref{support condition}) also holds then the self--generated
magnetic induction $\mathrm{B}_{\mathrm{int}}=\mathcal{S}(\mathrm{j}_{%
\mathrm{int}}^{\bot })$ is produced by some current $\mathrm{j}_{\mathrm{int}%
}^{\bot }\in \overline{\mathcal{J}}$ that is supported on the boundary $%
\partial \mathfrak{C}$ of $\mathfrak{C}$.
\end{theorem}

\noindent \textit{Proof. }(i) is Theorem \ref{lemma var prob3 copy(5)}
(iii). By Theorem \ref{lemma var prob3 copy(5)} (ii), note that $\mathrm{B}_{%
\mathrm{int}}\in \mathcal{B}$ is the unique maximizer of the variational
problem $\mathcal{P}_{\infty }$. By using Theorems \ref{theorem principal}
(ii)--(iii), \ref{thm limit pressure} (ii), \ref{lemma var prob3 copy(5)}
(i) and a simple contradiction argument, we prove the second assertion (ii).
Finally, (iii)--(iv) are consequences of Theorem \ref{lemma var prob3
copy(5)} (i)--(ii) and Lemmata \ref{lemma var prob3 copy(2)}--\ref{lemma var
prob3 copy(3)}. \hfill $\Box $

By Theorem \ref{theorem principal copy(1)}, the electron density
corresponding to minimizers of $\mathcal{F}_{l}^{(\epsilon )}$ is
space--homogeneous within the superconducting regime, in the limit $\epsilon
\rightarrow 0^{+}$ after $l\rightarrow \infty $. Indeed, the Mei{\ss }ner
effect corresponds here to the absence of magnetic induction inside the unit
box $\mathfrak{C}$, except within a $\epsilon $--neighborhood of the
boundary $\partial \mathfrak{C}$ of $\mathfrak{C}$. Therefore, in the limit $%
\epsilon \rightarrow 0^{+}$, the electron density $\mathrm{d}_{\beta }$,
defined for all $\mathfrak{t}\in \mathfrak{C}$ (a.e.) by (\ref{electron
density3}) for a magnetic induction $(\mathrm{B}_{\mathrm{int}}+\mathrm{B}_{%
\mathrm{ext}})|_{\mathfrak{C}}=0$, is constant in this case. This argument
justifies a posteriori the use of a purely magnetic energy in (\ref{magnetic
free energy}). See Remark \ref{dielectric}.

By Lemmata \ref{lemma var prob3 copy(2)}\ and \ref{lemma var prob3 copy(9)},
observe finally that the Euler--Lagrange equations associated with $\mathcal{%
P}_{\infty }$ yield the equality%
\begin{equation}
\mathrm{B}_{0}+\mathrm{B}_{\mathrm{ext}}=\mathrm{M}_{\beta }^{\bot }(\mathrm{%
B}_{0}+\mathrm{B}_{\mathrm{ext}})\equiv \mathrm{M}_{\beta }^{\bot }\quad
\text{a.e. in }\mathfrak{C}  \label{cool euler}
\end{equation}%
for any maximizer $\mathrm{B}_{0}\in \mathcal{B}$ of the variational problem
$\mathcal{P}_{\infty }$. Here, we denote as usual by $\mathrm{M}_{\beta
}^{\bot }=P^{\bot }\mathrm{M}_{\beta }$ the transverse component of the
(infinite volume) magnetization density $\mathrm{M}_{\beta }\equiv \mathrm{M}%
_{\beta }(\mathrm{B})$ defined a.e. on $\mathbb{R}^{3}$ for every $\mathrm{B}%
\in L^{2}$ by%
\begin{equation}
\mathrm{M}_{\beta ,\mathfrak{t}}\equiv \mathrm{M}_{\beta ,\mathfrak{t}}(%
\mathrm{B}):=\frac{\mathbf{1}[\mathfrak{t}\in \mathfrak{C}]\ \vartheta \sinh
\left( \beta h_{\mathfrak{t}}\right) }{\cosh \left( \beta h_{\mathfrak{t}%
}\right) +\mathrm{e}^{-\beta \lambda }\cosh \left( \beta g_{\mathrm{r}%
_{\beta }}\right) }\frac{\mathrm{B}\left( \mathfrak{t}\right) }{\left\vert
\mathrm{B}\left( \mathfrak{t}\right) \right\vert }  \label{magnetization0}
\end{equation}%
with $\mathrm{r}_{\beta }\in \lbrack 0,1/4)$ being solution of the
variational problem (\ref{var prob principal}), $g_{r}:=\{(\mu -\lambda
)^{2}+\gamma ^{2}r\}^{1/2}$ and $h_{\mathfrak{t}}:=\vartheta \left\vert
\mathrm{B}\left( \mathfrak{t}\right) \right\vert $ a.e. in the unit box $%
\mathfrak{C}$. By (\ref{magnetization00})--(\ref{magnetization}), $\mathrm{M}%
_{\beta }(\mathrm{B})$ is indeed the magnetization density if one applies a
fixed magnetic induction $\mathrm{B}$ on the system. See Section \ref{Fixed
Magnetic Induction} for more details.

Recall that units have been chosen so that the magnetic permeability of free
space equals $1$. As a consequence, Equation (\ref{cool euler}) implies that
the so--called magnetic field%
\begin{equation*}
\mathrm{H}:=\mathrm{B}_{0}+\mathrm{B}_{\mathrm{ext}}-\mathrm{M}_{\beta
}^{\bot }\in L^{2}
\end{equation*}%
is zero within the quantum system, i.e., $\mathrm{H}|_{\mathfrak{C}}=0$ a.e.
in the unit cubic box $\mathfrak{C}$. The latter is satisfied for every
maximizer $\mathrm{B}_{0}$ and in the whole phase diagram (not only in the
regime where the Mei{\ss }ner effect appears).

This suggests that $\mathrm{M}_{\beta }^{\bot }=-\mathrm{H}$ trivially holds
in the superconducting phase because $\mathrm{M}_{\beta }^{\bot }=0=\mathrm{H%
}$, see Theorem \ref{theorem principal copy(1)}. In other words, the Mei{\ss
}ner effect is not necessarily related to an effective magnetic
susceptibility equal to $-1$. A similar remark can be done about the
magnetic permeability of the quantum system in the superconducting phase.

\section{Universality of the Mei\ss ner effect\label{Section Universality}}

\noindent \textbf{1.} Conditions of Theorem \ref{theorem principal copy(1)}
are only sufficient and clearly not necessary for the Mei\ss ner effect.
See, e.g., Lemma \ref{lemma var prob3 copy(8)}. In particular, the
inequalities $\mu <-\vartheta ^{2}$ and $\gamma >|\mu -\lambda |\Gamma _{0}$
are far from being essential. The inequality $\gamma >2|\mu -\lambda |$ is
however necessary to get a superconducting phase. Therefore, Theorem \ref%
{theorem principal copy(1)} should be seen as an example where the
thermodynamics of the Mei\ss ner effect is rigorously proven from first
principles of quantum mechanics. \smallskip

\noindent \textbf{2.} Of course, the model $H_{l}:=T_{l}+\mathcal{M}_{l}$
under consideration is too simplified with respect to real superconductors,
as explained for instance in \cite{BruPedraAniko}. However, by combining
Grassmann integration, Brydges--Kennedy tree expansions and determinant
bounds, one should be able to show that the more realistic model%
\begin{equation}
H_{l,\mathfrak{e}}:=H_{l}+\sum\limits_{x,y\in \Lambda _{l}}\mathfrak{e}%
(x-y)\left( a_{x,\downarrow }^{\ast }a_{y,\downarrow }+a_{x,\uparrow }^{\ast
}a_{y,\uparrow }\right) \ ,  \label{model kinetic}
\end{equation}%
with hopping amplitude $\mathfrak{e}:\mathbb{Z}^{3}\rightarrow \mathbb{R}$
satisfying $\mathfrak{e}\left( -x\right) =\mathfrak{e}\left( x\right) $ and
\begin{equation*}
\left\Vert \mathfrak{e}\right\Vert _{1}:=\sum\limits_{x\in \mathbb{Z}%
^{3}}\left\vert \mathfrak{e}\left( x\right) \right\vert <\infty \ ,
\end{equation*}%
has essentially the same correlation functions as $H_{l}$ at low
temperatures, up to corrections of order $\Vert \mathfrak{e}\Vert _{1}$.
Indeed, by extending all the notation to the model $H_{l,\mathfrak{e}}$, one
gets the following generalization of Theorem \ref{theorem principal copy(1)}:

\begin{theorem}[Mei{\ss }ner effect at small hopping amplitude]
\label{Theorem inv trans}\mbox{ }\newline
Fix $\mu <-\vartheta ^{2}$, $\beta _{0}\in \mathbb{R}^{+}$ and $\mathrm{B}_{%
\mathrm{ext}}=\mathcal{S}_{0}(\mathrm{j}_{\mathrm{ext}})$ with $\mathrm{j}_{%
\mathrm{ext}}\in C_{0}^{\infty }\cap P^{\bot }\mathfrak{H}$. Then, there is $%
\gamma _{0}>\left\vert \mu \right\vert \Gamma _{0}^{+}$ such that, for all $%
\gamma \in \lbrack \gamma _{0},\infty )$, $\beta \in \lbrack \beta
_{0},2\beta _{0}]$, $\lambda \in \mathbb{R}$ and hopping amplitude $%
\mathfrak{e}$ such that $\lambda $ and $\Vert \mathfrak{e}\Vert _{1}$ are
sufficiently small, Assertions (i)--(iv) of Theorem \ref{theorem principal
copy(1)} hold true with
\begin{equation*}
\mathrm{r}_{\beta }(0)\geq \frac{1}{3}\left( \Gamma _{0}^{-2}-\gamma
^{-2}\mu ^{2}\right) >0
\end{equation*}%
replacing Inequality (\ref{inequality r beta}).
\end{theorem}

\noindent \textit{Proof. }A sketch of the proof is given in Section \ref%
{Section inv transl}. Various (partial) technical results used in that
section can be proven in a much more general setting. Therefore, we will
give the full proofs in separate papers.\hfill $\Box $

\begin{remark}[Mei{\ss }ner effect at small hopping amplitude and zero
temperature]
\label{dielectric copy(1)}\mbox{ }\newline
We conjecture that the above theorem holds true for all $\beta \geq \beta
_{0}$. Indeed, by using small/large (magnetic) field decompositions to
handle the variational problem $\mathfrak{B}_{\infty }^{(\epsilon )}$, one
shows that, for all $\epsilon \in \mathbb{R}^{+}$ and large $\gamma $,
\begin{multline*}
-\Vert \mathrm{B}_{\mathrm{int}}+\mathrm{B}_{\mathrm{ext}}\Vert
_{2}^{2}-\inf_{\mathrm{B}\in \mathcal{B}}\left\{ \Vert \mathrm{B}\Vert
_{2}^{2}-p_{l}\left( \mathfrak{T}_{\epsilon }\left( \mathrm{B}+\mathrm{B}_{%
\mathrm{int}}+\mathrm{B}_{\mathrm{ext}}\right) \right) \right\} \leq
\underset{l\rightarrow \infty }{\lim \inf }\mathcal{P}_{l}^{(\epsilon )}\leq
\underset{l\rightarrow \infty }{\lim \sup }\mathcal{P}_{l}^{(\epsilon )} \\
\leq -\frac{1}{2}\Vert \mathrm{B}_{\mathrm{int}}+\mathrm{B}_{\mathrm{ext}%
}\Vert _{2}^{2}-\inf_{\mathrm{B}\in \mathcal{B}}\left\{ \frac{1}{2}\Vert
\mathrm{B}\Vert _{2}^{2}-p_{l}\left( \mathfrak{T}_{\epsilon }\left( \mathrm{B%
}+\mathrm{B}_{\mathrm{int}}+\mathrm{B}_{\mathrm{ext}}\right) \right)
\right\} \ ,
\end{multline*}%
even in the presence of (small) hopping terms. Compare with Theorem \ref%
{coolthml1}. At large $\gamma $, this yields%
\begin{equation*}
\underset{\epsilon \rightarrow 0^{+}}{\lim }\underset{l\rightarrow \infty }{%
\lim \inf }\mathcal{P}_{l}^{(\epsilon )}=\underset{\epsilon \rightarrow 0^{+}%
}{\lim }\underset{l\rightarrow \infty }{\lim \sup }\mathcal{P}%
_{l}^{(\epsilon )}=\mathfrak{B}_{\infty }^{(0)}\ .
\end{equation*}%
On the other hand, Corollary \ref{corollary de funes copy(1)} says that $%
\mathrm{B}_{\mathrm{int}}$ is a critical point of the map $\mathcal{G}$ (\ref%
{map+sup}) from $\mathcal{B}$\ to $\mathbb{R}$ for all $\beta \geq \beta
_{0} $\ at large $\gamma $. We expect that, for large enough $\gamma $, $%
\mathrm{B}_{\mathrm{int}}$ is a global minimum of $\mathcal{G}$ and Theorem %
\ref{Theorem inv trans} would thus follow for $\beta \geq \beta _{0}$, by
the same arguments as in the special case $\beta \in \lbrack \beta
_{0},2\beta _{0}]$.
\end{remark}

Therefore, the Hamiltonian $H_{l}$ is a good model for certain kinds of
superconductors or ultra--cold Fermi gases in optical lattices for which the
strong coupling regime is justified.

Additionally, a similar study could have been done for the usual (reduced)
BCS model. In this model, the (screened) Coulomb repulsion is neglected,
i.e., $\lambda =0$, but the kinetic energy is taken into account:

\begin{theorem}[Mei{\ss }ner effect for $\protect\lambda =0$]
\label{Theorem inv trans copy(1)}\mbox{ }\newline
Fix $\mu <-\vartheta ^{2}$, $\lambda =0$, any hopping amplitude $\mathfrak{e}
$ such that $\Vert \mathfrak{e}\Vert _{1}<\infty $, and $\mathrm{B}_{\mathrm{%
ext}}=\mathcal{S}_{0}(\mathrm{j}_{\mathrm{ext}})$ with $\mathrm{j}_{\mathrm{%
ext}}\in C_{0}^{\infty }\cap P^{\bot }\mathfrak{H}$. Then, there are $\gamma
_{0},\beta _{0}\in \mathbb{R}^{+}$ such that, for all $\gamma \in \lbrack
\gamma _{0},\infty )$ and $\beta \in \lbrack \beta _{0},\infty )$, the
statements of Theorem \ref{Theorem inv trans} hold true.
\end{theorem}

\noindent \textit{Proof. }We follow Points 1--10 of Section \ref{Section inv
transl} and adapt them for this particular case: Points 1--4 and 7--8 are
exactly the same as in Section \ref{Section inv transl}. In Points 5--6 and
9--10 one uses the uniqueness of KMS states of quasi--free systems, which
yields the differentiability of the pressure limit $\mathrm{\tilde{p}}_{%
\mathfrak{e}}$ and continuity of the corresponding derivatives with respect
to the parameters. Recall that in the case $\lambda \neq 0$ these properties
follow from tree expansions and determinant bounds. In Point 10 we use that
the magnetization density $\mathrm{M}_{\beta }\equiv \mathrm{M}_{\beta }(%
\mathrm{B})$ is in the present case defined a.e. on $\mathbb{R}^{3}$ for
every $\mathrm{B}\in L^{2}$ by%
\begin{equation*}
\mathrm{M}_{\beta ,\mathfrak{t}}\equiv \mathrm{M}_{\beta ,\mathfrak{t}}(%
\mathrm{B}):=\frac{\mathbf{1}[\mathfrak{t}\in \mathfrak{C}]\vartheta }{%
2\left( 2\pi \right) ^{3}}\int_{\left[ -\pi ,\pi \right] ^{3}}\frac{\sinh
\left( \beta h_{\mathfrak{t}}\right) }{\cosh \left( \beta h_{\mathfrak{t}%
}\right) +\cosh \left( \beta \sqrt{(\mu -\mathfrak{\hat{e}}_{k})^{2}+\gamma
^{2}\mathrm{r}_{\beta }}\right) }\frac{\mathrm{B}\left( \mathfrak{t}\right)
}{\left\vert \mathrm{B}\left( \mathfrak{t}\right) \right\vert }\mathrm{d}%
^{3}k
\end{equation*}%
with magnetic strength $h_{\mathfrak{t}}:=\vartheta \left\vert \mathrm{B}%
\left( \mathfrak{t}\right) \right\vert $ a.e. for $\mathfrak{t}\in \mathfrak{%
C}$. Here, $\mathrm{r}_{\beta }$ is defined by (\ref{var blilbi}). \hfill $%
\Box $

\noindent \textbf{3.} In fact, the Mei\ss ner effect is directly related to
the existence of states minimizing the free--energy and having small
magnetization densities at fixed magnetic induction $\mathrm{B}$. It is the
case in our models within the superconducting regime where the magnetization
density $\mathrm{M}_{\beta }$ is generally exponentially small in the limit $%
\beta \rightarrow \infty $. See, e.g., (\ref{magnetizationbis}). Indeed, it
is only necessary to verify the inequality $\Vert \mathrm{M}_{\beta }(%
\mathrm{B})\Vert _{2}\leq \mathbf{m}\Vert \mathrm{B}\Vert _{2}$ with $%
\mathbf{m}<1$ and that the system can produce currents without increasing
the free--energy density in the thermodynamic limit. Then, assuming this
phenomenon to happen in real superconductors,\ details of the model are not
that important anymore. The self--generated magnetic induction $\mathrm{B}_{%
\mathrm{int}}=\mathcal{S}(\mathrm{j}_{\mathrm{int}}^{\bot })$ is in this
case the unique solution of the variational problem
\begin{equation}
\mathfrak{A}:=\frac{1}{2}\inf_{\mathrm{B}\in \mathcal{B}}\ \left\Vert
\mathrm{B}+\mathrm{B}_{\mathrm{ext}}\right\Vert _{2}^{2}\ ,
\label{var prob temp zero-0}
\end{equation}%
whereas the corresponding divergence--free current density $\mathrm{j}_{%
\mathrm{int}}^{\bot }$ is the unique minimizer of%
\begin{equation}
\mathfrak{J}:=\frac{1}{2}\inf_{j^{\bot }\in \overline{\mathcal{J}}}\
\left\Vert j^{\bot }+\mathrm{j}_{\mathrm{ext}}\right\Vert _{\mathfrak{H}%
}^{2}=\mathfrak{A}\ ,  \label{var prob temp zerobis-0}
\end{equation}%
where $\overline{\mathcal{J}}$ is the (norm) closure of the set $\mathcal{J}$
defined by (\ref{biosavat1}). See Equation (\ref{magnetic energy2}) and
Theorem \ref{lemma var prob3 copy(5)}. These variational problems are
studied in Lemmata \ref{lemma var prob3 copy(2)} and \ref{lemma var prob3
copy(3)}. See also the corresponding Euler--Lagrange equations (\ref{eq
tempezero1}) and (\ref{euler courants}).

Both variational problems can certainly be numerically studied in detail and
the resulting (self--generated) magnetic induction $\mathrm{B}_{\mathrm{int}%
} $ will correspond to the usual pictures found in textbooks on
superconductors to illustrate the Mei\ss ner effect. The study of $\mathfrak{%
A}$ and $\mathfrak{J}$ may moreover be of relevance in completely different
contexts, like in fluid dynamics where currents and magnetic inductions are
respectively replaced by vortex lines and velocity fields.

\begin{remark}[General superconducting domains]
\mbox{ }\newline
Results of Lemmata \ref{lemma var prob3 copy(2)} and \ref{lemma var prob3
copy(3)} as well as the Euler--Lagrange equations (\ref{eq tempezero1}) and (%
\ref{euler courants}) can be straightforwardly extended to all external
magnetic inductions $\mathrm{B}_{\mathrm{ext}}=\mathcal{S}(\mathrm{j}_{%
\mathrm{ext}})$ with $\mathrm{j}_{\mathrm{ext}}\in P^{\bot }\mathfrak{H}$,
and all bounded domains $\mathfrak{C}\subset \mathbb{R}^{3}$ with (for
example) piecewise smooth boundary $\partial \mathfrak{C}$.
\end{remark}

\noindent \textbf{4.} Observe that a suppression of the magnetic induction
in the box $\mathfrak{C}$ by minimizers of the magnetic free--energy density
can also appear at small enough inverse temperatures $\beta <\vartheta ^{-1}$%
, see Lemma \ref{lemma var prob3 copy(7)}. Indeed, for high temperatures,
the pressure $p_{\infty }\left( \mathrm{B}\right) $ mainly comes from its
entropic part and so, it does not depend much on the magnetic induction $%
\mathrm{B}$ in this regime. In particular, the magnetization density $%
\mathrm{M}_{\beta }$ becomes again small. On the other hand, as explained
above, the minimizer of $\mathfrak{A}$ vanishes inside the box $\mathfrak{C}$%
. See again Lemmata \ref{lemma var prob3 copy(2)}--\ref{lemma var prob3
copy(3)}.

From the physical point of view, the appearance of such a phenomenon at high
temperature is however questionable. Indeed, our analysis is based on the
possibility for the system to create any current density, see Theorem \ref%
{coolthml1 copy(2)}. It is proven by using patches of superconducting phases
with negligible volume (with respect to $\left\vert \Lambda \right\vert $).
As explained in Remark \ref{dynamics}, such a configuration should be
rapidly destroyed by the quantum dynamics of the system at (high)
temperatures where no global superconducting phase exists. In other words,
our results are physically well--founded for sufficiently low temperatures
where we can ensure the existence of a (global) superconducting phase
defined by $\mathbf{r}_{\beta }>0$ and for sufficiently weak external
magnetic inductions $\mathrm{B}_{\mathrm{ext}}$. Indeed, in order to
suppress magnetic inductions $\mathrm{B}_{\mathrm{ext}}$, the quantum system
has to produce currents via superconducting patches close to the boundary $%
\partial \mathfrak{C}$. Such patches are however rapidly destabilized by a
too strong magnetic induction $\mathrm{B}_{\mathrm{ext}}$ that succeeds to
penetrate the region close to $\partial \mathfrak{C}$. The latter is
suggested by Equation (\ref{local cooper pairs}). Indeed, this equation
shows that, for any $\mathrm{B}$ satisfying $|\mathrm{B}_{\mathfrak{t}}|\geq
\vartheta ^{-1}\mathrm{h}>\vartheta ^{-1}\mathrm{h}_{\mathrm{c}}$ a.e. on a
non--empty open set $\mathfrak{D}\subset \mathfrak{C}$, the local Cooper
pair condensate density $\mathrm{r}_{\beta ,\mathfrak{D}}=\mathcal{O}(%
\mathrm{e}^{-\beta (\mathrm{h}-\mathrm{h}_{\mathrm{c}})})$\ in the region $%
\mathfrak{D}$ must be exponentially small, as $\beta \rightarrow \infty $.
In particular, no superconducting current can be created within $\mathfrak{D}
$.

\section{Technical proofs\label{meisneer sect 2}}

\subsection{Vector Potentials\label{Vector potentials for all current
densities}}

We start this section by defining the vector potential $\mathcal{A}\left(
j\right) $ for all $j\in \mathfrak{H}$. It will become important while
proving the Mei{\ss }ner effect.

The vector potential $\mathcal{A}\left( j\right) $ associated with any
current density $j\in \mathfrak{H}$ is the distribution defined by
\begin{equation}
\mathcal{A}\left( j\right) \left( \varphi \right) :=\langle j,\varphi
\rangle _{\mathfrak{H}}\in \mathbb{R}\ ,\qquad \varphi \in C_{0}^{\infty }\ .
\label{vector potential10}
\end{equation}%
For any $j\in C_{0}^{\infty }$, $\mathcal{A}\left( j\right) $ can be seen as
a $C^{\infty }$--function, as usual. See (\ref{vector potential1}). For
convenience, we ignore this distinction and write%
\begin{equation}
\mathcal{A}\left( j\right) (\mathfrak{t})\equiv \frac{1}{4\pi }\int_{\mathbb{%
R}^{3}}\frac{j\left( \mathfrak{s}\right) }{\left\vert \mathfrak{t}-\mathfrak{%
s}\right\vert }\mathrm{d}^{3}\mathfrak{s}\ ,\quad \mathfrak{t}\in \mathbb{R}%
^{3}\ ,  \label{vector potential2}
\end{equation}%
for all $j\in C_{0}^{\infty }$ and
\begin{equation}
\mathcal{A}\left( j\right) \left( \varphi \right) \equiv \langle \mathcal{A}%
\left( j\right) ,\varphi \rangle _{2}\ ,\qquad j,\varphi \in C_{0}^{\infty
}\ .  \label{vector potential2bis}
\end{equation}

Using Fourier transform, one verifies the following for the weak Laplace
operator applied on $\mathcal{A}\left( j\right) $:
\begin{equation}
\lbrack -\Delta \mathcal{A}\left( j\right) ](\varphi ):=\mathcal{A}\left(
j\right) \left( -\Delta \varphi \right) =j\left( \varphi \right)
\label{vector potential3}
\end{equation}%
for $j\in \mathfrak{H}$ and $\varphi \in C_{0}^{\infty }$. Indeed, recall
that $\mathfrak{H}$ can be seen as a subspace of (tempered) distributions in
$W^{-1,2}(\mathbb{R}^{3};\mathbb{R}^{3})$. Similarly, the curl of the vector
potential distribution $\mathcal{A}\left( j\right) $ equals%
\begin{equation*}
\nabla \times \mathcal{A}\left( j\right) =\mathcal{S}(j^{\bot })\ ,\qquad
j\in \mathfrak{H}\ ,
\end{equation*}%
in the weak sense. See, e.g., (\ref{bio savart fourier}).

\subsection{Thermodynamics at Fixed Magnetic Induction\label{Fixed Magnetic
Induction}}

For all inverse temperatures $\beta \in \mathbb{R}^{+}$, we associate to the
Hamiltonian $H_{l}:=T_{l}+\mathcal{M}_{l}\in \mathcal{U}_{\Lambda _{l}}$ the
finite volume pressure $p_{l}\left( \mathrm{B}\right) $ defined by (\ref%
{pressure0}) at fixed magnetic induction $\mathrm{B}\in L^{2}(\mathfrak{C};%
\mathbb{R}^{3})$ with $\mathfrak{C}:=[-1/2,1/2]^{3}$. Its thermodynamic
limit $p_{\infty }\left( \mathrm{B}\right) $ is explicitly computed in two
main steps.

The first step consists in assuming that $\mathrm{B}\in C^{0}(\mathfrak{C};%
\mathbb{R}^{3})$ is a continuous magnetic induction in order to get the
pressure $p_{\infty }\left( \mathrm{B}\right) $ by using \cite[Theorem 4.1]%
{BruPedra-homog}. The infinite volume pressure $p_{\infty }\left( \mathrm{B}%
\right) $ is then given by the maximization of a functional $\mathfrak{F}$
defined on $\mathbb{R}_{0}^{+}$ by%
\begin{equation}
\mathfrak{F}(r)\equiv \mathfrak{F}\left( r,\mathrm{B}\right) :=\mu +\beta
^{-1}\ln 2-\gamma r+\beta ^{-1}\int_{\mathfrak{C}}\ln \mathrm{Trace}_{\wedge
\mathcal{H}_{\{0\}}}(\mathrm{e}^{-\beta u\left( r,\mathfrak{t}\right) })%
\mathrm{d}^{3}\mathfrak{t}  \label{expliciti function1}
\end{equation}%
with the one--site Hamiltonian defined by%
\begin{equation}
u\left( r,\mathfrak{t}\right) :=-\mu (n_{0,\uparrow }+n_{0,\downarrow
})+2\lambda n_{0,\uparrow }n_{0,\downarrow }-\gamma \sqrt{r}(a_{0,\uparrow
}^{\ast }a_{0,\downarrow }^{\ast }+a_{0,\downarrow }a_{0,\uparrow })-\mathrm{%
B}\left( \mathfrak{t}\right) \cdot \mathrm{M}^{0}  \label{one site hamil}
\end{equation}%
for all $r\in \mathbb{R}_{0}^{+}$ (i.e., $r\geq 0$) and $\mathfrak{t}\in
\mathfrak{C}$ (a.e.). Here, $\mathrm{M}^{0}:=(\mathrm{m}_{1}^{0},\mathrm{m}%
_{2}^{0},\mathrm{m}_{3}^{0})$ is defined via (\ref{magne1}). Indeed, one has:

\begin{lemma}[Pressure for continuous fields]
\label{lemma var pro}\mbox{ }\newline
For any $\mathrm{B}\in C^{0}(\mathfrak{C};\mathbb{R}^{3})$,
\begin{equation*}
p_{\infty }\left( \mathrm{B}\right) :=\underset{l\rightarrow \infty }{\lim }%
p_{l}\left( \mathrm{B}\right) =\underset{r\geq 0}{\sup }\ \mathfrak{F}(r,%
\mathrm{B})<\infty \ .
\end{equation*}
\end{lemma}

\noindent \textit{Proof. }If $\mathrm{B}\in C^{0}(\mathfrak{C};\mathbb{R}%
^{3})$ is a continuous magnetic induction then one can replace in $\mathcal{M%
}_{l}$ (\ref{Hamiltonian BCS-Hubbard-inhom0}) the mean value (\ref{mean
value}) of the magnetic induction $\mathrm{B}$ with $\mathrm{B}(x/(2l))$, in
order to compute the pressure $p_{\infty }\left( \mathrm{B}\right) $. The
latter results from \cite[Eq. (3.11)]{BruPedra2} and straightforward
estimates using the uniform continuity of $\mathrm{B}$ on the compact set $%
\mathfrak{C}$. Then, using the gauge symmetry of the model as well as a
change of variable $r=\gamma \tilde{r}$ in the variational problem given by
\cite[Theorem 4.1]{BruPedra-homog}, we arrive at the assertion. \hfill $\Box
$

The second step uses the density of the set $C^{0}(\mathfrak{C};\mathbb{R}%
^{3})$ in $L^{2}(\mathfrak{C};\mathbb{R}^{3})$ to compute $p_{\infty }\left(
\mathrm{B}\right) $ for all $\mathrm{B}\in L^{2}(\mathfrak{C};\mathbb{R}%
^{3}) $. Combined with Lemma \ref{lemma var pro} it leads to an explicit
expression of $p_{\infty }\left( \mathrm{B}\right) $ for all $\mathrm{B}\in
L^{2}(\mathfrak{C};\mathbb{R}^{3})$. This is resumed in the following
theorem which serves as a springboard to the rest of the paper.

\begin{theorem}[Infinite volume pressure -- I]
\label{thm limit pressure}\mbox{ }\newline
\emph{(i)} For $\mathrm{B}\in L^{2}(\mathfrak{C};\mathbb{R}^{3})$, the
pressure $p_{l}\left( \mathrm{B}\right) $ converges to
\begin{equation*}
p_{\infty }\left( \mathrm{B}\right) :=\underset{l\rightarrow \infty }{\lim }%
p_{l}\left( \mathrm{B}\right) =\underset{r\geq 0}{\sup }\ \mathfrak{F}\left(
r,\mathrm{B}\right) <\infty
\end{equation*}%
with $\mathfrak{F}$ defined by (\ref{expliciti function1})--(\ref{one site
hamil}). See also (\ref{expliciti function2}).\newline
\emph{(ii)} The family $\left\{ \mathrm{B}\mapsto p_{l}\left( \mathrm{B}%
\right) \right\} _{l\in \mathbb{N\cup }\left\{ \infty \right\} }$ of maps
from $L^{2}(\mathfrak{C};\mathbb{R}^{3})$ to $\mathbb{R}$ is uniformly
Lipschitz equicontinuous: For all $l\in \mathbb{N}\cup \left\{ \infty
\right\} $,
\begin{equation}
|p_{l}(\mathrm{B})-p_{l}(\mathrm{C})|\leq 2\sqrt{3}\vartheta \Vert \mathrm{B}%
-\mathrm{C}\Vert _{2}\ ,\qquad \mathrm{B},\mathrm{C}\in L^{2}(\mathfrak{C};%
\mathbb{R}^{3})\text{ }.  \label{L toto}
\end{equation}
\end{theorem}

\noindent \textit{Proof. }The uniform Lipschitz equicontinuity (\ref{L toto}%
) of the family $\left\{ \mathrm{B}\mapsto p_{l}\left( \mathrm{B}\right)
\right\} _{l\in \mathbb{N}}$ follows from (\ref{Hamiltonian
BCS-Hubbard-inhom0}) and (\ref{pressure0}) together with the Cauchy--Schwarz
inequality, $|\mathfrak{C}|=1$, $\Vert \rho \Vert =1$ and $\Vert \mathrm{m}%
_{j}^{x}\Vert \leq 2\vartheta $ for any $j\in \{1,2,3\}$ and all $x\in
\mathbb{Z}^{3}$, see (\ref{magne1}). As a consequence, by Lemma \ref{lemma
var pro} and the density of $C^{0}(\mathfrak{C};\mathbb{R}^{3})$ in the
separable Hilbert space $L^{2}(\mathfrak{C};\mathbb{R}^{3})$, the pressure $%
p_{l}\left( \mathrm{B}\right) $ converges to some value $p_{\infty }\left(
\mathrm{B}\right) \in \mathbb{R}$ for all $\mathrm{B}\in L^{2}(\mathfrak{C};%
\mathbb{R}^{3})$ and (\ref{L toto}) is also satisfied for $l=\infty $. In
particular, for any $\mathrm{B}\in L^{2}(\mathfrak{C};\mathbb{R}^{3})$,
there exists a sequence $\{\mathrm{B}^{(n)}\}_{n\in \mathbb{N}}\subset C^{0}(%
\mathfrak{C};\mathbb{R}^{3})$ converging in norm to $\mathrm{B}$ such that
\begin{equation}
p_{\infty }\left( \mathrm{B}\right) =\underset{n\rightarrow \infty }{\lim }%
p_{\infty }(\mathrm{B}^{(n)})=\underset{n\rightarrow \infty }{\lim }\left\{
\underset{r\geq 0}{\sup }\ \mathfrak{F}(r,\mathrm{B}^{(n)})\right\} \ .
\label{perssure1}
\end{equation}%
On the other hand, by \cite[Eq. (3.11)]{BruPedra2}, one easily verifies
that, for any $\mathrm{B},\mathrm{C}\in L^{2}(\mathfrak{C};\mathbb{R}^{3})$
and $\mathfrak{t}\in \mathfrak{C}$ (a.e.),%
\begin{equation}
\left\vert \ln \mathrm{Trace}_{\wedge \mathcal{H}_{\{0\}}}(\mathrm{e}%
^{-\beta u\left( r,\mathfrak{t},\mathrm{B}\left( \mathfrak{t}\right) \right)
})-\ln \mathrm{Trace}_{\wedge \mathcal{H}_{\{0\}}}(\mathrm{e}^{-\beta
u\left( r,\mathfrak{t},\mathrm{C}\left( \mathfrak{t}\right) \right)
})\right\vert \leq 2\sqrt{3}\vartheta \beta \left\vert \mathrm{B}\left(
\mathfrak{t}\right) -\mathrm{C}\left( \mathfrak{t}\right) \right\vert
\label{perssure1bis}
\end{equation}
with $u\left( r,\mathfrak{t}\right) \equiv u\left( r,\mathfrak{t},\mathrm{B}%
\left( \mathfrak{t}\right) \right) $ being defined by (\ref{one site hamil}%
). By (\ref{expliciti function1}) combined with the Cauchy--Schwarz
inequality and $|\mathfrak{C}|=1$, it follows that%
\begin{equation}
\left\vert \mathfrak{F}\left( r,\mathrm{B}\right) -\mathfrak{F}\left( r,%
\mathrm{C}\right) \right\vert \leq 2\sqrt{3}\vartheta \Vert \mathrm{B}-%
\mathrm{C}\Vert _{2}\ ,\quad \mathrm{B},\mathrm{C}\in L^{2}(\mathfrak{C};%
\mathbb{R}^{3})\ .  \label{continuity of FracF}
\end{equation}%
Combined with the limits (\ref{perssure1}), this last inequality in turn
implies Lemma \ref{lemma var pro} extended to all $\mathrm{B}\in L^{2}(%
\mathfrak{C};\mathbb{R}^{3})$, that is, (i) holds. \hfill $\Box $

The functional $\mathfrak{F}$ can explicitly be computed and one gets%
\begin{equation}
\mathfrak{F}(r)\equiv \mathfrak{F}(r,\mathrm{B})=\mu +\beta ^{-1}\ln
2-\gamma r+\beta ^{-1}\int_{\mathfrak{C}}\ln \left\{ \cosh \left( \beta h_{%
\mathfrak{t}}\right) +\mathrm{e}^{-\lambda \beta }\cosh \left( \beta
g_{r}\right) \right\} \mathrm{d}^{3}\mathfrak{t}  \label{expliciti function2}
\end{equation}%
with $g_{r}:=\{(\mu -\lambda )^{2}+\gamma ^{2}r\}^{1/2}$ for any $r\in
\mathbb{R}_{0}^{+}$ and magnetic strength $h_{\mathfrak{t}}:=\vartheta
\left\vert \mathrm{B}\left( \mathfrak{t}\right) \right\vert $ a.e. for $%
\mathfrak{t}\in \mathfrak{C}$. Its properties can thus be studied in detail,
exactly as in \cite[Section 7]{BruPedra1}.

In particular, for any $\beta ,\gamma ,\vartheta \in \mathbb{R}^{+}$, real
numbers $\mu ,\lambda \in \mathbb{R}\ $and $\mathrm{B}\in L^{2}(\mathfrak{C};%
\mathbb{R}^{3})$, it is clear that the supremum of the variational problem
in Theorem \ref{thm limit pressure} (i) is reached for an order parameter $%
r\in \mathbb{R}_{0}^{+}$ in some bounded set. In particular, there is always
$\mathrm{r}_{\beta }\equiv \mathrm{r}_{\beta }\left( \mathrm{B}\right) \in
\mathbb{R}_{0}^{+}$ such that%
\begin{equation}
\underset{r\geq 0}{\sup }\ \mathfrak{F}(r,\mathrm{B})=\mathfrak{F}(\mathrm{r}%
_{\beta },\mathrm{B})\ ,\quad \mathrm{B}\in L^{2}(\mathfrak{C};\mathbb{R}%
^{3})\ .  \label{var prob principal}
\end{equation}%
Up to (special) points $(\beta ,\mu ,\lambda ,\gamma ,\vartheta ,\mathrm{B})$
corresponding to a phase transition of first order, $\mathrm{r}_{\beta }$
should always be unique and continuous with respect to each parameter.

For small inverse temperatures $\beta <<1$, $\mathrm{r}_{\beta }=0$. See
arguments of \cite[Sections 2 and 7]{BruPedra1}.\ On the other hand, any
non--zero solution $\mathrm{r}_{\beta }$ of the variational problem of
Theorem \ref{thm limit pressure} (i) has to be solution of the gap equation
(or Euler--Lagrange equation):%
\begin{equation}
\int_{\mathfrak{C}}\frac{\sinh \left( \beta g_{\mathrm{r}_{\beta }}\right) }{%
\mathrm{e}^{\beta \lambda }\cosh \left( \beta h_{\mathfrak{t}}\right) +\cosh
\left( \beta g_{\mathrm{r}_{\beta }}\right) }\mathrm{d}^{3}\mathfrak{t}=%
\frac{2g_{\mathrm{r}_{\beta }}}{\gamma }\ .  \label{gap equation}
\end{equation}%
Because $\tanh (t)\leq 1$ for $t\in \mathbb{R}_{0}^{+},$ we then conclude
that
\begin{equation}
\mathrm{r}_{\beta }\leq \max \left\{ 0,\mathrm{r}_{\max }\right\} \mathrm{\
\ }\text{with}\mathrm{\ \ r}_{\max }:=\frac{1}{4}-\gamma ^{-2}\left( \mu
-\lambda \right) ^{2}.  \label{bounded r}
\end{equation}%
In particular, if $\gamma \leq 2|\mu -\lambda |$ then $\mathrm{r}_{\beta }=0$
for any $\beta \in \mathbb{R}^{+}$. However, at fixed $\beta ,\lambda ,\mu
,\vartheta ,\mathrm{B}$, there is $\gamma _{c}>2|\lambda -\mu |$ such that $%
\mathrm{r}_{\beta }>0$ for any $\gamma \geq \gamma _{c}$. The latter can
easily be seen like in \cite[Section 7]{BruPedra1}. In other words, the
domain of parameters $(\beta ,\mu ,\lambda ,\gamma ,\vartheta ,\mathrm{B})$
where $\mathrm{r}_{\beta }\in \mathbb{R}^{+}$ is non--empty.

To illustrate this, we give a regime where $\mathrm{r}_{\beta }$ becomes
strictly positive for sufficiently low temperatures and large BCS couplings:

\begin{lemma}[Superconducting phase -- I]
\label{lemma var prob3 copy(6)}\mbox{ }\newline
Let $R\in \mathbb{R}^{+}$, $\mu <-R\vartheta $ and $\gamma >\left\vert \mu
-\lambda \right\vert \Gamma _{0}$ with
\begin{equation*}
\Gamma _{0}:=\frac{4}{1-R\vartheta |\mu |^{-1}}>4\ .
\end{equation*}%
Then, there is $\beta _{0}\in \mathbb{R}^{+}$ such that, for all $\beta
>\beta _{0}$,
\begin{equation*}
\underset{\mathrm{B}\in b_{R}(0)}{\inf }\mathrm{r}_{\beta }(\mathrm{B})\geq
\Gamma _{0}^{-2}-\gamma ^{-2}\left( \mu -\lambda \right) ^{2}>0
\end{equation*}%
with
\begin{equation*}
b_{R}\left( 0\right) :=\left\{ \mathrm{B}\in L^{2}(\mathfrak{C};\mathbb{R}%
^{3}):\left\Vert \mathrm{B}\right\Vert _{2}\leq R\right\} \ .
\end{equation*}%
Moreover, we can choose $\beta _{0}\equiv \beta _{0}(\gamma )$ as a
decreasing function of $\gamma $.
\end{lemma}

\noindent \textit{Proof. }For any $\mathrm{B}\in b_{R}(0)$, note that $\Vert
\mathrm{B}\Vert _{1}\leq R$, by the Cauchy--Schwarz inequality. Then, for
any $\mu <0$, the set%
\begin{equation*}
\mathfrak{D}_{\mu }:=\left\{ \mathfrak{t}\in \mathfrak{C}:|h_{\mathfrak{t}%
}|\geq |\mu |\right\}
\end{equation*}%
satisfies $|\mathfrak{D}_{\mu }|\leq R\vartheta |\mu |^{-1}$ for all $%
\mathrm{B}\in b_{R}(0)$. For every $\mu <-R\vartheta $ and $\varepsilon \in
\mathbb{R}^{+}$, let
\begin{equation}
\Gamma _{\varepsilon }:=(1+\varepsilon )\Gamma _{0}>0\ .  \label{GammaGamma}
\end{equation}%
Since $\gamma >\left\vert \mu -\lambda \right\vert \Gamma _{0}$, we can
choose $\varepsilon \in \mathbb{R}^{+}$\ such that $\gamma >\left\vert \mu
-\lambda \right\vert \Gamma _{\varepsilon }$. It follows that%
\begin{equation}
\mathfrak{r}_{\varepsilon }:=\Gamma _{\varepsilon }^{-2}-\gamma ^{-2}\left(
\mu -\lambda \right) ^{2}>0\ .  \label{rminin}
\end{equation}%
Then, by (\ref{GammaGamma}), there is $\beta _{0}\in \mathbb{R}^{+}$ such
that, for all $\beta >\beta _{0}$,%
\begin{equation*}
\frac{\tanh \left( \beta g_{\mathfrak{r}_{\varepsilon }}\right) }{g_{%
\mathfrak{r}_{\varepsilon }}}=\frac{\Gamma _{\varepsilon }}{\gamma }\tanh
\left( \frac{\beta \gamma }{\Gamma _{\varepsilon }}\right) >\frac{%
4+\varepsilon }{\gamma \left( 1-R\vartheta |\mu |^{-1}\right) }\ .
\end{equation*}%
Observe that $\beta _{0}\equiv \beta _{0}(\gamma )$ can be taken as a
decreasing function of $\gamma $. The function $t^{-1}\tanh \left( \beta
t\right) $ is decreasing on $\mathbb{R}_{0}^{+}$. Therefore, we deduce from
the last inequality that
\begin{equation}
\left( 1-R\vartheta |\mu |^{-1}\right) \frac{\tanh \left( \beta g_{r}\right)
}{g_{r}}>\frac{4+\varepsilon }{\gamma }  \label{debile0}
\end{equation}%
for any $r\in \lbrack 0,\mathfrak{r}_{\varepsilon }]$, all $\beta >\beta
_{0} $ and fixed $\varepsilon \in \mathbb{R}^{+}$ such that\ $\gamma
>\left\vert \mu -\lambda \right\vert \Gamma _{\varepsilon }$.

Now, we compute that $\partial _{r}\mathfrak{F}\left( r,\mathrm{B}\right) >0$
is equivalent to%
\begin{equation*}
\int_{\mathfrak{C}}\frac{\sinh \left( \beta g_{r}\right) }{g_{r}\left(
\mathrm{e}^{\lambda \beta }\cosh \left( \beta h_{\mathfrak{t}}\right) +\cosh
\left( \beta g_{r}\right) \right) }\mathrm{d}^{3}\mathfrak{t}>\frac{2}{%
\gamma }\ .
\end{equation*}%
Using $|\mathfrak{D}_{\mu }|\leq R\vartheta |\mu |^{-1}$ and (\ref{debile0}),%
\begin{equation}
\int_{\mathfrak{C}}\frac{\sinh \left( \beta g_{r}\right) }{g_{r}\left(
\mathrm{e}^{\lambda \beta }\cosh \left( \beta h_{\mathfrak{t}}\right) +\cosh
\left( \beta g_{r}\right) \right) }\mathrm{d}^{3}\mathfrak{t}>\frac{%
4+\varepsilon }{2\gamma }  \label{debile1}
\end{equation}%
for any $\beta >\beta _{0}$, all $\mathrm{B}\in b_{R}(0)$, and $r\in \lbrack
0,\mathfrak{r}_{\varepsilon }]$. In particular,
\begin{equation*}
\partial _{r}\mathfrak{F}\left( r,\mathrm{B}\right) >0\ ,\qquad r\in \lbrack
0,\mathfrak{r}_{\varepsilon }]\ ,
\end{equation*}%
which yields $\mathrm{r}_{\beta }(\mathrm{B})\geq \mathfrak{r}_{\varepsilon
}>0$ for any $\beta >\beta _{0}$ and all magnetic inductions $\mathrm{B}\in
b_{R}(0)$. \hfill $\Box $

By using Griffiths arguments (see, e.g., \cite[Eq. (A.1)]{BruPedra1}) away
from critical points (defined by the existence of a first order phase
transition), one finds that, for any non--empty open region $\mathfrak{D}%
\subseteq \mathfrak{C}$, the Cooper pair condensate density%
\begin{equation*}
\mathrm{r}_{\beta ,\mathfrak{D}}:=\underset{l\rightarrow \infty }{\lim }\
\frac{1}{\left\vert \mathfrak{D}_{l}\right\vert ^{2}}\sum_{x,y\in \mathfrak{D%
}_{l}}\mathfrak{g}_{l}\left( a_{x,\uparrow }^{\ast }a_{x,\downarrow }^{\ast
}a_{y,\downarrow }a_{y,\uparrow }\right)
\end{equation*}%
with $\mathfrak{D}_{l}:=2l\mathfrak{D}\cap \Lambda _{l}$ equals
\begin{equation}
\mathrm{r}_{\beta ,\mathfrak{D}}=\frac{1}{\left\vert \mathfrak{D}\right\vert
}\int_{\mathfrak{D}}\mathrm{r}_{\beta ,\mathfrak{t}}\ \mathrm{d}^{3}%
\mathfrak{t}\in \lbrack 0,1/4]\ .  \label{local cooper pairs}
\end{equation}%
(Note that $\mathfrak{D}_{l}$ is non--empty for sufficiently large $l\in
\mathbb{N}$). Here, for all $\mathfrak{t}\in \mathfrak{C}$ (a.e.),%
\begin{equation}
\mathrm{r}_{\beta ,\mathfrak{t}}:=\frac{\gamma \mathrm{r}_{\beta }\sinh
\left( \beta g_{\mathrm{r}_{\beta }}\right) }{2g_{\mathrm{r}_{\beta }}\left(
\mathrm{e}^{\beta \lambda }\cosh \left( \beta h_{\mathfrak{t}}\right) +\cosh
\left( \beta g_{\mathrm{r}_{\beta }}\right) \right) }\ .
\label{local cooper pairs t}
\end{equation}%
The inequality $\mathrm{r}_{\beta ,\mathfrak{D}}\leq 1/4$ results from (\ref%
{gap equation}) and (\ref{bounded r}). In particular, for $\mathfrak{D}=%
\mathfrak{C}$,%
\begin{equation}
\mathrm{r}_{\beta ,\mathfrak{C}}\equiv \mathrm{r}_{\beta }\equiv \mathrm{r}%
_{\beta }\left( \mathrm{B}\right)  \label{cooper pair condensate density}
\end{equation}%
is the (global) Cooper pair condensate density, see (\ref{var prob principal}%
)--(\ref{gap equation}). When $\mathrm{r}_{\beta }\in \mathbb{R}^{+}$ is the
unique solution of the variational problem of (\ref{var prob principal}),
one obtains a s--wave superconducting phase with off--diagonal long range
order. As an example, see \cite[Theorems 3.1--3.3]{BruPedra1}.

In a similar way, we compute the three components
\begin{equation}
\mathrm{M}_{\beta ,\mathfrak{D}}:=\underset{l\rightarrow \infty }{\lim }\
\frac{1}{\left\vert \mathfrak{D}_{l}\right\vert }\sum_{x\in \mathfrak{D}_{l}}%
\mathfrak{g}_{l}\left( \mathrm{M}^{x}\right) \in \mathbb{R}^{3}
\label{magnetization00}
\end{equation}%
of the magnetization densities in the non--empty open region $\mathfrak{D}%
\subset \mathfrak{C}$. See (\ref{magne1}) and (\ref{magne2}). Away from
critical points,%
\begin{equation}
\mathrm{M}_{\beta ,\mathfrak{D}}=\frac{1}{\left\vert \mathfrak{D}\right\vert
}\int_{\mathfrak{D}}\mathrm{M}_{\beta ,\mathfrak{t}}\ \mathrm{d}^{3}%
\mathfrak{t}\in \lbrack -\vartheta ,\vartheta ]^{3}  \label{magnetization}
\end{equation}%
with $\vartheta \in \mathbb{R}^{+}$ and $\mathrm{M}_{\beta ,\mathfrak{t}%
}\equiv \mathrm{M}_{\beta ,\mathfrak{t}}\left( \mathrm{B}\right) $ defined
by (\ref{magnetization0}).

In particular, in the limit ($\beta \rightarrow \infty $) of low
temperatures,%
\begin{equation}
\left\vert \mathrm{M}_{\beta ,\mathfrak{D}}\right\vert =\mathcal{O}(\mathrm{e%
}^{-\beta (\mathrm{h}_{\mathrm{c}}-\mathrm{h})})\quad \text{and}\quad
\mathrm{r}_{\beta ,\mathfrak{D}}=\mathcal{O}(\mathrm{r}_{\beta })
\label{magnetizationbis}
\end{equation}%
whenever, for all $\mathfrak{t}\in \mathfrak{D}$ (a.e.),
\begin{equation}
h_{\mathfrak{t}}\leq \mathrm{h}<\mathrm{h}_{\mathrm{c}}\equiv \mathrm{h}_{%
\mathrm{c}}\left( \mathrm{B}\right) :=g_{\mathrm{r}_{\beta }}-\lambda
\label{magnetizationbisbis}
\end{equation}%
with $g_{\mathrm{r}_{\beta }}:=\{(\mu -\lambda )^{2}+\gamma ^{2}\mathrm{r}%
_{\beta }\}^{1/2}$. However, a strong and local magnetic induction such that
$h_{\mathfrak{t}}\geq \mathrm{h}>\mathrm{h}_{\mathrm{c}}$ (a.e.) on some
non--empty open set $\mathfrak{D}\subseteq \mathfrak{C}$ implies a strong
magnetization on $\mathfrak{D}_{l}:=2l\mathfrak{D}\cap \Lambda _{l}$, even
if a global superconducting phase exists, that is, even if $\mathrm{r}%
_{\beta }\in \mathbb{R}^{+}$. In this case, $\left\vert \mathrm{M}_{\beta ,%
\mathfrak{D}}\right\vert =\mathcal{O}(\vartheta )$, $\mathrm{r}_{\beta ,%
\mathfrak{D}}=\mathcal{O}(\mathrm{e}^{-\beta (\mathrm{h}-\mathrm{h}_{\mathrm{%
c}})})$ and the magnetic induction expels the Cooper pair condensate from
the (macroscopic) region $\mathfrak{D}\subseteq \mathfrak{C}$.

Meanwhile, away from critical points, the electron density
\begin{equation*}
\mathrm{d}_{\beta ,\mathfrak{D}}:=\underset{l\rightarrow \infty }{\lim }\
\frac{1}{\left\vert \mathfrak{D}_{l}\right\vert }\sum_{x\in \mathfrak{D}_{l}}%
\mathfrak{g}_{l}\left( n_{x,\uparrow }+n_{x,\downarrow }\right)
\end{equation*}%
with $\mathfrak{D}_{l}:=2l\mathfrak{D}\cap \Lambda _{l}$ equals
\begin{equation}
\mathrm{d}_{\beta ,\mathfrak{D}}:=\frac{1}{\left\vert \mathfrak{D}%
\right\vert }\int_{\mathfrak{D}}\mathrm{d}_{\beta ,\mathfrak{t}}\ \mathrm{d}%
^{3}\mathfrak{t}\in \lbrack 0,2]  \label{electron density2}
\end{equation}%
for any non--empty open region $\mathfrak{D}\subseteq \mathfrak{C}$, where,
for all $\mathfrak{t}\in \mathfrak{C}$ (a.e.),%
\begin{equation}
\mathrm{d}_{\beta ,\mathfrak{t}}:=1+\frac{\left( \mu -\lambda \right) \sinh
\left( \beta g_{\mathrm{r}_{\beta }}\right) }{g_{\mathrm{r}_{\beta }}\left(
e^{\beta \lambda }\cosh \left( \beta h_{\mathfrak{t}}\right) +\cosh \left(
\beta g_{\mathrm{r}_{\beta }}\right) \right) }\ .  \label{electron density3}
\end{equation}%
In particular, the electron density is space--homogeneous whenever the
magnetic induction $\mathrm{B}$ is a.e.\ constant in space within the unit
box $\mathfrak{C}$.

Apart from its physical interpretation (\ref{cooper pair condensate density}%
) as the (global) Cooper pair condensate density, the solution $\mathrm{r}%
_{\beta }$ is extremely useful because it allows a construction of
approximating minimizers of the free--energy in finite boxes. Indeed, let
\begin{equation*}
\bar{u}_{l}\left( r,\mathfrak{t}\right) \equiv \bar{u}_{l}\left( r,\mathfrak{%
t},\mathrm{B}\right) :=\int_{\mathfrak{C}}u\left( r,\mathfrak{t}+\frac{y}{2l}%
\right) \mathrm{d}^{3}y
\end{equation*}%
for all $r\in \mathbb{R}_{0}^{+}$ and $\mathfrak{t}\in \mathfrak{C}$. We
define such approximating states by (well--defined, cf. \cite[Theorem 11.2.]%
{Araki-Moriya}) product states of the form
\begin{equation}
\mathfrak{g}_{l,r}\equiv \mathfrak{g}_{l,r,\mathrm{B}}:=\underset{x\in
\Lambda _{l}}{\bigotimes }\omega _{\left( 2l\right) ^{-1}x,r}\in E_{\Lambda
_{l}}  \label{gibbs box0}
\end{equation}%
for all $l\in \mathbb{N}$ and $r\in \mathbb{R}_{0}^{+}$, where $\omega
_{\left( 2l\right) ^{-1}x,r}\equiv \omega _{\left( 2l\right) ^{-1}x,r,%
\mathrm{B}}$ is the (even) Gibbs state on $\mathcal{U}_{\{x\}}$ associated
with the one--site Hamiltonian $\alpha _{x}(\bar{u}_{l}(r,(2l)^{-1}x))$ and
thus defined by the density matrix
\begin{equation*}
\frac{\mathrm{e}^{-\beta \alpha _{x}(\bar{u}_{l}(r,(2l)^{-1}x))}}{\mathrm{%
Trace}_{\wedge \mathcal{H}_{\{x\}}}(\mathrm{e}^{-\beta \alpha _{x}(\bar{u}%
_{l}(r,(2l)^{-1}x))})}
\end{equation*}%
for all $x\in \mathbb{Z}^{3}$. Here, for every $x\in \mathbb{Z}^{3}$, $%
\alpha _{x}$ is the translation map from $\mathcal{U}_{\Lambda _{l}}$ to the
$C^{\ast }$--algebra $\mathcal{U}_{\Lambda _{l}+x}$ with identity $\mathbf{1}
$ and generators $\{a_{y+x,\mathrm{s}}\}_{y\in \Lambda _{l},\mathrm{s}\in
\{\uparrow ,\downarrow \}}$. More precisely, for every $x\in \mathbb{Z}^{3}$%
, $\alpha _{x}$ is the isomorphism of $C^{\ast }$--algebras uniquely defined
by the conditions%
\begin{equation*}
\alpha _{x}(a_{y,\mathrm{s}})=a_{y+x,\mathrm{s}}\ ,\qquad y\in \Lambda
_{l},\ \mathrm{s}\in \{\uparrow ,\downarrow \}\ .
\end{equation*}

Then, as suggested by \cite[Proposition 4.2]{BruPedra-homog}, for any $%
\mathrm{B}\in L^{2}(\mathfrak{C};\mathbb{R}^{3})$, the product states $\{%
\mathfrak{g}_{l,\mathrm{r}_{\beta }}\}_{l\in \mathbb{N}}$ minimize the
free--energy density functional $f_{l}$ of the system in the thermodynamic
limit $l\rightarrow \infty $. The proof is however more difficult than in
\cite[Proposition 4.2]{BruPedra-homog}.

Similar to Lemma \ref{lemma var pro}, we first consider continuous magnetic
inductions $\mathrm{B}\in C^{0}(\mathfrak{C};\mathbb{R}^{3})$:

\begin{lemma}[Approximating minimizers -- I]
\label{lemma free energy copy(2)}\mbox{ }\newline
For any $\mathrm{B}\in C^{0}(\mathfrak{C};\mathbb{R}^{3})$ and any solution $%
\mathrm{r}_{\beta }=\mathrm{r}_{\beta }(\mathrm{B})$ of (\ref{var prob
principal}),%
\begin{equation*}
\underset{l\rightarrow \infty }{\lim }\left\{ f_{l}(\mathrm{B},\mathfrak{g}%
_{l,\mathrm{r}_{\beta }})-\underset{\rho \in E_{\Lambda _{l}}}{\inf }%
f_{l}\left( \mathrm{B},\rho \right) \right\} =0\ .
\end{equation*}
\end{lemma}

\noindent \textit{Proof. }For every $r\in \mathbb{R}_{0}^{+}$, define the
continuous map $\mathfrak{p}_{r}$ from $\mathbb{R}$ to $\mathbb{R}$ by
\begin{equation*}
x\mapsto \mathfrak{p}_{r}(x):=\frac{\gamma \sinh \left( \beta g_{r}\right) }{%
2g_{r}\left( \mathrm{e}^{\beta \lambda }\cosh \left( \beta x\right) +\cosh
\left( \beta g_{r}\right) \right) }
\end{equation*}%
as well as $h_{\mathfrak{t}}\equiv h_{\mathfrak{t}}\left( \mathrm{B}\right)
:=\vartheta |\mathrm{B}(\mathfrak{t})|$ and%
\begin{equation}
\bar{h}_{\mathfrak{t},l}\equiv \bar{h}_{\mathfrak{t},l}\left( \mathrm{B}%
\right) :=\vartheta \left\vert \int_{\mathfrak{C}}\mathrm{B}\left( \mathfrak{%
t}+\frac{y}{2l}\right) \mathrm{d}^{3}y\right\vert \ ,\ \ \mathfrak{t}\in
\mathfrak{C}\ ,\ l\in \mathbb{N}\ .  \label{definition h bar}
\end{equation}%
By explicit computations, for any $l\in \mathbb{N}$, $x\in \Lambda _{l}$, $%
\mathrm{B}\in C^{0}(\mathfrak{C};\mathbb{R}^{3})$ and $r\in \mathbb{R}%
_{0}^{+}$,%
\begin{equation}
\omega _{\left( 2l\right) ^{-1}x,r}\left( a_{x,\downarrow }a_{x,\uparrow
}\right) =\sqrt{r}\mathfrak{p}_{r}(\bar{h}_{\left( 2l\right) ^{-1}x,l})\in
\mathbb{R}\ ,  \label{meisnner toto encore}
\end{equation}%
which, combined with the equicontinuity of $\mathfrak{p}_{r}$ and $\mathrm{B}
$ on compact sets, yields%
\begin{equation}
\underset{l\rightarrow \infty }{\lim }\left\{ \frac{1}{\left\vert \Lambda
_{l}\right\vert }\underset{x\in \Lambda _{l}}{\sum }\omega _{\left(
2l\right) ^{-1}x,r}\left( a_{x,\downarrow }a_{x,\uparrow }\right) \right\} =%
\sqrt{r}\int_{\mathfrak{C}}\mathfrak{p}_{r}(h_{\mathfrak{t}})\mathrm{d}^{3}%
\mathfrak{t}\ .  \label{gap equation 3-2}
\end{equation}%
In the same way, we show by explicit computations that
\begin{equation}
\underset{l\rightarrow \infty }{\lim }\left\{ \frac{1}{\left\vert \Lambda
_{l}\right\vert }\sum_{x\in \Lambda _{l}}\ln \mathrm{Trace}_{\wedge \mathcal{%
H}_{\{x\}}}(\mathrm{e}^{-\beta \alpha _{x}(\bar{u}_{l}(r,(2l)^{-1}x))})%
\right\} =\mathfrak{F}\left( r,\mathrm{B}\right) +\gamma r
\label{gap equation 3-1}
\end{equation}%
for any $\mathrm{B}\in C^{0}(\mathfrak{C};\mathbb{R}^{3})$ and $r\in \mathbb{%
R}_{0}^{+}$. Using the additivity of the von Neumann entropy of product
states and the passivity of Gibbs states,
\begin{equation}
\underset{x\in \Lambda _{l}}{\sum }\mathfrak{g}_{l,r}%
%TCIMACRO{\TeXButton{\Big (}{\Big (}}%
%BeginExpansion
\Big (%
%EndExpansion
\alpha _{x}(\bar{u}_{l}(r,(2l)^{-1}x))%
%TCIMACRO{\TeXButton{\Big )}{\Big )}}%
%BeginExpansion
\Big )%
%EndExpansion
-\beta ^{-1}S_{l}(\mathfrak{g}_{l,r})=-\beta ^{-1}\sum_{x\in \Lambda
_{l}}\ln \mathrm{Trace}_{\wedge \mathcal{H}_{\{x\}}}(\mathrm{e}^{-\beta
\alpha _{x}(\bar{u}_{l}(r,(2l)^{-1}x))})\ .  \label{gap equation 3}
\end{equation}%
Since
\begin{equation*}
H_{l}=\underset{x\in \Lambda _{l}}{\sum }%
%TCIMACRO{\TeXButton{\Big (}{\Big (}}%
%BeginExpansion
\Big (%
%EndExpansion
\alpha _{x}(\bar{u}_{l}(r,(2l)^{-1}x))+\gamma \sqrt{r}(a_{x,\uparrow }^{\ast
}a_{x,\downarrow }^{\ast }+a_{x,\downarrow }a_{x,\uparrow })%
%TCIMACRO{\TeXButton{\Big )}{\Big )}}%
%BeginExpansion
\Big )%
%EndExpansion
-\frac{\gamma }{\left\vert \Lambda _{l}\right\vert }\sum_{x,y\in \Lambda
_{l}}a_{x,\uparrow }^{\ast }a_{x,\downarrow }^{\ast }a_{y,\downarrow
}a_{y,\uparrow }\ ,
\end{equation*}%
we infer from (\ref{free energy density}) and (\ref{gap equation 3-2})--(\ref%
{gap equation 3}) that
\begin{equation}
\underset{l\rightarrow \infty }{\lim }f_{l}(\mathrm{B},\mathfrak{g}_{l,r})=-%
\mathfrak{F}\left( r,\mathrm{B}\right) -\gamma r\left( 1-\int_{\mathfrak{C}}%
\mathfrak{p}_{r}(h_{\mathfrak{t}})\mathrm{d}^{3}\mathfrak{t}\right) ^{2}
\label{lemma poil a grater 1}
\end{equation}%
for any $\mathrm{B}\in C^{0}(\mathfrak{C};\mathbb{R}^{3})$ and $r\in \mathbb{%
R}_{0}^{+}$. In particular, by using the gap equation (\ref{gap equation}),
\begin{equation}
\underset{l\rightarrow \infty }{\lim }f_{l}(\mathrm{B},\mathfrak{g}_{l,%
\mathrm{r}_{\beta }})=-\mathfrak{F}\left( \mathrm{r}_{\beta },\mathrm{B}%
\right) \ .  \label{lemma poil a grater 10}
\end{equation}%
The latter yields the lemma because of (\ref{pressure0}), (\ref{var prob
principal}) and Theorem \ref{thm limit pressure} (i). \hfill {}$\Box $

Similar to Theorem \ref{thm limit pressure} (i), we now extend Lemma \ref%
{lemma free energy copy(2)} to all $\mathrm{B}\in L^{2}(\mathfrak{C};\mathbb{%
R}^{3})$ by using the density of $C^{0}(\mathfrak{C};\mathbb{R}^{3})$ in $%
L^{2}(\mathfrak{C};\mathbb{R}^{3})$:

\begin{theorem}[Approximating minimizers -- II]
\label{lemma free energy copy(1)}\mbox{ }\newline
For any $\mathrm{B}\in L^{2}(\mathfrak{C};\mathbb{R}^{3})$ and any solution $%
\mathrm{r}_{\beta }=\mathrm{r}_{\beta }(\mathrm{B})$ of (\ref{var prob
principal}),%
\begin{equation*}
\underset{l\rightarrow \infty }{\lim }\left\{ f_{l}(\mathrm{B},\mathfrak{g}%
_{l,\mathrm{r}_{\beta }})-\underset{\rho \in E_{\Lambda _{l}}}{\inf }%
f_{l}\left( \mathrm{B},\rho \right) \right\} =0\ .
\end{equation*}
\end{theorem}

\noindent \textit{Proof. }We start by proving the norm equicontinuity of the
collection
\begin{equation}
\{\mathrm{B}\mapsto f_{l}(0,\mathfrak{g}_{l,r,\mathrm{B}})\}_{l\in \mathbb{N}%
}  \label{map a la con}
\end{equation}%
of maps from $L^{2}(\mathfrak{C};\mathbb{R}^{3})$ to $\mathbb{R}$. To this
end, we study, for all $\mathfrak{t}\in \mathfrak{C}$ and $l\in \mathbb{N}$,
the maps
\begin{equation*}
\mathrm{B}\mapsto \mathrm{d}_{\mathfrak{t,}l}\left( \mathrm{B}\right) :=%
\frac{\mathrm{e}^{-\beta \bar{u}_{l}(r,\mathfrak{t},\mathrm{B})}}{\mathrm{%
Trace}_{\wedge \mathcal{H}_{\{0\}}}(\mathrm{e}^{-\beta \bar{u}_{l}(r,%
\mathfrak{t},\mathrm{B})})}
\end{equation*}%
from $L^{2}(\mathfrak{C};\mathbb{R}^{3})$ to the real space of self--adjoint
elements of $\mathcal{U}_{\{0\}}$. Let $\Vert -\Vert _{\mathrm{Tr}}$ be the
trace norm of $\mathcal{U}_{\{0\}}$. Observe that%
\begin{equation*}
\left\Vert \mathrm{d}_{\mathfrak{t,}l}\left( \mathrm{B}\right) -\mathrm{d}_{%
\mathfrak{t,}l}\left( \mathrm{C}\right) \right\Vert _{\mathrm{Tr}}\leq \frac{%
2\left\Vert \mathrm{e}^{-\beta \bar{u}_{l}(r,\mathfrak{t},\mathrm{B})}-%
\mathrm{e}^{-\beta \bar{u}_{l}(r,\mathfrak{t},\mathrm{C})}\right\Vert _{%
\mathrm{Tr}}}{\left\Vert \mathrm{e}^{-\beta \bar{u}_{l}(r,\mathfrak{t},%
\mathrm{B})}\right\Vert _{\mathrm{Tr}}}
\end{equation*}%
for any $\mathrm{B},\mathrm{C}\in L^{2}(\mathfrak{C};\mathbb{R}^{3})$. Using
Duhamel's formula
\begin{equation*}
\mathrm{e}^{A_{1}}-\mathrm{e}^{A_{2}}=\int_{0}^{1}\mathrm{e}^{\tau
A_{1}}\left( A_{1}-A_{2}\right) \mathrm{e}^{\left( 1-\tau \right) A_{2}}%
\mathrm{d}^{3}\tau \ ,
\end{equation*}%
$\Vert A_{1}\Vert \leq \Vert A_{1}\Vert _{\mathrm{Tr}}$ and $\Vert
A_{1}A_{2}\Vert _{\mathrm{Tr}}\leq \Vert A_{1}\Vert \Vert A_{2}\Vert _{%
\mathrm{Tr}}$ for any $A_{1},A_{2}\in \mathcal{U}_{\{0\}}$, we then find that%
\begin{equation}
\left\Vert \mathrm{d}_{\mathfrak{t,}l}\left( \mathrm{B}\right) -\mathrm{d}_{%
\mathfrak{t,}l}\left( \mathrm{C}\right) \right\Vert _{\mathrm{Tr}}\leq
2\beta \left\Vert \int_{\mathfrak{C}}(\mathrm{B}-\mathrm{C})\left( \mathfrak{%
t}+\frac{y}{2l}\right) \mathrm{d}^{3}y\cdot \mathrm{M}^{0}\right\Vert _{%
\mathrm{Tr}}\int_{0}^{1}\mathfrak{n}\left( \mathrm{B},\mathrm{C},\tau
\right) \mathrm{d}^{3}\tau  \label{encore poil a gratter1-con}
\end{equation}%
with
\begin{equation*}
\mathfrak{n}\left( \mathrm{B},\mathrm{C},\tau \right) :=\frac{\left\Vert
\mathrm{e}^{-\beta \tau \bar{u}_{l}(r,\mathfrak{t},\mathrm{B})}\right\Vert
\left\Vert \mathrm{e}^{-\beta \left( 1-\tau \right) \bar{u}_{l}(r,\mathfrak{t%
},\mathrm{C})}\right\Vert }{\left\Vert \mathrm{e}^{-\beta \bar{u}_{l}(r,%
\mathfrak{t},\mathrm{B})}\right\Vert }\ .
\end{equation*}%
Straightforward computations show that
\begin{equation*}
\left\Vert \mathrm{e}^{-\beta \tau \bar{u}_{l}(r,\mathfrak{t},\mathrm{B}%
)}\right\Vert =\mathrm{e}^{\beta \tau \left( \mu +\max \left\{ g_{r}-\lambda
,\bar{h}_{\mathfrak{t},l}\left( \mathrm{B}\right) \right\} \right) }
\end{equation*}%
for any $\mathrm{B}\in L^{2}(\mathfrak{C};\mathbb{R}^{3})$ and all $\tau \in
\lbrack 0,1]$. Thus,%
\begin{equation*}
\mathfrak{n}\left( \mathrm{B},\mathrm{C},\tau \right) =\mathrm{e}^{\beta
\left( 1-\tau \right) \left( \max \left\{ g_{r}-\lambda ,\bar{h}_{\mathfrak{t%
},l}\left( \mathrm{C}\right) \right\} -\max \left\{ g_{r}-\lambda ,\bar{h}_{%
\mathfrak{t},l}\left( \mathrm{B}\right) \right\} \right) }
\end{equation*}%
and, by (\ref{encore poil a gratter1-con}) and $\Vert \mathrm{m}%
_{j}^{x}\Vert \leq 2\vartheta $ for any $j\in \{1,2,3\}$ and all $x\in
\mathbb{Z}^{3}$,%
\begin{equation*}
\left\Vert \mathrm{d}_{\mathfrak{t,}l}\left( \mathrm{B}\right) -\mathrm{d}_{%
\mathfrak{t,}l}\left( \mathrm{C}\right) \right\Vert _{\mathrm{Tr}}\leq 4%
\sqrt{3}\beta \bar{h}_{\mathfrak{t},l}\left( \mathrm{B}-\mathrm{C}\right)
\int_{0}^{1}\mathfrak{n}\left( \mathrm{B},\mathrm{C},\tau \right) \mathrm{d}%
^{3}\tau
\end{equation*}%
with $\bar{h}_{\mathfrak{t},l}$ defined by (\ref{definition h bar}).

If $\bar{h}_{\mathfrak{t},l}(\mathrm{B}-\mathrm{C})\leq 1$, then we deduce
from the last two assertions that
\begin{equation*}
\left\Vert \mathrm{d}_{\mathfrak{t,}l}\left( \mathrm{B}\right) -\mathrm{d}_{%
\mathfrak{t,}l}\left( \mathrm{C}\right) \right\Vert _{\mathrm{Tr}}\leq 4%
\sqrt{3}\left( \mathrm{e}^{\beta }-1\right) \bar{h}_{\mathfrak{t},l}\left(
\mathrm{B}-\mathrm{C}\right) \ .
\end{equation*}%
On the other hand, for any $\mathrm{B},\mathrm{C}\in L^{2}(\mathfrak{C};%
\mathbb{R}^{3})$,
\begin{equation*}
\left\Vert \mathrm{d}_{\mathfrak{t,}l}\left( \mathrm{B}\right) -\mathrm{d}_{%
\mathfrak{t,}l}\left( \mathrm{C}\right) \right\Vert _{\mathrm{Tr}}\leq
\left\Vert \mathrm{d}_{\mathfrak{t,}l}\left( \mathrm{B}\right) \right\Vert _{%
\mathrm{Tr}}+\left\Vert \mathrm{d}_{\mathfrak{t,}l}\left( \mathrm{C}\right)
\right\Vert _{\mathrm{Tr}}=2\ .
\end{equation*}

Therefore, for all $\mathrm{B},\mathrm{C}\in L^{2}(\mathfrak{C};\mathbb{R}%
^{3})$,%
\begin{equation*}
\left\Vert \mathrm{d}_{\mathfrak{t,}l}\left( \mathrm{B}\right) -\mathrm{d}_{%
\mathfrak{t,}l}\left( \mathrm{C}\right) \right\Vert _{\mathrm{Tr}}\leq 4%
\sqrt{3}\mathrm{e}^{\beta }\bar{h}_{\mathfrak{t},l}\left( \mathrm{B}-\mathrm{%
C}\right) \ .
\end{equation*}%
We now use the product structure (\ref{gibbs box0}) of $\mathfrak{g}_{l,r}$,
the uniform norm Lipschitz continuity of the von Neumann entropy and the
last bound to deduce the existence of a finite constant $D\in \mathbb{R}^{+}$
such that%
\begin{equation*}
\left\vert f_{l}(0,\mathfrak{g}_{l,r,\mathrm{B}})-f_{l}(0,\mathfrak{g}_{l,r,%
\mathrm{C}})\right\vert \leq \frac{D}{\left\vert \Lambda _{l}\right\vert }%
\sum_{x\in \Lambda _{l}}\bar{h}_{\left( 2l\right) ^{-1}x,l}\left( \mathrm{B}-%
\mathrm{C}\right)
\end{equation*}%
for all $\mathrm{B},\mathrm{C}\in L^{2}(\mathfrak{C};\mathbb{R}^{3})$ and
all $l\in \mathbb{N}$. By definition of $\bar{h}_{\mathfrak{t},l}$ and the
Cauchy--Schwarz inequality, we thus find that
\begin{equation}
\left\vert f_{l}(0,\mathfrak{g}_{l,r,\mathrm{B}})-f_{l}(0,\mathfrak{g}_{l,r,%
\mathrm{C}})\right\vert \leq D\Vert \mathrm{B}-\mathrm{C}\Vert _{2}
\label{map a la con equi}
\end{equation}%
for all $\mathrm{B},\mathrm{C}\in L^{2}(\mathfrak{C};\mathbb{R}^{3})$ and
all $l\in \mathbb{N}$. In other words, the collection (\ref{map a la con})
is norm equicontinuous.

We want now to prove from the last inequality that the collection
\begin{equation}
\{\mathrm{B}\mapsto f_{l}(\mathrm{B},\mathfrak{g}_{l,r,\mathrm{B}})\}_{l\in
\mathbb{N}}  \label{map a la con super}
\end{equation}%
of maps from $L^{2}(\mathfrak{C};\mathbb{R}^{3})$ to $\mathbb{R}$ is also
norm equicontinuous. So, we need to show the norm equicontinuity of the
family
\begin{equation}
\left\{ \mathrm{B}\mapsto \langle \mathrm{B},\mathfrak{m}_{l}\left( \mathrm{B%
}\right) \rangle _{2}\right\} _{l\in \mathbb{N}}  \label{map a la condebi}
\end{equation}%
of maps from $L^{2}(\mathfrak{C};\mathbb{R}^{3})$ to $\mathbb{R}$, where%
\begin{equation}
\mathfrak{m}_{l}\left( \mathrm{B}\right) \left( \mathfrak{t}\right)
:=\sum_{x\in \Lambda _{l}}\mathbf{1}\left[ 2l\mathfrak{t}\in (\mathfrak{C}+x)%
\right] \mathfrak{g}_{l,r,\mathrm{B}}(\mathrm{M}^{x})\ .
\label{map a la condebi+}
\end{equation}%
Indeed, for any $\mathrm{B},\mathrm{C}\in L^{2}(\mathfrak{C};\mathbb{R}^{3})$%
,
\begin{equation}
\left\vert \langle \mathrm{B},\mathfrak{m}_{l}\left( \mathrm{B}\right)
\rangle _{2}-\langle \mathrm{C},\mathfrak{m}_{l}\left( \mathrm{C}\right)
\rangle _{2}\right\vert \leq 2\sqrt{3}\vartheta \Vert \mathrm{B}-\mathrm{C}%
\Vert _{2}+\Vert \mathrm{C}\Vert _{2}\Vert \mathfrak{m}_{l}\left( \mathrm{B}%
\right) -\mathfrak{m}_{l}\left( \mathrm{C}\right) \Vert _{2}\ ,
\label{meissner plus}
\end{equation}%
using the Cauchy--Schwarz inequality and $\Vert \mathrm{m}_{j}^{x}\Vert \leq
2\vartheta $. On the other hand, for any $\mathrm{B},\mathrm{C}\in L^{2}(%
\mathfrak{C};\mathbb{R}^{3})$,%
\begin{equation}
\Vert \mathfrak{m}_{l}\left( \mathrm{B}\right) -\mathfrak{m}_{l}\left(
\mathrm{C}\right) \Vert _{2}^{2}=\frac{1}{\left\vert \Lambda _{l}\right\vert
}\sum_{x\in \Lambda _{l}}\left\vert \omega _{\left( 2l\right) ^{-1}x,r,%
\mathrm{B}}(\mathrm{M}^{x})-\omega _{\left( 2l\right) ^{-1}x,r,\mathrm{C}}(%
\mathrm{M}^{x})\right\vert ^{2}\ .  \label{poila grater00}
\end{equation}%
By explicit computations, for any $\mathrm{B}\in L^{2}(\mathfrak{C};\mathbb{R%
}^{3})$,
\begin{equation}
\omega _{\left( 2l\right) ^{-1}x,r,\mathrm{B}}(\mathrm{M}^{x})=\mathfrak{q}%
_{r}(\bar{h}_{\left( 2l\right) ^{-1}x,l}\left( \mathrm{B}\right) )\vartheta
\int_{\mathfrak{C}}\mathrm{B}\left( \frac{x+y}{2l}\right) \mathrm{d}^{3}y\ .
\label{poila grater1}
\end{equation}%
Here, for any $x\in \mathbb{R}^{+}$ and $r\in \mathbb{R}_{0}^{+}$,%
\begin{equation*}
\mathfrak{q}_{r}(x):=\frac{\vartheta \sinh \left( \beta x\right) }{x\left(
\cosh \left( \beta x\right) +\mathrm{e}^{-\beta \lambda }\cosh \left( \beta
g_{r}\right) \right) }\ ,
\end{equation*}%
whereas at $x=0$,
\begin{equation*}
\mathfrak{q}_{r}(0):=\frac{\vartheta \beta }{1+\mathrm{e}^{-\beta \lambda
}\cosh \left( \beta g_{r}\right) }\ .
\end{equation*}

Assume that $\bar{h}_{\left( 2l\right) ^{-1}x,l}(\mathrm{C})\leq 2$. Notice
that there is a finite constant $D\in \mathbb{R}^{+}$ such that $|\mathfrak{q%
}_{r}(x)|\leq D$ for all $x\in \mathbb{R}_{0}^{+}$ and%
\begin{equation*}
\left\vert \mathfrak{q}_{r}(x)-\mathfrak{q}_{r}(y)\right\vert \leq
D\left\vert x-y\right\vert \ ,\quad x,y\in \mathbb{R}_{0}^{+}\ ,
\end{equation*}%
by the mean value theorem. Then, by (\ref{poila grater1}), for any $\mathrm{B%
}\in L^{2}(\mathfrak{C};\mathbb{R}^{3})$,%
\begin{equation*}
\left\vert \omega _{\left( 2l\right) ^{-1}x,r,\mathrm{B}}(\mathrm{M}%
^{x})-\omega _{\left( 2l\right) ^{-1}x,r,\mathrm{C}}(\mathrm{M}%
^{x})\right\vert \leq 2D\left\vert \bar{h}_{\left( 2l\right) ^{-1}x,l}\left(
\mathrm{B}\right) -\bar{h}_{\left( 2l\right) ^{-1}x,l}\left( \mathrm{C}%
\right) \right\vert +D\vartheta \left\vert \int_{\mathfrak{C}}\left( \mathrm{%
B}-\mathrm{C}\right) \left( \frac{x+y}{2l}\right) \mathrm{d}^{3}y\right\vert
\ .
\end{equation*}%
Using Jensen's inequality, $(a+b)^{2}\leq 2a^{2}+2b^{2}$ and $%
(|a|-|b|)^{2}\leq |a-b|^{2}$, we then deduce from the last upper bound that%
\begin{equation}
\left\vert \omega _{\left( 2l\right) ^{-1}x,r,\mathrm{B}}(\mathrm{M}%
^{x})-\omega _{\left( 2l\right) ^{-1}x,r,\mathrm{C}}(\mathrm{M}%
^{x})\right\vert ^{2}\leq 10D^{2}\vartheta ^{2}\int_{\mathfrak{C}}\left\vert
\left( \mathrm{B}-\mathrm{C}\right) \left( \frac{x+y}{2l}\right) \right\vert
^{2}\mathrm{d}^{3}y\ ,  \label{toto meissner}
\end{equation}%
provided that $\bar{h}_{\left( 2l\right) ^{-1}x,l}(\mathrm{C})\leq 2$.

Assume now that $\bar{h}_{\left( 2l\right) ^{-1}x,l}(\mathrm{C})\geq 2$ and $%
\bar{h}_{\mathfrak{t},l}(\mathrm{B}-\mathrm{C})\leq 1$. Remark that there is
a finite constant $D\in \mathbb{R}^{+}$ such that, for all $x\geq 2$ and $%
\left\vert x-y\right\vert \leq 1$,
\begin{equation*}
\left\vert \mathfrak{q}_{r}(x)-\mathfrak{q}_{r}(y)\right\vert \leq
D\left\vert x\right\vert ^{-1}\left\vert x-y\right\vert \ ,
\end{equation*}%
again by the mean value theorem. Similar to (\ref{toto meissner}), one then
gets that%
\begin{equation}
\left\vert \omega _{\left( 2l\right) ^{-1}x,r,\mathrm{B}}(\mathrm{M}%
^{x})-\omega _{\left( 2l\right) ^{-1}x,r,\mathrm{C}}(\mathrm{M}%
^{x})\right\vert ^{2}\leq 4D^{2}\vartheta ^{2}\int_{\mathfrak{C}}\left\vert
\left( \mathrm{B}-\mathrm{C}\right) \left( \frac{x+y}{2l}\right) \right\vert
^{2}\mathrm{d}^{3}y\ ,  \label{toto meissner2}
\end{equation}%
provided that $\bar{h}_{\left( 2l\right) ^{-1}x,l}(\mathrm{C})\geq 2$ and $%
\bar{h}_{\mathfrak{t},l}(\mathrm{B}-\mathrm{C})\leq 1$.

In the same way, we observe that
\begin{equation}
\left\vert \omega _{\left( 2l\right) ^{-1}x,r,\mathrm{B}}(\mathrm{M}%
^{x})-\omega _{\left( 2l\right) ^{-1}x,r,\mathrm{C}}(\mathrm{M}%
^{x})\right\vert ^{2}\leq 48\vartheta ^{2}\leq 48\vartheta ^{4}\int_{%
\mathfrak{C}}\left\vert \left( \mathrm{B}-\mathrm{C}\right) \left( \frac{x+y%
}{2l}\right) \right\vert ^{2}\mathrm{d}^{3}y  \label{toto meissner3}
\end{equation}%
whenever $\bar{h}_{\mathfrak{t},l}(\mathrm{B}-\mathrm{C})\geq 1$.

We then infer from (\ref{toto meissner})--(\ref{toto meissner3}) the
existence of a finite constant $D\in \mathbb{R}^{+}$ so that, for any $%
\mathrm{B},\mathrm{C}\in L^{2}(\mathfrak{C};\mathbb{R}^{3})$ and all $x\in
\Lambda _{l}$,%
\begin{equation*}
\left\vert \omega _{\left( 2l\right) ^{-1}x,r,\mathrm{B}}(\mathrm{M}%
^{x})-\omega _{\left( 2l\right) ^{-1}x,r,\mathrm{C}}(\mathrm{M}%
^{x})\right\vert ^{2}\leq D\int_{\mathfrak{C}}\left\vert \left( \mathrm{B}-%
\mathrm{C}\right) \left( \frac{x+y}{2l}\right) \right\vert ^{2}\mathrm{d}%
^{3}y\ .
\end{equation*}%
Hence, we deduce from (\ref{poila grater00}) that
\begin{equation}
\Vert \mathfrak{m}_{l}\left( \mathrm{B}\right) -\mathfrak{m}_{l}\left(
\mathrm{C}\right) \Vert _{2}^{2}\leq D\Vert \mathrm{B}-\mathrm{C}\Vert
_{2}^{2}  \label{map a la con equi2}
\end{equation}%
for any $\mathrm{B},\mathrm{C}\in L^{2}(\mathfrak{C};\mathbb{R}^{3})$.

By (\ref{map a la con equi}), (\ref{meissner plus}) and (\ref{map a la con
equi2}), the families (\ref{map a la con}) and (\ref{map a la condebi}) are
norm equicontinuous and so is the collection (\ref{map a la con super}).
Using this, (\ref{continuity of FracF}) and the uniform Lipschitz continuity
of the function $\mathfrak{p}_{r}$ together with the density of $C^{0}(%
\mathfrak{C};\mathbb{R}^{3})\subset L^{2}(\mathfrak{C};\mathbb{R}^{3})$,
Equation (\ref{lemma poil a grater 1}) holds for all $\mathrm{B}\in L^{2}(%
\mathfrak{C};\mathbb{R}^{3})$ and $r\in \mathbb{R}_{0}^{+}$. By the gap
equation (\ref{gap equation}), one gets (\ref{lemma poil a grater 10}) for
all $\mathrm{B}\in L^{2}(\mathfrak{C};\mathbb{R}^{3})$ which implies the
assertion because of (\ref{pressure0}), (\ref{var prob principal}) and
Theorem \ref{thm limit pressure} (i).\hfill {}$\Box $

Therefore, the sequence $\{\mathfrak{g}_{l,\mathrm{r}_{\beta }}\}_{l\in
\mathbb{N}}$ of approximating minimizers is a good starting point to
construct the states $\rho _{l}\in E_{\Lambda _{l}}$ of Theorem \ref%
{coolthml1 copy(2)}. To this end, we define from $\{\mathfrak{g}_{l,\mathrm{r%
}_{\beta }}\}_{l\in \mathbb{N}}$ states manifesting some current in
subregions of the box $\Lambda _{l}$ with very small volumes with respect to
the total volume $|\Lambda _{l}|=(2l)^{3}$. The latter is done as follows:

Take two positive real numbers $\eta ^{\bot },\eta \in \mathbb{R}^{+}$ such
that
\begin{equation}
0<\eta ^{\bot }<\eta <1  \label{condition base}
\end{equation}%
and define the small elementary box%
\begin{equation*}
\mathfrak{G}_{l}:=\mathbb{Z}^{3}\cap \{[-\ell ^{\eta },\ell ^{\eta }]\times
\lbrack -\ell ^{\eta ^{\bot }},\ell ^{\eta ^{\bot }}]^{2}\}\subset \Lambda
_{l}
\end{equation*}%
with $\ell :=l-1$ for $l>1$. (Note indeed that $\Lambda _{l}:=\{\mathbb{Z}%
\cap \lbrack -l,l-1]\}^{3}$.) More conditions on the constants $\eta ^{\bot
},\eta $ will be fixed later. We denote now by $[t]$ the integer part of $%
t\in \mathbb{R}_{0}^{+}$ to define the set%
\begin{equation*}
\mathcal{R}_{l}:=\{(k_{1},k_{2},k_{3})\in \mathbb{Z}^{3}:|k_{1}|\leq \lbrack
\ell ^{1-\eta }]\ ,\ |k_{2,3}|\leq \lbrack \ell ^{1-\eta ^{\bot }}]\}\ .
\end{equation*}%
For any $k\in \mathcal{R}_{l}$, we define the translated elementary boxes
\begin{equation}
\mathfrak{G}_{l,k}:=\mathfrak{G}_{l}+(k_{1}[2\ell ^{\eta }],k_{2}[2\ell
^{\eta ^{\bot }}],k_{3}[2\ell ^{\eta ^{\bot }}])\subset \Lambda _{l}\ .
\label{elementary box}
\end{equation}%
To create currents from $\mathrm{I}_{l}^{x,y}$ (\ref{curents observable}) we
perform some gauge transformation inside these elementary boxes. Indeed, for
any $k\in \mathcal{R}_{l}$ and $t\in \{0,1\}$, we use the automorphism $%
U_{k,t}$ of the $C^{\ast }$--algebra $\mathcal{U}_{\Lambda _{l+[\ell ^{\eta
}]}}$ uniquely defined by%
\begin{eqnarray*}
\forall x &\notin &\mathfrak{G}_{l,k}\ ,\ \mathrm{s}\in \{\uparrow
,\downarrow \}\ ,\qquad U_{k,t}\left( a_{x,\mathrm{s}}\right) :=\mathrm{e}%
^{it\pi /2}a_{x,\mathrm{s}}\ , \\
\forall x &\in &\mathfrak{G}_{l,k}\ ,\ \mathrm{s}\in \{\uparrow ,\downarrow
\}\ ,\qquad U_{k,t}\left( a_{x,\mathrm{s}}\right) :=\mathrm{e}^{i\mathrm{C}%
_{k}x_{1}/(2l)}a_{x,\mathrm{s}}\ .
\end{eqnarray*}%
The real parameter $\mathrm{C}_{k}\in \mathbb{R}$ will be chosen as a
function of the current density $\mathrm{j}$ to be produced by the system.

Take now any solution $\mathrm{r}_{\beta }=\mathrm{r}_{\beta }(\mathrm{B})$
of (\ref{var prob principal}) for $\mathrm{B}\in L^{2}(\mathfrak{C};\mathbb{R%
}^{3})$ and any even state $\varpi \in E_{\left\{ 0\right\} }$ satisfying
\begin{equation}
\varpi _{0}\left( a_{0,\downarrow }a_{0,\uparrow }\right) =1\ .
\label{local superconducting}
\end{equation}%
Such a state exists because $-1$ and $1$ both belong to the spectrum of $%
\mathop{\rm Re}\left( a_{0,\downarrow }a_{0,\uparrow }\right) $. We denote
by $\varpi _{x}:=\varpi _{0}\circ \alpha _{-x}$ the corresponding translated
state on $\mathcal{U}_{\left\{ x\right\} }$ for any $x\in \mathbb{Z}^{d}$.
For any $k\in \mathcal{R}_{l}$, define the state%
\begin{equation}
\nu _{k}:=\frac{1}{2}\left\{ \left( \underset{x\in \Lambda _{l+[\ell ^{\eta
}]}\backslash \mathfrak{G}_{l,k}}{\otimes }\omega _{\left( 2l\right) ^{-1}x,%
\mathrm{r}_{\beta }}\right) \bigotimes \left( \underset{x\in \mathfrak{G}%
_{l,k}}{\otimes }\varpi _{x}\right) \right\} \circ \left(
U_{k,0}+U_{k,1}\right)  \label{definition nu_k}
\end{equation}%
satisfying $\nu _{k}\left( \mathrm{I}_{l}^{x,y}\right) =0$ whenever\ $%
\left\{ x,y\right\} \cap \left( \Lambda _{l}\backslash \mathfrak{G}%
_{l,k}\right) \neq \emptyset $. Note indeed that, for any $x\in \Lambda
_{l+[\ell ^{\eta }]}$,
\begin{equation*}
\omega _{\left( 2l\right) ^{-1}x,\mathrm{r}_{\beta }}(a_{x,\downarrow
}a_{x,\uparrow })\in \mathbb{R}\ ,
\end{equation*}%
see (\ref{meisnner toto encore}) which clearly holds for all $\mathrm{B}\in
L^{2}(\mathfrak{C};\mathbb{R}^{3})$. Finally, for any $l\in \mathbb{N}$, we
set
\begin{equation}
\rho _{l}:=\frac{1}{\left\vert \mathcal{R}_{l}\right\vert \left\vert
\mathfrak{G}_{l}\right\vert }\underset{k\in \mathcal{R}_{l},\ x\in \mathfrak{%
G}_{l}}{\sum }\nu _{k}\circ \alpha _{x}|_{E_{\Lambda _{l}}}\in E_{\Lambda
_{l}}\ .  \label{carlos mini}
\end{equation}%
This state implies currents, in general. Moreover, observe that the term%
\begin{equation}
\underset{x\in \Lambda _{l}\backslash \mathfrak{G}_{l,k}}{\otimes }\omega
_{\left( 2l\right) ^{-1}x,\mathrm{r}_{\beta }}=\mathfrak{g}_{l,\mathrm{r}%
_{\beta }}|_{\mathcal{U}_{\Lambda _{l}\backslash \mathfrak{G}_{l,k}}}
\label{definition nu_kbis}
\end{equation}%
in the definition (\ref{definition nu_k}) of $\nu _{k}$ is the restriction
of the approximating minimizer $\mathfrak{g}_{l,\mathrm{r}_{\beta }}$ (\ref%
{gibbs box0}) on $\mathcal{U}_{\Lambda _{l}\backslash \mathfrak{G}_{l,k}}$.
Therefore, observing that $H_{l}$ is gauge invariant, one can infer from
Theorem \ref{lemma free energy copy(1)} that $\{\rho _{l}\}_{l\in \mathbb{N}%
} $ is also a sequence of approximating minimizers:

\begin{lemma}[Approximating minimizers -- III]
\label{lemma free energy}\mbox{ }\newline
For any $\mathrm{B}\in L^{2}(\mathfrak{C};\mathbb{R}^{3})$ and any set $\{%
\mathrm{C}_{k}\}_{k\in \mathcal{R}_{l}}\subset \mathbb{R}$,%
\begin{equation*}
\underset{l\rightarrow \infty }{\lim }\left\{ f_{l}(\mathrm{B},\rho _{l})-%
\underset{\rho \in E_{\Lambda _{l}}}{\inf }f_{l}\left( \mathrm{B},\rho
\right) \right\} =0\ .
\end{equation*}
\end{lemma}

\noindent \textit{Proof. }Note that (\ref{condition base}) implies that, for
all $k\in \mathcal{R}_{l}$,
\begin{equation*}
|\mathfrak{G}_{l,k}|=|\mathfrak{G}_{l}|=o(\left\vert \Lambda _{l}\right\vert
)\ .
\end{equation*}%
Therefore, we deduce from (\ref{free energy density}), Theorem \ref{lemma
free energy copy(1)}, (\ref{definition nu_k}) and (\ref{definition nu_kbis})
that
\begin{equation*}
\underset{l\rightarrow \infty }{\lim }\left\{ f_{l}(\mathrm{B},\nu _{k}\circ
\alpha _{x}|_{E_{\Lambda _{l}}})-\underset{\rho \in E_{\Lambda _{l}}}{\inf }%
f_{l}\left( \mathrm{B},\rho \right) \right\} =0\ ,
\end{equation*}%
uniformly for all $k\in \mathcal{R}_{l}$ and $x\in \mathfrak{G}_{l}$. Since
the state $\rho _{l}$ is a convex combination of states $\{\nu _{k}\circ
\alpha _{x}|_{E_{\Lambda _{l}}}\}_{k\in \mathcal{R}_{l},x\in \mathfrak{G}%
_{l}}$, the assertion then follows from the convexity of the free--energy
density $f_{l}$.\hfill $\Box $

We use now the sequence $\{\rho _{l}\}_{l\in \mathbb{N}}$ of approximating
minimizers to create a one--component current arbitrarily close to any
prescribed smooth function $\mathrm{j}_{1}\in C_{0}^{\infty }(\mathfrak{C};%
\mathbb{R})$ at large $l\in \mathbb{N}$. With respect to Theorem \ref%
{coolthml1 copy(2)}, the function $\mathrm{j}_{1}$ is the first coordinate
of some smooth current density $\mathrm{j}\in C^{\infty }(\mathfrak{C};%
\mathbb{R}^{3})$.

We now fix the real parameters $\{\mathrm{C}_{k}\}_{k\in \mathcal{R}_{l}}$
as follows: Observing that
\begin{equation*}
\int_{-l^{\eta }}^{l^{\eta }}\int_{-l^{\eta }}^{l^{\eta }}\left(
z_{1}-y_{1}\right) ^{2}\mathrm{d}z_{1}\ \mathrm{d}y_{1}=\frac{8l^{4\eta }}{3}%
\ ,
\end{equation*}%
we remark that%
\begin{equation}
\mathrm{K}:=\underset{l\rightarrow \infty }{\lim }%
%TCIMACRO{\TeXButton{\Big \{}{\Big \{}}%
%BeginExpansion
\Big \{%
%EndExpansion
l^{-4\eta -4\eta ^{\bot }}\sum\limits_{y,z\in \mathfrak{G}_{l}}\left(
z_{1}-y_{1}\right) ^{2}%
%TCIMACRO{\TeXButton{\Big \}}{\Big \}}}%
%BeginExpansion
\Big \}%
%EndExpansion
=\frac{2^{7}}{3}\ .  \label{definition de K}
\end{equation}%
Then, we define the constant $\mathrm{C}_{k}$ by
\begin{equation}
\mathrm{C}_{k}:=\frac{8}{\mathrm{K}\gamma }l^{6-4\eta -4\eta ^{\bot }}%
\mathrm{j}_{1}\left( \frac{k_{1}[\ell ^{\eta }]}{2l},\frac{k_{2}[\ell ^{\eta
^{\bot }}]}{2l},\frac{k_{3}[\ell ^{\eta ^{\bot }}]}{2l}\right)  \label{Ck}
\end{equation}%
for all $k\in \mathcal{R}_{l}$ and any $\mathrm{j}_{1}\in C_{0}^{\infty }(%
\mathfrak{C};\mathbb{R})$. With this choice of parameters, as $l\rightarrow
\infty $, one produces indeed the current density $(\mathrm{j}_{1}(\mathfrak{%
t}),0,0)$ for $\mathfrak{t}\in \mathbb{R}^{3}$:

\begin{lemma}[One--component currents]
\label{lemma1 current}\mbox{ }\newline
Assume that (\ref{condition base}) holds with $3\eta +4\eta ^{\bot }>6$ and $%
\mathrm{j}_{1}\in C_{0}^{\infty }(\mathfrak{C};\mathbb{R})$. Then, for any $%
\varkappa <\min \{1-\eta ,\eta -\eta ^{\bot }\}$,%
\begin{equation*}
\underset{\mathfrak{t}\in \mathbb{R}^{3}}{\sup }\
%TCIMACRO{\TeXButton{\big |}{\big |}}%
%BeginExpansion
\big |%
%EndExpansion
j_{\rho _{l}}^{(l)}(\mathfrak{t})-\left( \mathrm{j}_{1}\left( \mathfrak{t}%
\right) ,0,0\right)
%TCIMACRO{\TeXButton{\big |}{\big |}}%
%BeginExpansion
\big |%
%EndExpansion
=o(l^{-\varkappa })
\end{equation*}%
with the rescaled current density $j_{\rho _{l}}^{(l)}$ defined by (\ref%
{rescaled current density}).
\end{lemma}

\noindent \textit{Proof. }By (\ref{definition nu_k}), observe that, for any $%
k\in \mathcal{R}_{l}$ and $x_{0}\in \mathfrak{G}_{l}$, the state $\nu
_{k,x_{0}}:=\nu _{k}\circ \alpha _{x_{0}}|_{E_{\Lambda _{l}}}\in E_{\Lambda
_{l}}$ creates a current density $j_{\nu _{k,x_{0}}}\left( x\right) $ at $%
x\in \mathbb{R}^{3}$ which is equal to
\begin{equation}
j_{\nu _{k,x_{0}}}\left( x\right) =\frac{4\gamma }{\left\vert \Lambda
_{l}\right\vert }\sum\limits_{y,z\in \mathfrak{G}_{l,k},\ y\neq z}\frac{y-z}{%
\left\vert y-z\right\vert ^{3}}\xi \left( \frac{x+x_{0}-\frac{y+z}{2}}{%
\left\vert y-z\right\vert }\right) \sin \left[ \mathrm{C}_{k}\left(
y_{1}-z_{1}\right) l^{-1}\right] \ .  \label{carlos_currents1}
\end{equation}%
See Equation (\ref{carlos_currents}). Because of (\ref{Ck}), note that the
condition%
\begin{equation*}
\delta :=3\eta +4\eta ^{\bot }-6>0
\end{equation*}%
implies that, for any $k\in \mathcal{R}_{l}$,%
\begin{equation}
l^{\eta }\mathrm{C}_{k}=\mathcal{O}(l^{-\delta })  \label{condition sympa}
\end{equation}%
vanishes in the thermodynamic limit $l\rightarrow \infty $. As a
consequence, we can deduce from (\ref{carlos_currents1}) that%
\begin{equation}
2l\ j_{\nu _{k,x_{0}}}\left( x\right) =\frac{4\gamma }{\left\vert \Lambda
_{l}\right\vert }\sum\limits_{y,z\in \mathfrak{G}_{l,k},\ y\neq z}\frac{y-z}{%
\left\vert y-z\right\vert ^{3}}\xi \left( \frac{x+x_{0}-\frac{y+z}{2}}{%
\left\vert y-z\right\vert }\right) \left[ 2\mathrm{C}_{k}\left(
y_{1}-z_{1}\right) +\mathcal{O}(l^{-2}|\mathrm{C}_{k}\left(
y_{1}-z_{1}\right) |^{3})\right] ,  \label{carlos_currents2}
\end{equation}%
because $\left\vert z_{1}-y_{1}\right\vert \leq l^{\eta }$ for all $y,z\in
\mathfrak{G}_{l,k}$. Note that the factor $2l$ above is related to the
definition of the rescaled current density (\ref{rescaled current density}).
The current density functional $\rho \mapsto j_{\rho }$ of the box $\Lambda
_{l}$ defined by (\ref{carlos_currents}) is affine. It follows that the
current density induced by the approximating minimizer $\rho _{l}$ (\ref%
{carlos mini}) at $x\in \mathbb{R}^{3}$ equals
\begin{equation}
j_{\rho _{l}}\left( x\right) =\frac{1}{\left\vert \mathcal{R}_{l}\right\vert
\left\vert \mathfrak{G}_{l}\right\vert }\underset{k\in \mathcal{R}_{l}\ ,\
x_{0}\in \mathfrak{G}_{l}}{\sum }j_{\nu _{k,x_{0}}}\left( x\right) \ .
\label{carlos_currents3}
\end{equation}%
Since $\xi \in C_{0}^{\infty }(\mathbb{R}^{3};\mathbb{R})$ is a smooth and
compactly supported function, we infer from (\ref{condition sympa}) that the
norm of the vector%
\begin{equation*}
\frac{4\gamma }{l^{2}\left\vert \Lambda _{l}\right\vert \left\vert \mathcal{R%
}_{l}\right\vert \left\vert \mathfrak{G}_{l}\right\vert }\underset{k\in
\mathcal{R}_{l}\ ,\ x_{0}\in \mathfrak{G}_{l}}{\sum }\ \sum\limits_{y,z\in
\mathfrak{G}_{l,k},\ y\neq z}\frac{y-z}{\left\vert y-z\right\vert ^{3}}\xi
\left( \frac{x+x_{0}-\frac{y+z}{2}}{\left\vert y-z\right\vert }\right)
\left\vert \mathrm{C}_{k}\left( y_{1}-z_{1}\right) \right\vert ^{3}
\end{equation*}%
converges to zero as $l\rightarrow \infty $ faster than $l^{-1}$, uniformly
for $x\in \mathbb{R}^{3}$. Using the explicit parameters (\ref{Ck}), the
smoothness of $\mathrm{j}_{1}$ as well as (\ref{carlos_currents2}), we can
rewrite (\ref{carlos_currents3}) as
\begin{equation}
j_{\rho _{l}}^{(l)}(\mathfrak{t})=\frac{1}{\left\vert \mathcal{R}%
_{l}\right\vert \left\vert \mathfrak{G}_{l}\right\vert }\underset{k\in
\mathcal{R}_{l}}{\sum }\ \underset{2l\mathfrak{s}_{k}\in \mathfrak{G}_{l,k}}{%
\sum }\chi _{_{l}}\left( \mathfrak{t}-\mathfrak{s}_{k}\right) \left( \mathrm{%
j}_{1}\left( \mathfrak{s}_{k}\right) +\mathcal{O}\left( l^{\eta -1}\right)
\right) +o(l^{\eta -1})\ ,  \label{convolution product}
\end{equation}%
uniformly for all $\mathfrak{t}\in \mathbb{R}^{3}$. Here, the function $\chi
_{_{l}}\in C_{0}^{\infty }$ is defined, for all $\mathfrak{t}\in \mathbb{R}%
^{3}$, by%
\begin{equation*}
\chi _{_{l}}\left( \mathfrak{t}\right) :=\frac{8}{\mathrm{K}}l^{3-4\eta
-4\eta ^{\bot }}\sum\limits_{y,z\in \mathfrak{G}_{l}}\frac{y-z}{\left\vert
y-z\right\vert ^{3}}\left( y_{1}-z_{1}\right) \xi \left( \frac{2l\mathfrak{t}%
-\frac{y+z}{2}}{\left\vert y-z\right\vert }\right) \ .
\end{equation*}

By (\ref{carlos1}), note that%
\begin{equation*}
\int_{\mathbb{R}^{3}}\chi _{_{l}}\left( \mathfrak{t}\right) \mathrm{d}^{3}%
\mathfrak{t}=\frac{1}{\mathrm{K}}l^{-4\eta -4\eta ^{\bot
}}\sum\limits_{y,z\in \mathfrak{G}_{l}}\left( y-z\right) \left(
y_{1}-z_{1}\right) \ .
\end{equation*}%
Observe also that we have chosen the constant $\mathrm{K}$ (\ref{definition
de K}) in the definition of $\mathrm{C}_{k}$ to have exactly the limit%
\begin{equation*}
\underset{l\rightarrow \infty }{\lim }\int_{\mathbb{R}^{3}}\chi
_{_{l}}\left( \mathfrak{t}\right) \mathrm{d}^{3}\mathfrak{t}=\left(
1,0,0\right) \ .
\end{equation*}%
More precisely,
\begin{equation*}
\int_{\mathbb{R}^{3}}\chi _{_{l}}\left( \mathfrak{t}\right) \mathrm{d}^{3}%
\mathfrak{t}-\left( 1,0,0\right) =\mathcal{O}(l^{\eta ^{\bot }-\eta })\ .
\end{equation*}%
Since $\xi \in C_{0}^{\infty }(\mathbb{R}^{3};\mathbb{R})$ is compactly
supported, the support $\mathrm{supp}(\chi _{_{l}})$ of $\chi _{_{l}}\in
C_{0}^{\infty }$ has radius
\begin{equation}
\sup \left\{ \left\vert x\right\vert :\chi _{_{l}}\left( x\right) \neq
0\right\} =\mathcal{O}(l^{\eta -1)})  \label{important inequality2}
\end{equation}%
and belongs to a sufficiently large box $[-L,L]^{3}$, $L>0$, for all $l\in
\mathbb{N}$:
\begin{equation}
\lbrack -L,L]^{3}\supset \ \underset{l\in \mathbb{N}}{\bigcup }\ \mathrm{supp%
}(\chi _{_{l}})\ .  \label{important inequality2bis}
\end{equation}%
In fact, the sequence $\{\chi _{_{l}}\}_{l\in \mathbb{N}}$, seen as a family
of distributions, converges to the delta function, as $l\rightarrow \infty $.

Now therefore, the right hand side of (\ref{convolution product})
approximates the convolution $\chi _{_{l}}\ast \mathrm{j}_{1}\left(
\mathfrak{t}\right) $ since the sum is a Riemann sum. By (\ref{convolution
product}), it is then straightforward to verify the existence of a constant $%
D$ not depending on (a sufficiently large) $l\in \mathbb{N}$ and $\mathfrak{t%
}\in \mathfrak{C}$ such that%
\begin{equation}
%TCIMACRO{\TeXButton{\big |}{\big |}}%
%BeginExpansion
\big |%
%EndExpansion
j_{\rho _{l}}^{(l)}(\mathfrak{t})-\left( \mathrm{j}_{1}\left( \mathfrak{t}%
\right) ,0,0\right)
%TCIMACRO{\TeXButton{\big |}{\big |}}%
%BeginExpansion
\big |%
%EndExpansion
\leq D\underset{\mathfrak{s}\in \mathrm{supp}(\chi _{_{l}})}{\sup }%
%TCIMACRO{\TeXButton{\big |}{\big |}}%
%BeginExpansion
\big |%
%EndExpansion
\mathrm{j}_{1}\left( \mathfrak{s}+\mathfrak{t}\right) -\mathrm{j}_{1}\left(
\mathfrak{t}\right)
%TCIMACRO{\TeXButton{\big |}{\big |}}%
%BeginExpansion
\big |%
%EndExpansion
+D(l^{\eta -1}+l^{\eta ^{\bot }-\eta })\ .  \label{important inequality}
\end{equation}%
The continuity of $\mathrm{j}_{1}\in C_{0}^{\infty }(\mathfrak{C};\mathbb{R}%
) $ implies its equicontinuity on any compact set. Hence, by (\ref{important
inequality2})--(\ref{important inequality2bis}),
\begin{equation}
\underset{\mathfrak{t}\in \mathfrak{C}}{\sup }\ \underset{\mathfrak{s}\in
\mathrm{supp}(\chi _{_{l}})}{\sup }%
%TCIMACRO{\TeXButton{\big |}{\big |}}%
%BeginExpansion
\big |%
%EndExpansion
\mathrm{j}_{1}\left( \mathfrak{s}+\mathfrak{t}\right) -\mathrm{j}_{1}\left(
\mathfrak{t}\right)
%TCIMACRO{\TeXButton{\big |}{\big |}}%
%BeginExpansion
\big |%
%EndExpansion
=\mathcal{O}(l^{\eta -1})\ .  \label{important inequalitybis}
\end{equation}%
The lemma is then a consequence of (\ref{important inequality})--(\ref%
{important inequalitybis}).\hfill $\Box $

Lemma \ref{lemma1 current} can be extended to all current densities $\mathrm{%
j}\in C^{\infty }(\mathfrak{C};\mathbb{R}^{3})$. See Theorem \ref{coolthml1
copy(2)}. Meanwhile, as explained in\ Section \ref{Section Magnetic Energy},
for every approximating minimizer $\rho _{l}$ (\ref{definition nu_k})--(\ref%
{carlos mini}), the current density $j_{\rho _{l}}^{(l)}\in C_{0}^{\infty }$
can be decomposed into longitudinal and transverse components $(j_{\rho
_{l}}^{(l)})^{\parallel }=P^{\parallel }j_{\rho _{l}}^{(l)}$ and $(j_{\rho
_{l}}^{(l)})^{\perp }=P^{\perp }j_{\rho _{l}}^{(l)}$, respectively. So, we
conclude this section by showing that the energy norm of $(j_{\rho
_{l}}^{(l)})^{\parallel }$ is negligible as $l\rightarrow \infty $ whenever $%
\nabla \cdot \mathrm{j}=0$:

\begin{lemma}[Energy norm estimates]
\label{propositioncool}\mbox{ }\newline
Assume that (\ref{condition base}) holds with $3\eta +4\eta ^{\bot }>6$ and $%
\mathrm{j}_{1}\in C_{0}^{\infty }(\mathfrak{C};\mathbb{R})$. Then, for any $%
\varkappa <\min \{1-\eta ,\eta -\eta ^{\bot }\}$,
\begin{equation*}
\Vert \mathrm{j}^{\bot }-(j_{\rho _{l}}^{(l)})^{\bot }\Vert _{\mathfrak{H}%
}\leq \Vert \mathrm{j}-j_{\rho _{l}}^{(l)}\Vert _{\mathfrak{H}%
}=o(l^{-\varkappa })\ .
\end{equation*}%
In particular, if $\mathrm{j}$ is divergence--free (i.e., $\mathrm{j}=%
\mathrm{j}^{\bot }$), then
\begin{equation*}
\Vert (j_{\rho _{l}}^{(l)})^{\parallel }\Vert _{\mathfrak{H}%
}=o(l^{-\varkappa })\ .
\end{equation*}
\end{lemma}

\noindent \textit{Proof. }First, observe that%
\begin{equation}
\left\Vert j\right\Vert _{\mathfrak{H}}^{2}\leq \left\Vert j\right\Vert
_{2}^{2}+\int_{\{k\in \mathbb{C}\text{ }:\text{ }\left\vert k\right\vert
^{2}\leq 1\}}\frac{|F[j](k)|^{2}}{\left\vert k\right\vert ^{2}}\mathrm{d}%
^{3}k  \label{inequality}
\end{equation}%
for all $j\in L^{2}\cap \mathfrak{H}$, where $F[j]$ is the Fourier transform
of $j$. Therefore, the assertion follows from Theorem \ref{coolthml1 copy(2)}
together with the fact that $P^{\parallel }$, $P^{\bot }$ are mutually
orthogonal projections. Recall that, for some sufficiently large $L\in
\mathbb{R}^{+}$ and all $l\in \mathbb{N}$, the support (\ref{support}) of $%
j_{\rho _{l}}^{(l)}$ is contained in the box $[-L,L]^{3}$.\hfill $\Box $

\subsection{Thermodynamics with Self--Generated Magnetic Inductions\label%
{section Self--Generated}}

We analyze now the thermodynamics corresponding to the magnetic free--energy
density functionals $\mathcal{F}_{l}^{(\epsilon )}$ defined by (\ref%
{magnetic free energy}) on the sets $E_{\Lambda _{l}}$ of states for all $%
l\in \mathbb{N}$ and $\epsilon \in \mathbb{R}^{+}$. By contrast with the
previous section, the magnetic induction $\mathrm{B}=B_{\rho }^{(l)}$ (\ref%
{def.B.rho}) is now self--generated by the system in the state $\rho \in
E_{\Lambda _{l}}$.

We first need to compute the thermodynamic limit $\mathcal{P}_{\infty
}^{(\epsilon )}$ of the magnetic pressure (\ref{magnetic pressure}), that
is,
\begin{equation*}
\mathcal{P}_{l}^{(\epsilon )}:=-\inf_{\rho \in E_{\Lambda _{l}}^{\bot }}%
\mathcal{F}_{l}^{(\epsilon )}(\rho )\ ,\qquad l\in \mathbb{N}\ ,\ \epsilon
\in \mathbb{R}^{+},
\end{equation*}%
where $E_{\Lambda _{l}}^{\bot }$ is defined by (\ref{new set of states}).
This requires various arguments and we present them in several lemmata.

Recall Equation (\ref{important equality}) which is actually satisfied for
all $\mathrm{B}\in L^{2}$:
\begin{equation}
\langle \mathrm{B},m_{\rho }^{(l)}\rangle _{2}=f_{l}\left( 0,\rho \right)
-f_{l}\left( \mathfrak{T}_{\epsilon }\mathrm{B},\rho \right) \ ,\quad
\mathrm{B}\in L^{2},\ \rho \in E_{\Lambda _{l}}\ .
\label{important equality2}
\end{equation}%
Here, $\mathfrak{T}_{\epsilon }$ is the Hilbert--Schmidt operator defined,
for any $\epsilon \in \mathbb{R}^{+}$, by
\begin{equation}
\mathfrak{T}_{\epsilon }\mathrm{B}:=\mathbf{1}\left[ \mathfrak{t}\in
\mathfrak{C}\right] \ (\xi _{\epsilon }\ast \mathrm{B})\ ,\qquad \mathrm{B}%
\in L^{2}\ .  \label{operator compact}
\end{equation}%
See also (\ref{xi eps m}). In particular, $\mathfrak{T}_{\epsilon }$ has
Hilbert--Schmidt norm equal to $\epsilon ^{-3/2}\Vert \xi \Vert _{2}$ but
its operator norm satisfies $\Vert \mathfrak{T}_{\epsilon }\Vert \leq 1$,
because of $\Vert \xi _{\epsilon }\Vert _{1}=1$ and Young's inequality. We
also add that
\begin{equation}
\underset{\epsilon \rightarrow 0^{+}}{\lim }\left\Vert \left( \mathfrak{T}%
_{\epsilon }-\mathbf{1}\left[ \mathfrak{t}\in \mathfrak{C}\right] \right)
\mathrm{B}\right\Vert _{2}=0\ ,\qquad \mathrm{B}\in L^{2}\ .
\label{strong convergence}
\end{equation}%
The latter can easily be proven for all $\mathrm{B}\in C_{0}^{\infty }$ by
direct estimates. Then, one uses the density of $C_{0}^{\infty }$ in $L^{2}$
as well as $\Vert \mathfrak{T}_{\epsilon }\Vert \leq 1$ for any $\epsilon
\in \mathbb{R}^{+}$ to get (\ref{strong convergence}), i.e., the strong
convergence of $\mathfrak{T}_{\epsilon }$ as $\epsilon \rightarrow 0^{+}$
towards the (non--compact) operator $\mathfrak{T}_{0}$ defined by%
\begin{equation}
\mathfrak{T}_{0}\mathrm{B}:=\mathbf{1}\left[ \mathfrak{t}\in \mathfrak{C}%
\right] \ \mathrm{B}\ ,\qquad \mathrm{B}\in L^{2}\ .
\label{operator non compact}
\end{equation}%
Obviously, $\Vert \mathfrak{T}_{0}\Vert \leq 1$.

The first step is to study the collection $\left\{ \mathrm{B}\mapsto
p_{l}\left( \mathfrak{T}_{\epsilon }\mathrm{B}\right) \right\} _{l\in
\mathbb{N\cup \{\infty \}}}$ of maps from $L^{2}$ to $\mathbb{R}$ at any
fixed $\epsilon \in \mathbb{R}^{+}$. Indeed, since $\mathfrak{T}_{\epsilon }$
is a compact operator for every $\epsilon \in \mathbb{R}^{+}$, such maps
have much stronger continuity properties than the maps $\mathrm{B}\mapsto
p_{l}\left( \mathrm{B}\right) $ analyzed for all $l\in \mathbb{N\cup
\{\infty \}}$ in Theorem \ref{thm limit pressure}. An important additional
feature at any $\epsilon \in \mathbb{R}^{+}$ is the weak equicontinuity of
the collection $\left\{ \mathrm{B}\mapsto p_{l}\left( \mathfrak{T}_{\epsilon
}\mathrm{B}\right) \right\} _{l\in \mathbb{N}}$ of maps on any ball%
\begin{equation}
b_{R}\left( 0\right) :=\left\{ \mathrm{B}\in L^{2}:\left\Vert \mathrm{B}%
\right\Vert _{2}\leq R\right\}  \label{ball}
\end{equation}%
of radius $R\in \mathbb{R}^{+}$ centered at $0$. The latter is a consequence
of the following lemma:

\begin{lemma}[Magnetic interaction energy]
\label{lemma equicontinuity}\mbox{ }\newline
The family $\{\mathrm{B}\mapsto \langle \mathrm{B},m_{\rho }^{(l)}\rangle
_{2}\}_{l\in \mathbb{N},\rho \in E_{\Lambda _{l}}}$ of maps from $%
b_{R}\left( 0\right) $ to $\mathbb{R}$ is equicontinuous in the weak
topology.
\end{lemma}

\noindent \textit{Proof. }Since $\mathfrak{T}_{\epsilon }$ is a
Hilbert--Schmidt operator satisfying $\Vert \mathfrak{T}_{\epsilon }\Vert
\leq 1$ for every $\epsilon \in \mathbb{R}^{+}$, it is compact and its
singular value decomposition is
\begin{equation*}
\mathfrak{T}_{\epsilon }=\overset{\infty }{\sum_{n=1}}\lambda
_{n}|v_{n}\rangle \langle w_{n}|\ ,
\end{equation*}%
where $\{v_{n}\}_{n=1}^{\infty },\{w_{n}\}_{n=1}^{\infty }$ are orthonormal
bases of $L^{2}$ and $\{\lambda _{n}\}_{n=1}^{\infty }\subset \lbrack 0,1]$
is a set of real numbers satisfying%
\begin{equation*}
\overset{\infty }{\sum_{n=1}}\lambda _{n}^{2}<\infty \ .
\end{equation*}%
Take any $\varepsilon \in \mathbb{R}^{+}$. Then, there is $N\in \mathbb{N}$\
such that
\begin{equation*}
\left\Vert \mathfrak{T}_{\epsilon }-\overset{N}{\sum_{n=1}}\lambda
_{n}|v_{n}\rangle \langle w_{n}|\right\Vert \leq \frac{\varepsilon }{8R\sqrt{%
3}\vartheta }\ .
\end{equation*}%
Choose now $\mathrm{B}\in L^{2}$ and $\delta :=\varepsilon /(4N\sqrt{3}%
\vartheta )$. Meanwhile, remark that $\Vert \rho \Vert =1$ and $\Vert
\mathrm{m}_{j}^{x}\Vert \leq 2\vartheta $ for any $j\in \{1,2,3\}$ and all $%
x\in \mathbb{Z}^{3}$, see (\ref{magne1}). Therefore, by (\ref{important
equality2}) and the Cauchy--Schwarz inequality,
\begin{equation*}
\left\vert \langle \mathrm{C}-\mathrm{B},m_{\rho }^{(l)}\rangle
_{2}\right\vert \leq \varepsilon \ ,\quad \mathrm{B}\in b_{R}\left( 0\right)
\ ,\ \mathrm{C}\in \mathcal{V}_{\delta }\left( \mathrm{B}\right) \ ,
\end{equation*}%
where $\mathcal{V}_{\delta }\left( \mathrm{B}\right) $ is the weak
neighborhood%
\begin{equation*}
\mathcal{V}_{\delta }\left( \mathrm{B}\right) :=%
%TCIMACRO{\TeXButton{\Big \{}{\Big \{}}%
%BeginExpansion
\Big \{%
%EndExpansion
\mathrm{C}\in b_{R}\left( 0\right) :\ \underset{n\in \left\{ 1,\ldots
,N\right\} }{\sup }\left\vert \langle \mathrm{C}-\mathrm{B},w_{n}\rangle
_{2}\right\vert \leq \delta
%TCIMACRO{\TeXButton{\Big \}}{\Big \}}}%
%BeginExpansion
\Big \}%
%EndExpansion
\ .
\end{equation*}%
In other words, the maps $\mathrm{B}\mapsto \langle \mathrm{B},m_{\rho
}^{(l)}\rangle _{2}$ from $b_{R}\left( 0\right) $ to $\mathbb{R}$ are
equicontinuous in the weak topology for all $l\in \mathbb{N}$ and $\rho \in
E_{\Lambda _{l}}$ . \hfill $\Box $

We now use Lemma \ref{lemma equicontinuity} to prove a stronger version of
Ascoli's theorem \cite[Theorem A.5]{Rudin} for the weak equicontinuous
family $\{p_{l}(\mathfrak{T}_{\epsilon }\mathrm{B})\}_{l\in \mathbb{N}}$ at
fixed $\epsilon \in \mathbb{R}^{+}$: $p_{l}\left( \mathfrak{T}_{\epsilon }%
\mathrm{B}\right) $ converges to $p_{\infty }\left( \mathfrak{T}_{\epsilon }%
\mathrm{B}\right) $ as $l\rightarrow \infty $ (and not only along a
subsequence), uniformly for any $\mathrm{B}\in b_{R}\left( 0\right) $.

\begin{lemma}[Uniform convergence of pressures]
\label{lemma equicontinuity copy(1)}\mbox{ }\newline
For any $\epsilon \in \mathbb{R}^{+}$, the sequence $\left\{ p_{l}\left(
\mathfrak{T}_{\epsilon }\mathrm{B}\right) \right\} _{l\in \mathbb{N}}$ is a
uniformly Cauchy sequence on any ball $b_{R}\left( 0\right) \subset L^{2}$
of arbitrary radius $R\in \mathbb{R}^{+}$ centered at $0$.
\end{lemma}

\noindent \textit{Proof. }For any $R\in \mathbb{R}^{+}$, the ball $%
b_{R}\left( 0\right) $ is weakly compact in $L^{2}$ (Banach--Alaoglu
theorem) and the weak topology is metrizable on $b_{R}\left( 0\right) $,
see, e.g., \cite[Theorem 10.10]{BruPedra2}. Denote by $d_{R}$ any metric on $%
b_{R}\left( 0\right) $ generating the weak topology. Define also by
\begin{equation*}
\mathfrak{b}_{\delta }\left( \mathrm{B}\right) :=\left\{ \mathrm{C}\in
b_{R}\left( 0\right) :d_{R}(\mathrm{B},\mathrm{C})<\delta \right\}
\end{equation*}%
the weak ball of radius $\delta \in \mathbb{R}^{+}$ centered at $\mathrm{B}%
\in L^{2}$. Balls $\mathfrak{b}_{\delta }\left( \mathrm{B}\right) $ are
clearly weakly open sets in $b_{R}\left( 0\right) $. Thus, using the weak
compactness of $b_{R}\left( 0\right) $ as well as the weak density of $C^{0}$
in $L^{2}$, for any $\delta \in \mathbb{R}^{+}$, there is a finite number $%
N_{\delta }\in \mathbb{N}$ of continuous centers $\{\mathrm{B}%
^{(n)}\}_{n=1}^{N_{\delta }}\subset b_{R}\left( 0\right) \cap C^{0}$ such
that%
\begin{equation}
b_{R}\left( 0\right) =\ \underset{n=1}{\overset{N_{\delta }}{\bigcup }}%
\mathfrak{b}_{\delta }(\mathrm{B}^{(n)})\ .  \label{ball1}
\end{equation}

Fix $\epsilon \in \mathbb{R}^{+}$.\ From Lemma \ref{lemma equicontinuity},
the collection $\left\{ \mathrm{B}\mapsto p_{l}\left( \mathfrak{T}_{\epsilon
}\mathrm{B}\right) \right\} _{l\in \mathbb{N}}$ of maps from $b_{R}\left(
0\right) $ to $\mathbb{R}$ is equicontinuous in the weak topology. See also (%
\ref{pressure0}) and (\ref{important equality2}). By the weak compactness
and metrizability of the ball $b_{R}\left( 0\right) $, the family $\left\{
\mathrm{B}\mapsto p_{l}\left( \mathfrak{T}_{\epsilon }\mathrm{B}\right)
\right\} _{l\in \mathbb{N}}$ is uniformly equicontinuous in the weak
topology: For any $\varepsilon \in \mathbb{R}^{+}$ there is $\delta \in
\mathbb{R}^{+}$ such that, for all $l\in \mathbb{N\cup \{\infty \}}$, $%
\mathrm{B}\in b_{R}\left( 0\right) $ and $\mathrm{C}\in \mathfrak{b}_{\delta
}\left( \mathrm{B}\right) $,
\begin{equation}
\left\vert p_{l}\left( \mathfrak{T}_{\epsilon }\mathrm{B}\right)
-p_{l}\left( \mathfrak{T}_{\epsilon }\mathrm{C}\right) \right\vert \leq
\frac{\varepsilon }{3}\ .  \label{ball2}
\end{equation}%
By Lemma \ref{lemma var pro}, for any $\varepsilon \in \mathbb{R}^{+}$,
there is $L\in \mathbb{R}^{+}$ such that, for any $n\in \{1,\ldots
,N_{\delta }\}$ and integers $l_{1},l_{2}>L$,
\begin{equation}
\left\vert p_{l_{1}}(\mathfrak{T}_{\epsilon }\mathrm{B}^{(n)})-p_{l_{2}}(%
\mathfrak{T}_{\epsilon }\mathrm{B}^{(n)})\right\vert \leq \frac{\varepsilon
}{3}\ .  \label{ball3}
\end{equation}%
By (\ref{ball1}), (\ref{ball2}) and (\ref{ball3}), for any $\varepsilon \in
\mathbb{R}^{+}$, there is $L\in \mathbb{R}^{+}$ such that, for all $\mathrm{B%
}\in b_{R}\left( 0\right) $ and integers $l_{1},l_{2}>L$,
\begin{equation*}
\left\vert p_{l_{1}}\left( \mathfrak{T}_{\epsilon }\mathrm{B}\right)
-p_{l_{2}}\left( \mathfrak{T}_{\epsilon }\mathrm{B}\right) \right\vert \leq
\varepsilon \ .
\end{equation*}%
\hfill $\Box $

We now use Lemmata \ref{lemma equicontinuity} and \ref{lemma equicontinuity
copy(1)} to deduce a stronger version of Theorem \ref{thm limit pressure}:

\begin{theorem}[Infinite volume pressure -- II]
\label{thm limit pressure copy(1)}\mbox{ }\newline
Let $b_{R}\left( 0\right) \subset L^{2}$ be any ball of radius $R\in \mathbb{%
R}^{+}$ centered at $0$, see (\ref{ball}). Then, for any $\epsilon \in
\mathbb{R}^{+}$, one has:\newline
\emph{(i)} The pressure $p_{l}\left( \mathfrak{T}_{\epsilon }\mathrm{B}%
\right) $ converges to $p_{\infty }\left( \mathfrak{T}_{\epsilon }\mathrm{B}%
\right) $ uniformly on $b_{R}\left( 0\right) $, as $l\rightarrow \infty $.
See Theorem \ref{thm limit pressure} (i). \newline
\emph{(ii)} The family $\left\{ \mathrm{B}\mapsto p_{l}\left( \mathfrak{T}%
_{\epsilon }\mathrm{B}\right) \right\} _{l\in \mathbb{N\cup }\left\{ \infty
\right\} }$ of maps from $b_{R}\left( 0\right) $ to $\mathbb{R}$ is
equicontinuous in the weak topology.
\end{theorem}

\noindent \textit{Proof. }Both assertions (i) and (ii) are direct
consequences of Lemmata \ref{lemma equicontinuity} and \ref{lemma
equicontinuity copy(1)} combined with (\ref{pressure0}) and (\ref{important
equality2}). \hfill $\Box $

Recall that $\mathfrak{T}_{\epsilon }$ is defined, for any $\epsilon \in
\mathbb{R}_{0}^{+}$, by (\ref{operator compact}) and (\ref{operator non
compact}) and always satisfy $\Vert \mathfrak{T}_{\epsilon }\Vert \leq 1$.
We now study the variational problems defined, for all $\epsilon \in \mathbb{%
R}_{0}^{+}$ and $l\in \mathbb{N}\cup \left\{ \infty \right\} $, by
\begin{equation}
\mathfrak{B}_{l}^{(\epsilon )}:=\inf_{\mathrm{B}\in \mathcal{B}}\left\{
\frac{1}{2}\Vert \mathrm{B}+\mathrm{B}_{\mathrm{ext}}\Vert
_{2}^{2}-p_{l}\left( \mathfrak{T}_{\epsilon }\mathrm{B}+\mathfrak{T}%
_{\epsilon }\mathrm{B}_{\mathrm{ext}}\right) \right\}  \label{B-infnite}
\end{equation}%
with $\mathcal{B}$ defined by (\ref{definition B}). We break this further
preliminary analysis in three short Lemmata. Note that Lemmata \ref{lemma
var prob2} and \ref{lemma var prob3} both exclude the case $\epsilon =0$.

\begin{lemma}[Variational problems $\mathfrak{B}_{l}^{(\protect\epsilon )}$
-- I]
\label{lemma var prob1}\mbox{ }\newline
For any $\mathrm{B}_{\mathrm{ext}}\in L^{2}$, there is $R\in \mathbb{R}^{+}$
such that, for all $\epsilon \in \mathbb{R}_{0}^{+}$ and $l\in \mathbb{N}%
\cup \left\{ \infty \right\} $,%
\begin{equation*}
\mathfrak{B}_{l}^{(\epsilon )}=\inf_{\mathrm{B}\in \mathcal{B}\cap
b_{R}\left( 0\right) }\left\{ \frac{1}{2}\Vert \mathrm{B}+\mathrm{B}_{%
\mathrm{ext}}\Vert _{2}^{2}-p_{l}\left( \mathfrak{T}_{\epsilon }\mathrm{B}+%
\mathfrak{T}_{\epsilon }\mathrm{B}_{\mathrm{ext}}\right) \right\} \ ,
\end{equation*}%
where $b_{R}\left( 0\right) \subset L^{2}$ is the ball (\ref{ball}) of
radius $R\in \mathbb{R}^{+}$ centered at $0$.
\end{lemma}

\noindent \textit{Proof. }The assertion is a direct consequence of $\Vert
\mathfrak{T}_{\epsilon }\Vert \leq 1$ together with the uniform Lipschitz
continuity of the collection $\left\{ \mathrm{B}\mapsto p_{l}\left( \mathrm{B%
}\right) \right\} _{l\in \mathbb{N\cup }\left\{ \infty \right\} }$ of maps
from $L^{2}(\mathfrak{C};\mathbb{R}^{3})$ to $\mathbb{R}$, see Theorem \ref%
{thm limit pressure} (ii). We omit the details.\hfill $\Box $

\begin{lemma}[Variational problems $\mathfrak{B}_{l}^{(\protect\epsilon )}$
-- II]
\label{lemma var prob2}\mbox{ }\newline
For any $\mathrm{B}_{\mathrm{ext}}\in L^{2}$, all $\epsilon \in \mathbb{R}%
^{+}$ and $l\in \mathbb{N}\cup \left\{ \infty \right\} $, there is a
sequence $\{\mathrm{B}_{\epsilon }^{(l,n)}\}_{n\in \mathbb{N}}\subset
\mathcal{S}_{0}(\mathcal{J})$ converging in norm to $\mathrm{B}_{\epsilon
}^{(l)}\in \mathcal{B}$ as $n\rightarrow \infty $ such that
\begin{equation*}
\underset{n\rightarrow \infty }{\lim }\left\{ \frac{1}{2}\Vert \mathrm{B}%
_{\epsilon }^{(l,n)}+\mathrm{B}_{\mathrm{ext}}\Vert _{2}^{2}-p_{l}(\mathfrak{%
T}_{\epsilon }(\mathrm{B}_{\epsilon }^{(l,n)}+\mathrm{B}_{\mathrm{ext}%
}))\right\} =\frac{1}{2}\Vert \mathrm{B}_{\epsilon }^{(l)}+\mathrm{B}_{%
\mathrm{ext}}\Vert _{2}^{2}-p_{l}(\mathfrak{T}_{\epsilon }(\mathrm{B}%
_{\epsilon }^{(l)}+\mathrm{B}_{\mathrm{ext}}))=\mathfrak{B}_{l}^{(\epsilon
)}.
\end{equation*}
\end{lemma}

\noindent \textit{Proof. }The map $\mathrm{B}\mapsto \Vert \mathrm{B}\Vert
_{2}$ is lower semi--continuous in the weak topology, whereas $\mathrm{B}%
\mapsto p_{l}\left( \mathfrak{T}_{\epsilon }\mathrm{B}\right) $ is weakly
continuous on any ball $b_{R}\left( 0\right) $ for all $\epsilon \in \mathbb{%
R}^{+}$ and $l\in \mathbb{N}\cup \left\{ \infty \right\} $, see Theorem \ref%
{thm limit pressure copy(1)} (ii). Using these properties together with
Lemma \ref{lemma var prob1}, the weak closure $\mathcal{B}$ of $\mathcal{S}%
_{0}(\mathcal{J})$ and the weak compactness of $b_{R}\left( 0\right) $, we
deduce the existence of (a possibly non--unique minimizer) $\mathrm{B}%
_{\epsilon }^{(l)}\in \mathcal{B}$ such that
\begin{equation*}
\mathfrak{B}_{l}^{(\epsilon )}=\frac{1}{2}\Vert \mathrm{B}_{\epsilon }^{(l)}+%
\mathrm{B}_{\mathrm{ext}}\Vert _{2}^{2}-p_{l}(\mathfrak{T}_{\epsilon }%
\mathrm{B}_{\epsilon }^{(l)}+\mathfrak{T}_{\epsilon }\mathrm{B}_{\mathrm{ext}%
})
\end{equation*}%
for any $\epsilon \in \mathbb{R}^{+}$ and $l\in \mathbb{N}\cup \left\{
\infty \right\} $.

Meanwhile, the set $\mathcal{S}_{0}(\mathcal{J})$ with $\mathcal{J}$ defined
by (\ref{biosavat1}) is a convex subset of the Hilbert space $L^{2}$.
Therefore, we infer from \cite[Theorem 3.12]{Rudin} that its weak closure $%
\mathcal{B}$ (\ref{definition B}) coincides with the norm closure of $%
\mathcal{S}_{0}(\mathcal{J})$. In particular, for any $\epsilon \in \mathbb{R%
}^{+}$ and $l\in \mathbb{N}\cup \left\{ \infty \right\} $, there is a
sequence $\{\mathrm{B}_{\epsilon }^{(l,n)}\}_{n\in \mathbb{N}}\subset
\mathcal{S}_{0}(\mathcal{J})$ converging in norm to $\mathrm{B}_{\epsilon
}^{(l)}\in \mathcal{B}$, as $n\rightarrow \infty $. Since, by Theorem \ref%
{thm limit pressure} (ii), the maps $\mathrm{B}\mapsto \Vert \mathrm{B}\Vert
_{2}$ and $\mathrm{B}\mapsto p_{l}\left( \mathfrak{T}_{\epsilon }\mathrm{B}%
\right) $ are both norm continuous, we deduce that
\begin{equation*}
\underset{n\rightarrow \infty }{\lim }\left\{ \frac{1}{2}\Vert \mathrm{B}%
_{\epsilon }^{(l,n)}+\mathrm{B}_{\mathrm{ext}}\Vert _{2}^{2}-p_{l}(\mathfrak{%
T}_{\epsilon }\mathrm{B}_{\epsilon }^{(l,n)}+\mathfrak{T}_{\epsilon }\mathrm{%
B}_{\mathrm{ext}})\right\} =\frac{1}{2}\Vert \mathrm{B}_{\epsilon }^{(l)}+%
\mathrm{B}_{\mathrm{ext}}\Vert _{2}^{2}-p_{l}(\mathfrak{T}_{\epsilon }%
\mathrm{B}_{\epsilon }^{(l)}+\mathfrak{T}_{\epsilon }\mathrm{B}_{\mathrm{ext}%
})
\end{equation*}%
for any $\epsilon \in \mathbb{R}^{+}$ and $l\in \mathbb{N}\cup \left\{
\infty \right\} $.\hfill $\Box $

\begin{lemma}[Variational problems $\mathfrak{B}_{l}^{(\protect\epsilon )}$
-- III]
\label{lemma var prob3}\mbox{ }\newline
For any $\mathrm{B}_{\mathrm{ext}}\in L^{2}$ and $\epsilon \in \mathbb{R}%
^{+} $, $\lim_{l\rightarrow \infty }\mathfrak{B}_{l}^{(\epsilon )}=\mathfrak{%
B}_{\infty }^{(\epsilon )}$.
\end{lemma}

\noindent \textit{Proof.} By Lemmata \ref{lemma var prob1}--\ref{lemma var
prob2}, there are $R\in \mathbb{R}^{+}$ and minimizers $\mathrm{B}_{\epsilon
}^{(l)}\in \mathcal{B}$ of $\mathfrak{B}_{l}^{(\epsilon )}$ satisfying $%
\mathrm{B}_{\epsilon }^{(l)}\in b_{R}\left( 0\right) $ for all $\epsilon \in
\mathbb{R}^{+}$ and $l\in \mathbb{N}\cup \left\{ \infty \right\} $.
Therefore, the lemma follows from the uniform convergence of $p_{l}\left(
\mathfrak{T}_{\epsilon }\mathrm{B}\right) $ towards $p_{\infty }\left(
\mathfrak{T}_{\epsilon }\mathrm{B}\right) $ on $b_{R}\left( 0\right) $, see
Theorem \ref{thm limit pressure copy(1)} (i). \hfill $\Box $

Even if the map $\mathrm{B}\mapsto \Vert \mathrm{B}\Vert _{2}$ is lower
semi--continuous in the weak topology and although Lemma \ref{lemma var
prob1} also holds for $\epsilon =0$ and $l=\infty $, the existence of
minimizer(s) of the variational problem $\mathfrak{B}_{\infty }^{(0)}$ is
far from being clear. Indeed, one can check that the map $\mathrm{B}\mapsto
p_{\infty }(\mathrm{B}+\mathrm{B}_{\mathrm{ext}})$ is not upper
semi--continuous in the weak topology. Nevertheless, $\mathfrak{B}_{\infty
}^{(0)}$ can be obtained from $\mathfrak{B}_{\infty }^{(\epsilon )}$ by
taking the limit $\epsilon \rightarrow 0^{+}$:

\begin{lemma}[Variational problems $\mathfrak{B}_{\infty }^{(\protect%
\epsilon )}$ -- I]
\label{lemma var prob3 copy(10)}\mbox{ }\newline
For any $\mathrm{B}_{\mathrm{ext}}\in L^{2}$, $\lim_{\epsilon \rightarrow
0^{+}}\mathfrak{B}_{\infty }^{(\epsilon )}=\mathfrak{B}_{\infty }^{(0)}$.
\end{lemma}

\noindent \textit{Proof.} Take any sequence $\{\mathrm{B}_{0,n}\}_{n\in
\mathbb{N}}\subset \mathcal{B}$ of approximating minimizers of $\mathfrak{B}%
_{\infty }^{(0)}$, that is,
\begin{equation}
\mathfrak{B}_{\infty }^{(0)}=\underset{n\rightarrow \infty }{\lim }\left\{
\frac{1}{2}\Vert \mathrm{B}_{0,n}+\mathrm{B}_{\mathrm{ext}}\Vert
_{2}^{2}-p_{\infty }(\mathrm{B}_{0,n}+\mathrm{B}_{\mathrm{ext}})\right\} \ .
\label{new var pb1}
\end{equation}%
Then, for any $\epsilon \in \mathbb{R}^{+}$ and every $n\in \mathbb{N}$, $%
\mathfrak{B}_{l}^{(\epsilon )}$ is by definition bounded from above by%
\begin{equation}
\mathfrak{B}_{\infty }^{(\epsilon )}\leq \frac{1}{2}\Vert \mathrm{B}_{0,n}+%
\mathrm{B}_{\mathrm{ext}}\Vert _{2}^{2}-p_{\infty }(\mathfrak{T}_{\epsilon }(%
\mathrm{B}_{0,n}+\mathrm{B}_{\mathrm{ext}}))\ .  \label{new var pb2}
\end{equation}%
The operator $\mathfrak{T}_{\epsilon }$\ converges in the strong topology to
$\mathfrak{T}_{0}$, as $\epsilon \rightarrow 0^{+}$. See (\ref{strong
convergence}) and (\ref{operator non compact}). Moreover, the map $\mathrm{B}%
\mapsto p_{\infty }\left( \mathrm{B}\right) $ is uniformly Lipschitz
continuous, by Theorem \ref{thm limit pressure} (ii). It follows that
\begin{equation*}
\underset{\epsilon \rightarrow 0^{+}}{\lim }p_{\infty }(\mathfrak{T}%
_{\epsilon }(\mathrm{B}_{0,n}+\mathrm{B}_{\mathrm{ext}}))=p_{\infty }(%
\mathrm{B}_{0,n}+\mathrm{B}_{\mathrm{ext}})\ .
\end{equation*}%
Combining this with (\ref{new var pb1}) and (\ref{new var pb2}), we then
obtain the upper bound
\begin{equation}
\underset{\epsilon \rightarrow 0^{+}}{\lim \sup }\ \mathfrak{B}_{\infty
}^{(\epsilon )}\leq \mathfrak{B}_{\infty }^{(0)}
\label{upper bound de funes}
\end{equation}%
for any $\mathrm{B}_{\mathrm{ext}}\in L^{2}$.

On the other hand, we note that the map
\begin{equation}
x\mapsto \mathfrak{h}_{r}\left( x\right) :=\beta ^{-1}\ln \left\{ \cosh
\left( \beta \vartheta \left\vert x\right\vert \right) +\mathrm{e}^{-\lambda
\beta }\cosh \left( \beta g_{r}\right) \right\}  \label{magnetization0bis}
\end{equation}%
from $\mathbb{R}^{3}$ to $\mathbb{R}^{+}$ is a convex function at any fixed $%
(\beta ,\mu ,\lambda ,\gamma ,\vartheta ,r)$. Using this together with $%
\Vert \xi _{\epsilon }\Vert _{1}=1$ and Jensen's inequality, we find that,
for any $\mathrm{B}\in L^{2}$ and $\mathfrak{t}\in \mathfrak{C}$ (a.e),
\begin{equation*}
\mathfrak{h}_{r}\left( \left( \xi _{\epsilon }\ast \mathrm{B}\right) \left(
\mathfrak{t}\right) \right) \leq \xi _{\epsilon }\ast \left( \mathfrak{h}%
_{r}\circ \mathrm{B}\right) \left( \mathfrak{t}\right) \ ,
\end{equation*}%
which in turn implies
\begin{equation}
\mathfrak{F}(r,\xi _{\epsilon }\ast \mathrm{B})\leq \mu +\beta ^{-1}\ln
2-\gamma r+\int_{\mathfrak{C}}\xi _{\epsilon }\ast \left( \mathfrak{h}%
_{r}\circ \mathrm{B}\right) \left( \mathfrak{t}\right) \mathrm{d}^{3}%
\mathfrak{t}  \label{inequality de funes1}
\end{equation}%
for any $\epsilon \in \mathbb{R}^{+}$, $r\in \mathbb{R}_{0}^{+}$ and $%
\mathrm{B}\in L^{2}$, see (\ref{expliciti function2}). Using Fubini's
theorem, for any $\epsilon \in (0,\epsilon _{\xi })$, $r\in \mathbb{R}%
_{0}^{+}$ and $\mathrm{B}\in L^{2}$, we get the equality%
\begin{equation}
\int_{\mathfrak{C}}\xi _{\epsilon }\ast \left( \mathfrak{h}_{r}\circ \mathrm{%
B}\right) \left( \mathfrak{t}\right) \mathrm{d}^{3}\mathfrak{t}=\int_{%
\mathbb{R}^{3}}\xi _{\epsilon }\left( \mathfrak{s}\right) \int_{\mathfrak{C}%
\backslash \mathfrak{C}_{\epsilon }}\mathfrak{h}_{r}\left( \mathrm{B}\left(
\mathfrak{t}-\mathfrak{s}\right) \right) \mathrm{d}^{3}\mathfrak{t}\ \mathrm{%
d}^{3}\mathfrak{s}+\int_{\mathbb{R}^{3}}\xi _{\epsilon }\left( \mathfrak{s}%
\right) \int_{\mathfrak{C}_{\epsilon }}\mathfrak{h}_{r}\left( \mathrm{B}%
\left( \mathfrak{t}-\mathfrak{s}\right) \right) \mathrm{d}^{3}\mathfrak{t}\
\mathrm{d}^{3}\mathfrak{s}\ ,  \label{inequality de funes1+1}
\end{equation}%
where, for any $\epsilon <\epsilon _{\xi }$,
\begin{equation}
\mathfrak{C}_{\epsilon }:=%
%TCIMACRO{\TeXButton{\Big \{}{\Big \{}}%
%BeginExpansion
\Big \{%
%EndExpansion
\mathfrak{t}\in \mathfrak{C}:\ \inf \left\{ \left\vert \mathfrak{t}-%
\mathfrak{s}\right\vert :\mathfrak{s}\in \partial \mathfrak{C}\right\}
>\epsilon R_{\xi }%
%TCIMACRO{\TeXButton{\Big \}}{\Big \}}}%
%BeginExpansion
\Big \}%
%EndExpansion
\ .  \label{radius2}
\end{equation}%
Here, $\partial \mathfrak{C}$ is the boundary of $\mathfrak{C}$ and $%
\epsilon _{\xi }:=1/(2R_{\xi })$ with $R_{\xi }$ being the radius of the
support of the function $\xi \in C_{0}^{\infty }$, see (\ref{xi eps m}) and (%
\ref{radius1}). By $\Vert \xi _{\epsilon }\Vert _{1}=1$ together with the
Cauchy--Schwarz inequality and
\begin{equation*}
\left\vert \mathfrak{h}_{r}(x)\right\vert \leq D\left( \left\vert
x\right\vert +1\right) \ ,\qquad x\in \mathbb{R}^{3},
\end{equation*}%
for some finite constant $D\in \mathbb{R}^{+}$, the absolute value of the
first integral in the right hand side of (\ref{inequality de funes1+1}) is
bounded by
\begin{equation}
D(|\mathfrak{C}\backslash \mathfrak{C}_{\epsilon }|^{1/2}\Vert \mathrm{B}%
\Vert _{2}+|\mathfrak{C}\backslash \mathfrak{C}_{\epsilon }|)
\label{inequality de funes1+2}
\end{equation}%
for any $\epsilon \in (0,\epsilon _{\xi })$, $r\in \mathbb{R}_{0}^{+}$ and $%
\mathrm{B}\in L^{2}$. Meanwhile, using similar arguments,
\begin{equation}
\left\vert \int_{\mathbb{R}^{3}}\xi _{\epsilon }\left( \mathfrak{s}\right)
\int_{\mathfrak{C}_{\epsilon }}\mathfrak{h}_{r}\left( \mathrm{B}\left(
\mathfrak{t}-\mathfrak{s}\right) \right) \mathrm{d}^{3}\mathfrak{t}\ \mathrm{%
d}^{3}\mathfrak{s}-\int_{\mathfrak{C}}\mathfrak{h}_{r}\left( \mathrm{B}%
\left( \mathfrak{t}\right) \right) \mathrm{d}^{3}\mathfrak{t}\right\vert
\leq D(|\mathfrak{C}\backslash \mathfrak{C}_{\epsilon }|^{1/2}\Vert \mathrm{B%
}\Vert _{2}+|\mathfrak{C}\backslash \mathfrak{C}_{\epsilon }|)\ .
\label{inequality de funes1+3}
\end{equation}%
From (\ref{inequality de funes1})--(\ref{inequality de funes1+3}) and
Theorem \ref{thm limit pressure} (i), we thus deduce that%
\begin{equation}
p_{\infty }(\mathfrak{T}_{\epsilon }(\mathrm{B}+\mathrm{B}_{\mathrm{ext}%
}))\leq p_{\infty }(\mathrm{B}+\mathrm{B}_{\mathrm{ext}})+2D(|\mathfrak{C}%
\backslash \mathfrak{C}_{\epsilon }|^{1/2}\Vert \mathrm{B}\Vert _{2}+|%
\mathfrak{C}\backslash \mathfrak{C}_{\epsilon }|)  \label{de funes 0}
\end{equation}%
for any $\epsilon \in (0,\epsilon _{\xi })$, $\mathrm{B}_{\mathrm{ext}}\in
L^{2}$ and all $\mathrm{B}\in b_{R}\left( 0\right) $ with $R\in \mathbb{R}%
^{+}$ being any fixed radius. As a consequence, for any $\mathrm{B}_{\mathrm{%
ext}}\in L^{2}$, there is $R\in \mathbb{R}^{+}$ such that, for all $\epsilon
\in (0,\epsilon _{\xi })$,%
\begin{equation}
\mathfrak{B}_{\infty }^{(\epsilon )}\geq \mathfrak{B}_{\infty }^{(0)}-2D(R|%
\mathfrak{C}\backslash \mathfrak{C}_{\epsilon }|^{1/2}+|\mathfrak{C}%
\backslash \mathfrak{C}_{\epsilon }|)\ ,  \label{lower bound de funes}
\end{equation}%
because of Lemma \ref{lemma var prob1}.

Since $|\mathfrak{C}\backslash \mathfrak{C}_{\epsilon }|=\mathcal{O}%
(\epsilon )$, we therefore combine the lower bound (\ref{lower bound de
funes}) in the limit $\epsilon \rightarrow 0^{+}$ with the upper bound (\ref%
{upper bound de funes}) to arrive at the assertion.\hfill $\Box $

We can now deduce that minimizers of $\mathfrak{B}_{\infty }^{(\epsilon )}$
are approximating minimizers of the variational problem $\mathfrak{B}%
_{\infty }^{(0)}$:

\begin{lemma}[Variational problems $\mathfrak{B}_{\infty }^{(\protect%
\epsilon )}$ -- II]
\label{lemma var prob3 copy(12)}\mbox{ }\newline
Let $\mathrm{B}_{\mathrm{ext}}\in L^{2}$. Then, any family $\{\mathrm{B}%
_{\epsilon }\}_{\epsilon \in \mathbb{R}^{+}}\subset \mathcal{B}$ of
minimizers $\mathrm{B}_{\epsilon }$ of $\mathfrak{B}_{\infty }^{(\epsilon )}$
minimizes $\mathfrak{B}_{\infty }^{(0)}$ in the limit $\epsilon \rightarrow
0^{+}$.
\end{lemma}

\noindent \textit{Proof. }Take any family $\{\mathrm{B}_{\epsilon
}\}_{\epsilon \in \mathbb{R}^{+}}\subset \mathcal{B}\cap b_{R}\left(
0\right) $ of minimizers $\mathrm{B}_{\epsilon }$ of $\mathfrak{B}_{\infty
}^{(\epsilon )}$, see Lemmata \ref{lemma var prob1}--\ref{lemma var prob2}.
By Lemma \ref{lemma var prob3 copy(10)}\ and (\ref{de funes 0}),%
\begin{equation}
\underset{\epsilon \rightarrow 0^{+}}{\lim }\left\{ \frac{1}{2}\Vert \mathrm{%
B}_{\epsilon }+\mathrm{B}_{\mathrm{ext}}\Vert _{2}^{2}-p_{\infty }(\mathrm{B}%
_{\epsilon }+\mathrm{B}_{\mathrm{ext}})\right\} =\mathfrak{B}_{\infty
}^{(0)}\ .  \label{de funes encore 00}
\end{equation}%
In other words, $\{\mathrm{B}_{\epsilon }\}_{\epsilon \in \mathbb{R}^{+}}$
is a family of approximating minimizers of $\mathfrak{B}_{\infty }^{(0)}$%
.\hfill $\Box $

\begin{remark}
\label{lemma var prob3 copy(11)}\mbox{ }\newline
By Lemma \ref{lemma var prob1}, the Banach--Alaoglu theorem and the
separability of $L^{2}$, any family $\{\mathrm{B}_{\epsilon }\}_{\epsilon
\in \mathbb{R}^{+}}$ of minimizers of $\mathfrak{B}_{\infty }^{(\epsilon )}$
converges in the weak topology and along a subsequence to some $\mathrm{B}%
_{0}\in \mathcal{B}\cap b_{R}\left( 0\right) $, as $\epsilon \rightarrow
0^{+}$. In general, $\mathrm{B}_{0}$ may not be a minimizer of $\mathfrak{B}%
_{\infty }^{(0)}$. Sufficient conditions to ensure that $\mathrm{B}_{0}$ is
a minimizer of $\mathfrak{B}_{\infty }^{(0)}$ are given in Theorem \ref%
{lemma var prob3 copy(5)}.
\end{remark}

We are now in position to obtain the magnetic pressures $\mathcal{P}_{\infty
}^{(\epsilon )}$ as the variational problems $-\mathfrak{B}_{\infty
}^{(\epsilon )}$ for all $\epsilon \in \mathbb{R}_{0}^{+}$. We start by
considering the case $\epsilon \in \mathbb{R}^{+}$. The case $\epsilon =0$
will then be a direct consequence of Lemma \ref{lemma var prob3 copy(10)}.

\begin{theorem}[Infinite volume magnetic pressure]
\label{coolthml1}\mbox{ }\newline
Let $\mathrm{B}_{\mathrm{ext}}=\mathcal{S}_{0}(\mathrm{j}_{\mathrm{ext}})$
with $\mathrm{j}_{\mathrm{ext}}\in C_{0}^{\infty }\cap P^{\bot }\mathfrak{H}$%
. Then,%
\begin{equation*}
\mathcal{P}_{\infty }^{(\epsilon )}:=\lim_{l\rightarrow \infty }\mathcal{P}%
_{l}^{(\epsilon )}=-\mathfrak{B}_{\infty }^{(\epsilon )}\ ,\qquad \epsilon
\in \mathbb{R}^{+}.
\end{equation*}
\end{theorem}

\noindent \textit{Proof. }By (\ref{new set of states}), note first that
\begin{equation}
\mathcal{P}_{l}^{(\epsilon )}\leq -\inf_{j\in \mathcal{J}_{l}}\left\{ \frac{1%
}{2}\Vert \mathcal{S}_{0}(j)+\mathrm{B}_{\mathrm{ext}}\Vert
_{2}^{2}-p_{l}\left( \mathfrak{T}_{\epsilon }(\mathcal{S}_{0}(j)+\mathrm{B}_{%
\mathrm{ext}})\right) \right\}  \label{extra1}
\end{equation}%
with $\mathcal{J}_{l}$ being the set defined by
\begin{equation}
\mathcal{J}_{l}:=\{j\in \mathfrak{H}:\Vert j^{\parallel }\Vert _{\mathfrak{H}%
}\leq l^{-\varkappa }\}\cap C_{0}^{\infty }(\mathfrak{C};\mathbb{R}^{3})\ .
\label{courant J_l}
\end{equation}%
Since $\ker \mathcal{S}=P^{\parallel }\mathfrak{H}$, i.e., $\mathcal{S}(j)=%
\mathcal{S}(j^{\bot })$, we thus infer from (\ref{definition B}), (\ref%
{B-infnite}) and (\ref{extra1}) that $\mathcal{P}_{l}^{(\epsilon )}\leq -%
\mathfrak{B}_{l}^{(\epsilon )}$ for any $\epsilon \in \mathbb{R}^{+}$ and $%
l\in \mathbb{N}$. In particular, in the limit $l\rightarrow \infty $ one
gets $\mathcal{P}_{\infty }^{(\epsilon )}\leq -\mathfrak{B}_{\infty
}^{(\epsilon )}$ for any $\epsilon \in \mathbb{R}^{+}$, using Lemma \ref%
{lemma var prob3}. It remains to show that, for any $\epsilon \in \mathbb{R}%
^{+}$, $-\mathfrak{B}_{\infty }^{(\epsilon )}$ is a lower bound of the
magnetic pressure $\mathcal{P}_{\infty }^{(\epsilon )}$.

By Lemma \ref{lemma var prob2}, there is a norm convergent sequence $\{%
\mathrm{B}_{\epsilon }^{(n)}\}_{n\in \mathbb{N}}\subset \mathcal{S}_{0}(%
\mathcal{J})$ such that%
\begin{equation}
\mathfrak{B}_{\infty }^{(\epsilon )}=\underset{n\rightarrow \infty }{\lim }%
\left\{ \frac{1}{2}\Vert \mathrm{B}_{\epsilon }^{(n)}+\mathrm{B}_{\mathrm{ext%
}}\Vert _{2}^{2}-p_{\infty }(\mathfrak{T}_{\epsilon }(\mathrm{B}_{\epsilon
}^{(n)}+\mathrm{B}_{\mathrm{ext}}))\right\} \ .  \label{norm limit0}
\end{equation}%
Moreover, for any $n\in \mathbb{N}$, there is by definition a current
density $\mathrm{j}_{\epsilon }^{(n)}\in \mathcal{J}$ generating the
magnetic induction $\mathrm{B}_{\epsilon }^{(n)}=\mathcal{S}_{0}(\mathrm{j}%
_{\epsilon }^{(n)})$. Therefore, by Lemmata \ref{lemma free energy}, \ref%
{propositioncool} and \ref{lemma equicontinuity} together with (\ref%
{projection important}) and (\ref{magnetic energy2}), for any fixed $n\in
\mathbb{N}$, there is a sequence $\{\rho _{l}\}_{l\in \mathbb{N}}$ of
quasi--divergence--free states $\rho _{l}\in E_{\Lambda _{l}}^{\bot }$
satisfying%
\begin{equation}
\underset{l\rightarrow \infty }{\lim }\Vert B_{\rho _{l}}^{(l)}-\mathrm{B}%
_{\epsilon }^{(n)}\Vert _{2}=0\ ,\qquad n\in \mathbb{N}\ ,
\label{norm limit}
\end{equation}%
and
\begin{equation}
\underset{l\rightarrow \infty }{\lim }\left\{ f_{l}(\mathfrak{T}_{\epsilon
}B_{\rho _{l}}^{(l)},\rho _{l})-\underset{\rho \in E_{\Lambda _{l}}}{\inf }%
f_{l}(\mathfrak{T}_{\epsilon }(\mathrm{B}_{\epsilon }^{(n)}+\mathrm{B}_{%
\mathrm{ext}}),\rho )\right\} =0\ .  \label{norm limitbis}
\end{equation}%
Hence, the lower bound $\lim_{l\rightarrow \infty }\mathcal{P}%
_{l}^{(\epsilon )}\geq -\mathfrak{B}_{\infty }^{(\epsilon )}$ for any $%
\epsilon \in \mathbb{R}^{+}$ is a direct consequence of (\ref{important
equality}), (\ref{magnetic pressure}), (\ref{norm limit0}), (\ref{norm limit}%
) and (\ref{norm limitbis}) together with Theorem \ref{thm limit pressure}
(i). \hfill $\Box $

\begin{corollary}[Magnetic pressure for $\protect\epsilon =0$]
\label{corollary de funes}\mbox{ }\newline
Let $\mathrm{B}_{\mathrm{ext}}=\mathcal{S}_{0}(\mathrm{j}_{\mathrm{ext}})$
with $\mathrm{j}_{\mathrm{ext}}\in C_{0}^{\infty }\cap P^{\bot }\mathfrak{H}$%
. Then,
\begin{equation*}
\mathcal{P}_{\infty }:=\underset{\epsilon \rightarrow 0^{+}}{\lim }\mathcal{P%
}_{\infty }^{(\epsilon )}=-\mathfrak{B}_{\infty }^{(0)}\ .
\end{equation*}
\end{corollary}

\noindent \textit{Proof. }See Lemma \ref{lemma var prob3 copy(10)} and
Theorem \ref{coolthml1}.\hfill $\Box $

It remains to establish the relation between the solutions of the
variational problem $\mathfrak{B}_{\infty }^{(\epsilon )}$ for $\epsilon \in
\mathbb{R}^{+}$ and the sets $\mathbb{B}_{\epsilon }^{(\pm )}$ of all weak ($%
-$) and norm ($+$) cluster points of self--generated magnetic inductions $%
B_{\omega _{\epsilon ,l}}^{(l)}$, see (\ref{set of magnetic inductions}).
This result is a relatively direct corollary of Theorems \ref{thm limit
pressure copy(1)} and \ref{coolthml1}.

\begin{corollary}[Magnetic inductions]
\label{lemma var prob3 copy(1)}\mbox{ }\newline
Let $\epsilon \in \mathbb{R}^{+}$ and $\mathrm{B}_{\mathrm{ext}}=\mathcal{S}%
_{0}(\mathrm{j}_{\mathrm{ext}})$ with $\mathrm{j}_{\mathrm{ext}}\in
C_{0}^{\infty }\cap P^{\bot }\mathfrak{H}$. Then, $\mathbb{B}_{\epsilon
}^{(+)}=\mathbb{B}_{\epsilon }^{(-)}\subset \mathcal{B}$ is a set of
minimizers of $\mathfrak{B}_{\infty }^{(\epsilon )}$.
\end{corollary}

\noindent \textit{Proof. }The inclusion $\mathbb{B}_{\epsilon
}^{(+)}\subseteq \mathbb{B}_{\epsilon }^{(-)}$ is clear and $\mathbb{B}%
_{\epsilon }^{(-)}\neq \emptyset $, by weak compactness of balls. Take any $%
\mathrm{B}_{\epsilon }\in \mathbb{B}_{\epsilon }^{(-)}$. By definition of $%
\mathbb{B}_{\epsilon }^{(-)}$, there is a subsequence $\{l_{n}\}_{n\in
\mathbb{N}}$ such that $B_{\omega _{\epsilon ,l_{n}}}^{(l_{n})}$ converges
in the weak topology to $\mathrm{B}_{\epsilon }\in \mathbb{B}_{\epsilon }$,
as $n\rightarrow \infty $. Note that $\{B_{\omega _{\epsilon
,l_{n}}}^{(l_{n})}\}_{n\in \mathbb{N}}\subset \mathcal{S}_{0}(\mathcal{J}%
_{l})$ with $\mathcal{J}_{l}$ being the set defined by (\ref{courant J_l}).
Since $\ker \mathcal{S}=P^{\parallel }\mathfrak{H}$, we have $\{B_{\omega
_{\epsilon ,l_{n}}}^{(l_{n})}\}_{n\in \mathbb{N}}\subset \mathcal{S}_{0}(%
\mathcal{J})$. Therefore,
\begin{equation*}
\mathbb{B}_{\epsilon }^{(-)}\subset \mathcal{B}:=\overline{\mathcal{S}_{0}(%
\mathcal{J})}\ ,
\end{equation*}
see (\ref{definition B}). Using Theorem \ref{thm limit pressure copy(1)}
(ii), the weak lower semi--continuity of the map $\mathrm{B}\mapsto \Vert
\mathrm{B}\Vert _{2}$ as well as Theorem \ref{coolthml1}, we also deduce
that $\mathrm{B}_{\epsilon }\in \mathbb{B}_{\epsilon }^{(-)}$ must be a
solution of the variational problem $\mathfrak{B}_{\infty }^{(\epsilon )}$
and
\begin{equation}
\underset{n\rightarrow \infty }{\lim }\Vert B_{\omega _{\epsilon
,l_{n}}}^{(l_{n})}\Vert _{2}=\Vert \mathrm{B}_{\epsilon }\Vert _{2}\ .
\label{limite utile}
\end{equation}%
To prove the latter, use the equality%
\begin{equation}
\Vert B_{\omega _{\epsilon ,l_{n}}}^{(l_{n})}+\mathrm{B}\Vert _{2}^{2}=\Vert
B_{\omega _{\epsilon ,l_{n}}}^{(l_{n})}\Vert _{2}^{2}+\Vert \mathrm{B}\Vert
_{2}^{2}+2\langle B_{\omega _{\epsilon ,l_{n}}}^{(l_{n})},\mathrm{B}\rangle
_{2}  \label{a la con plus}
\end{equation}%
for $\mathrm{B}=\mathrm{B}_{\mathrm{ext}}$, as well as the weak continuity
of the map
\begin{equation*}
\mathrm{B}\mapsto 2\langle \mathrm{B},\mathrm{B}_{\mathrm{ext}}\rangle
_{2}+p_{\infty }(\mathfrak{T}_{\epsilon }(\mathrm{B}+\mathrm{B}_{\mathrm{ext}%
}))\ ,
\end{equation*}%
see Theorem \ref{thm limit pressure copy(1)} (ii). It follows that $\mathbb{B%
}_{\epsilon }^{(-)}$ is a set of minimizers of $\mathfrak{B}_{\infty
}^{(\epsilon )}$. Using (\ref{limite utile}) and (\ref{a la con plus}) with $%
\mathrm{B}=-\mathrm{B}_{\epsilon }$, we deduce that $B_{\omega _{\epsilon
,l_{n}}}^{(l_{n})}$ converges in norm to $\mathrm{B}_{\epsilon }$, as $%
l\rightarrow \infty $. In other words, $\mathbb{B}_{\epsilon
}^{(-)}\subseteq \mathbb{B}_{\epsilon }^{(+)}$. \hfill $\Box $

By Lemma \ref{lemma var prob3 copy(12)}, this corollary also links in the
limit $\epsilon \rightarrow 0^{+}$ the sets $\{\mathbb{B}_{\epsilon }^{(\pm
)}\}_{\epsilon \in \mathbb{R}^{+}}$ to the approximating minimizers of the
variational problem $\mathfrak{B}_{\infty }^{(0)}$.

We analyze now in detail the variational $\mathfrak{B}_{\infty }^{(\epsilon
)}$ for all $\epsilon \in \mathbb{R}_{0}^{+}$. In the limit $\beta
\rightarrow \infty $ of low temperatures, recall (\ref{magnetizationbis}),
that is, $\left\vert \mathrm{M}_{\beta ,\mathfrak{D}}\right\vert =\mathcal{O}%
(\mathrm{e}^{-\beta (\mathrm{h}_{\mathrm{c}}-\mathrm{h})})$ whenever (\ref%
{magnetizationbisbis}) is satisfied. It means that, as $\beta \rightarrow
\infty $, the pressure $p_{\infty }\left( \mathrm{B}\right) $ does not
depend much on magnetic inductions $\mathrm{B}\in \mathcal{B}$ that satisfy (%
\ref{magnetizationbisbis}) on $\mathfrak{C}$. Therefore, we first study the
variational problem (\ref{var prob temp zero-0}), that is,
\begin{equation}
\mathfrak{A}:=\frac{1}{2}\inf_{\mathrm{B}\in \mathcal{B}}\ \left\Vert
\mathrm{B}+\mathrm{B}_{\mathrm{ext}}\right\Vert _{2}^{2}\ .
\label{var prob temp zero}
\end{equation}

\begin{lemma}[Variational problem $\mathfrak{A}$]
\label{lemma var prob3 copy(2)}\mbox{ }\newline
Let $\mathrm{B}_{\mathrm{ext}}=\mathcal{S}_{0}(\mathrm{j}_{\mathrm{ext}})$
with $\mathrm{j}_{\mathrm{ext}}\in C_{0}^{\infty }\cap P^{\bot }\mathfrak{H}$%
. Then, there is a unique minimizer $\mathrm{B}_{\mathrm{int}}\in \mathcal{B}
$ of $\mathfrak{A}$. The latter fulfills $\mathrm{B}_{\mathrm{int}}=-\mathrm{%
B}_{\mathrm{ext}}$ a.e. in $\mathfrak{C}$.
\end{lemma}

\noindent \textit{Proof. }By \cite[Theorem 3.12]{Rudin}, recall that the
weak closure $\mathcal{B}$ (\ref{definition B}) coincides with the norm
closure of $\mathcal{S}_{0}(\mathcal{J})$. By linearity of the Biot--Savart
operator $\mathcal{S}$, we then conclude that $\mathcal{B}\subset P^{\bot
}L^{2}$, equipped with the $L^{2}$--scalar product, is a sub--Hilbert space
of $L^{2}$. As a consequence, by strict convexity and weak lower
semi--continuity of the map $\mathrm{B}\mapsto \Vert \mathrm{B}\Vert _{2}$,
there is a unique minimizer $\mathrm{B}_{\mathrm{int}}\in \mathcal{B}$
satisfying the Euler--Lagrange equations%
\begin{equation}
\langle \mathrm{B}_{\mathrm{int}}+\mathrm{B}_{\mathrm{ext}},\mathrm{B}%
\rangle _{2}=0\ ,\qquad \mathrm{B}\in \mathcal{B}\ .  \label{eq tempezero1}
\end{equation}

Now, since the space $C_{0}^{\infty }(\mathfrak{C};\mathbb{R}^{3})$\textit{\
}is dense in $L^{2}(\mathfrak{C};\mathbb{R}^{3})$, it suffices to prove (\ref%
{eq tempezero1}) for all $B\in C_{0}^{\infty }(\mathfrak{C};\mathbb{R}^{3})$
(instead of $\mathrm{B}\in \mathcal{B}$). Take
\begin{equation*}
j^{\bot }:=\nabla \times B\in C_{0}^{\infty }(\mathfrak{C};\mathbb{R}^{3})
\end{equation*}
for any $B\in C_{0}^{\infty }(\mathfrak{C};\mathbb{R}^{3})$. Then, clearly, $%
\nabla \cdot j^{\bot }=0$ and thus $j^{\bot }\in \mathcal{J}$. Moreover, as
explained in Remark \ref{remark projection magn}, $\mathcal{S}_{0}(j^{\bot
})=P^{\bot }B$ and $P^{\bot }B\in \mathcal{B}$. Since, by definition of the
Biot--Savart operator, $\mathrm{B}_{\mathrm{int}}\in P^{\bot }L^{2}$ and, by
assumption, $\mathrm{B}_{\mathrm{ext}}\in P^{\bot }L^{2}$, we infer from (%
\ref{eq tempezero1}) at $\mathrm{B}=P^{\bot }B$ that
\begin{equation}
\langle \mathrm{B}_{\mathrm{int}}+\mathrm{B}_{\mathrm{ext}},P^{\bot
}B\rangle _{2}=\langle \mathrm{B}_{\mathrm{int}}+\mathrm{B}_{\mathrm{ext}%
},B\rangle _{2}=0  \label{eq tempezero2}
\end{equation}%
for any $B\in C_{0}^{\infty }(\mathfrak{C};\mathbb{R}^{3})$. Indeed, $%
P^{\parallel }$ and $P^{\bot }$ are mutually orthogonal projections. See,
e.g., (\ref{projection important}).\hfill $\Box $

Since $\mathcal{B}$ (\ref{definition B}) is a closed space with respect to
the $L^{2}$--norm (cf. \cite[Theorem 3.12]{Rudin}), we use (\ref{magnetic
energy2}) to observe that $\mathcal{B}=\mathcal{S}(\overline{\mathcal{J}})$,
where
\begin{equation*}
\overline{\mathcal{J}}\subseteq P^{\bot }\mathfrak{H}\cap \overline{%
C_{0}^{\infty }(\mathfrak{C};\mathbb{R}^{3})}^{\left\Vert -\right\Vert _{%
\mathfrak{H}}}
\end{equation*}%
is the (norm) closure of the set $\mathcal{J}$. As a consequence, Equations (%
\ref{magnetic energy2}) and (\ref{var prob temp zero}) yield (\ref{var prob
temp zerobis-0}), that is,
\begin{equation}
\mathfrak{A}=\mathfrak{J}:=\frac{1}{2}\inf_{j^{\bot }\in \overline{\mathcal{J%
}}}\ \left\Vert j^{\bot }+\mathrm{j}_{\mathrm{ext}}\right\Vert _{\mathfrak{H}%
}^{2}\ .  \label{var prob temp zerobis}
\end{equation}%
In particular, there is a one--to--one map from minimizers of (\ref{var prob
temp zero}) and minimizers of (\ref{var prob temp zerobis}). By (\ref%
{magnetic energy2bis}) and (\ref{eq tempezero1}), the unique minimizer $%
\mathrm{j}_{\mathrm{int}}^{\bot }\in \overline{\mathcal{J}}$ satisfies the
Euler--Lagrange equations%
\begin{equation}
\langle \mathrm{j}_{\mathrm{int}}^{\bot }+\mathrm{j}_{\mathrm{ext}},j^{\bot
}\rangle _{\mathfrak{H}}=0\ ,\qquad j^{\bot }\in \overline{\mathcal{J}}\ .
\label{euler courants}
\end{equation}%
The latter implies that $\mathrm{j}_{\mathrm{int}}^{\bot }$ is a
distribution supported on the boundary $\partial \mathfrak{C}$ of $\mathfrak{%
C}$, provided (\ref{support condition}) holds:

\begin{lemma}[Variational problem $\mathfrak{J}$]
\label{lemma var prob3 copy(3)}\mbox{ }\newline
Let $\mathrm{B}_{\mathrm{ext}}=\mathcal{S}_{0}(\mathrm{j}_{\mathrm{ext}})$
with $\mathrm{j}_{\mathrm{ext}}\in C_{0}^{\infty }\cap P^{\bot }\mathfrak{H}$%
. Assume additionally (\ref{support condition}), that is, $\mathrm{supp}(%
\mathrm{j}_{\mathrm{ext}})\subset \mathbb{R}\backslash \mathfrak{C}$. Then,
there is a unique minimizer $\mathrm{j}_{\mathrm{int}}^{\bot }\in \overline{%
\mathcal{J}}$ of $\mathfrak{J}$ which, as a distribution, is supported on
the boundary $\partial \mathfrak{C}$ of $\mathfrak{C}$.
\end{lemma}

\noindent \textit{Proof. }Uniqueness and existence is a direct consequence
of Lemma \ref{lemma var prob3 copy(2)}, as explained after (\ref{var prob
temp zerobis}). Now, for any $\phi \in C_{0}^{\infty }(\mathfrak{C};\mathbb{R%
}^{3})$ with support $\mathrm{supp}(\phi )\subset \mathfrak{C}$, we apply (%
\ref{euler courants}) to $j^{\bot }:=\nabla \times \nabla \times \phi $ as
well as (\ref{inequality idiote}) in Fourier space to deduce that
\begin{equation}
\langle \mathrm{j}_{\mathrm{int}}^{\bot }+\mathrm{j}_{\mathrm{ext}},-\Delta
\phi \rangle _{\mathfrak{H}}=0\ ,\qquad \phi \in C_{0}^{\infty }(\mathfrak{C}%
;\mathbb{R}^{3})\ .  \label{euler courants2}
\end{equation}%
The current density $\mathrm{j}_{\mathrm{ext}}$ and the minimizer $\mathrm{j}%
_{\mathrm{int}}^{\bot }$ both create vector potentials respectively equal to
$\mathcal{A}\left( \mathrm{j}_{\mathrm{ext}}\right) $ and $\mathcal{A}\left(
\mathrm{j}_{\mathrm{int}}^{\bot }\right) $, see (\ref{vector potential10})--(%
\ref{vector potential2bis}). Since $\mathrm{j}_{\mathrm{ext}}\in
C_{0}^{\infty }\cap P^{\bot }\mathfrak{H}$ is by assumption supported on $%
\mathbb{R}\backslash \mathfrak{C}$, $-\Delta \mathcal{A}\left( \mathrm{j}_{%
\mathrm{ext}}\right) =0$ (in the strong sense) on\ the unit box $\mathfrak{C}
$, which, together with (\ref{euler courants2}), implies that
\begin{equation}
-\Delta \mathcal{A}\left( \mathrm{j}_{\mathrm{int}}^{\bot }\right) =0\
,\qquad \mathrm{on\ }C_{0}^{\infty }(\mathfrak{C};\mathbb{R}^{3})\ .
\label{euler courants4}
\end{equation}%
Combining this equality with (\ref{vector potential3}) and $\mathrm{j}_{%
\mathrm{int}}^{\bot }\in \overline{\mathcal{J}}$, we arrive at the
assertion.\hfill $\Box $

Therefore, we deduce from Lemmata \ref{lemma var prob3 copy(2)} and \ref%
{lemma var prob3 copy(3)} that the solution $\mathrm{B}_{\mathrm{int}}=%
\mathcal{S}(\mathrm{j}_{\mathrm{int}}^{\bot })$ of the variational problem $%
\mathfrak{A}$ (\ref{var prob temp zero}) comes from surface currents $%
\mathrm{j}_{\mathrm{int}}^{\bot }\in \overline{\mathcal{J}}$ which
annihilate a.e. all the total magnetic induction inside the bulk $\mathfrak{C%
}$. We take advantage of this property to analyze the full variational
problem $\mathfrak{B}_{\infty }^{(\epsilon )}$ (\ref{B-infnite}).

Like in Lemma \ref{lemma var prob3 copy(2)}, we start with a first
consequence of the Euler--Lagrange equations associated with $\mathfrak{B}%
_{\infty }^{(\epsilon )}$ for any $\epsilon \in \mathbb{R}_{0}^{+}$. Recall
that $\mathfrak{B}_{\infty }^{(\epsilon )}$ has minimizer(s) $\mathrm{B}%
_{\epsilon }\in \mathcal{B}$ for all $\epsilon \in \mathbb{R}^{+}$ (cf.
Lemma \ref{lemma var prob2}), but the existence of minimizer(s) of the
variational problem $\mathfrak{B}_{\infty }^{(0)}$ is unclear.

\begin{lemma}[Variational problems $\mathfrak{B}_{\infty }^{(\protect%
\epsilon )}$ -- III]
\label{lemma var prob3 copy(9)}\mbox{ }\newline
Let $\epsilon \in \mathbb{R}_{0}^{+}$ and $\mathrm{B}_{\mathrm{ext}}=%
\mathcal{S}_{0}(\mathrm{j}_{\mathrm{ext}})$ with $\mathrm{j}_{\mathrm{ext}%
}\in C_{0}^{\infty }\cap P^{\bot }\mathfrak{H}$. Assume that $\mathrm{B}%
_{\epsilon }$ is a minimizer of $\mathfrak{B}_{\infty }^{(\epsilon )}$. Then,%
\begin{equation*}
\mathrm{B}_{\epsilon }-\mathrm{B}_{\mathrm{int}}=\mathrm{M}_{\beta }^{\bot }(%
\mathfrak{T}_{\epsilon }\mathrm{B}_{\epsilon }+\mathfrak{T}_{\epsilon }%
\mathrm{B}_{\mathrm{ext}})\quad \text{a.e. in }\mathfrak{C}\text{ .}
\end{equation*}%
Here, $\mathrm{B}_{\mathrm{int}}$ is the unique minimizer of $\mathfrak{A}$
(Lemma \ref{lemma var prob3 copy(2)}), whereas $\mathrm{M}_{\beta }^{\bot
}=P^{\bot }\mathrm{M}_{\beta }\in C^{\infty }$ is the transverse component
of the magnetization density $\mathrm{M}_{\beta }\equiv \mathrm{M}_{\beta }(%
\mathrm{B})$ defined on $\mathbb{R}^{3}$ by (\ref{magnetization0}) for all $%
\mathrm{B}\in L^{2}$.
\end{lemma}

\noindent \textit{Proof. }For any $r\in \mathbb{R}_{0}^{+}$ and all $\mathrm{%
B},\mathrm{C}\in L^{2}$, the map $t\mapsto \mathfrak{F}\left( r,\mathrm{B}+t%
\mathrm{C}\right) $ from $\mathbb{R}$ to $\mathbb{R}$ is differentiable.
Explicit computations show that
\begin{equation}
\partial _{t}\mathfrak{F}\left( r,\mathrm{B}+t\mathrm{C}\right)
|_{t=0}=\int_{\mathbb{R}^{3}}\frac{\mathbf{1}[\mathfrak{t}\in \mathfrak{C}]\
\vartheta \sinh \left( \beta h_{\mathfrak{t}}\right) }{\cosh \left( \beta h_{%
\mathfrak{t}}\right) +\mathrm{e}^{-\beta \lambda }\cosh \left( \beta
g_{r}\right) }\frac{\mathrm{B}\left( \mathfrak{t}\right) }{\left\vert
\mathrm{B}\left( \mathfrak{t}\right) \right\vert }\cdot \mathrm{C}\left(
\mathfrak{t}\right) \ \mathrm{d}^{3}\mathfrak{t}\ .  \label{eq sup trivial 1}
\end{equation}%
On the other hand, by (\ref{B-infnite}) combined with Theorem \ref{thm limit
pressure} (i),
\begin{equation}
\mathfrak{B}_{\infty }^{(\epsilon )}=\underset{r\geq 0}{\inf }\inf_{\mathrm{B%
}\in \mathcal{B}}\left\{ \frac{1}{2}\Vert \mathrm{B}+\mathrm{B}_{\mathrm{ext}%
}\Vert _{2}^{2}-\mathfrak{F}\left( r,\mathfrak{T}_{\epsilon }\mathrm{B}+%
\mathfrak{T}_{\epsilon }\mathrm{B}_{\mathrm{ext}}\right) \right\}
\label{eq sup trivial 2}
\end{equation}%
for any $\epsilon \in \mathbb{R}_{0}^{+}$. Therefore, by (\ref%
{magnetization0}), (\ref{var prob principal}), (\ref{eq sup trivial 1}) and (%
\ref{eq sup trivial 2}), the corresponding Euler--Lagrange equations
associated with $\mathfrak{B}_{\infty }^{(\epsilon )}$ read: For all $%
\mathrm{B}\in \mathcal{B}$, all $\mathrm{r}_{\beta }\in \mathbb{R}_{0}^{+}$
solution of (\ref{var prob principal}), and any minimizer $\mathrm{B}%
_{\epsilon }$ of $\mathfrak{B}_{\infty }^{(\epsilon )}$ (provided it exists
when $\epsilon =0$),
\begin{equation}
\langle \mathrm{B}_{\epsilon }+\mathrm{B}_{\mathrm{ext}},\mathrm{B}\rangle
_{2}=\langle \mathrm{M}_{\beta }(\mathfrak{T}_{\epsilon }\mathrm{B}%
_{\epsilon }+\mathfrak{T}_{\epsilon }\mathrm{B}_{\mathrm{ext}}),\mathrm{B}%
\rangle _{2}\ .  \label{euler B1}
\end{equation}%
(Note that $\mathrm{M}_{\beta }$ depends on $\mathrm{r}_{\beta }$.) Using (%
\ref{eq tempezero1}) and (\ref{euler B1}), we arrive at the equality%
\begin{equation}
\langle \mathrm{B}_{\epsilon }-\mathrm{B}_{\mathrm{int}},\mathrm{B}\rangle
_{2}=\langle \mathrm{M}_{\beta }(\mathfrak{T}_{\epsilon }\mathrm{B}%
_{\epsilon }+\mathfrak{T}_{\epsilon }\mathrm{B}_{\mathrm{ext}}),\mathrm{B}%
\rangle _{2}\ ,\quad \mathrm{B}\in \mathcal{B}\ ,  \label{euler B2}
\end{equation}%
from which one easily shows the assertion. See proof of Lemma \ref{lemma var
prob3 copy(2)} for more details. \hfill $\Box $

Lemma \ref{lemma var prob3 copy(9)} or Equation (\ref{euler B2}) yields a
general estimate on the $L^{2}$--norm of solutions of $\mathfrak{B}_{\infty
}^{(\epsilon )}$ and $\mathfrak{A}$:

\begin{lemma}[Variational problems $\mathfrak{B}_{\infty }^{(\protect%
\epsilon )}$ -- IV]
\label{lemma var prob3 copy(4)}\mbox{ }\newline
Let $\epsilon \in \mathbb{R}_{0}^{+}$ and $\mathrm{B}_{\mathrm{ext}}=%
\mathcal{S}_{0}(\mathrm{j}_{\mathrm{ext}})$ with $\mathrm{j}_{\mathrm{ext}%
}\in C_{0}^{\infty }\cap P^{\bot }\mathfrak{H}$. Assume that $\mathrm{B}%
_{\epsilon }$ is a minimizer of $\mathfrak{B}_{\infty }^{(\epsilon )}$.
Then, $\mathrm{B}_{\epsilon }-\mathrm{B}_{\mathrm{int}}\in b_{\vartheta
}\left( 0\right) $. Moreover, if $\epsilon <\epsilon _{\xi }$ (cf. (\ref%
{radius1})) then
\begin{equation*}
\Vert \mathrm{B}_{\epsilon }-\mathrm{B}_{\mathrm{int}}\Vert _{2}\leq \Vert
\mathrm{M}_{\beta }\left( \mathfrak{T}_{\epsilon }(\mathrm{B}_{\epsilon }-%
\mathrm{B}_{\mathrm{int}})\right) \Vert _{2}+\vartheta |\mathfrak{C}%
\backslash \mathfrak{C}_{\epsilon }|^{1/2}
\end{equation*}%
with $\vartheta |\mathfrak{C}\backslash \mathfrak{C}_{\epsilon }|^{1/2}=%
\mathcal{O}(\sqrt{\epsilon })$, see (\ref{radius2}).
\end{lemma}

\noindent \textit{Proof. }Assume that $\mathrm{B}_{\epsilon }$ is a
minimizer of $\mathfrak{B}_{\infty }^{(\epsilon )}$ for some $\epsilon \in
\mathbb{R}_{0}^{+}$. We already know that such a minimizer exists for all $%
\epsilon \in \mathbb{R}^{+}$ (cf. Lemma \ref{lemma var prob2}). Since $%
\mathrm{B}_{\epsilon },\mathrm{B}_{\mathrm{int}}\in \mathcal{B}$, it follows
from (\ref{euler B2}) applied to
\begin{equation*}
\mathrm{B}=\mathrm{\tilde{B}}_{\epsilon }:=\mathrm{B}_{\epsilon }-\mathrm{B}%
_{\mathrm{int}}\in \mathcal{B}
\end{equation*}%
that
\begin{equation}
\Vert \mathrm{\tilde{B}}_{\epsilon }\Vert _{2}^{2}=\int_{\mathfrak{C}}%
\mathrm{M}_{\beta ,\mathfrak{t}}(\mathfrak{T}_{\epsilon }(\mathrm{B}%
_{\epsilon }+\mathrm{B}_{\mathrm{ext}}))\cdot \mathrm{\tilde{B}}_{\epsilon
}\ \mathrm{d}^{3}\mathfrak{t}\ .  \label{euler B3}
\end{equation}%
In particular, by (\ref{magnetization0}) and the Cauchy--Schwarz inequality,
we obtain $\Vert \mathrm{\tilde{B}}_{\epsilon }\Vert _{2}\leq \vartheta $.
In other words, $\mathrm{B}_{\epsilon }-\mathrm{B}_{\mathrm{int}}\in
b_{\vartheta }\left( 0\right) $, see (\ref{ball}).

Take now $\epsilon <\epsilon _{\xi }$. Then, $\mathfrak{C}_{\epsilon
}\subseteq \mathfrak{C}$ is a non--empty set, see (\ref{radius2}). By Lemma %
\ref{lemma var prob3 copy(2)}, $\mathfrak{T}_{\epsilon }\mathrm{B}_{\mathrm{%
int}}=-\mathfrak{T}_{\epsilon }\mathrm{B}_{\mathrm{ext}}$ a.e. in $\mathfrak{%
C}_{\epsilon }$. We thus rewrite (\ref{euler B3}) as%
\begin{equation}
\Vert \mathrm{\tilde{B}}_{\epsilon }\Vert _{2}^{2}=\int_{\mathfrak{C}%
\backslash \mathfrak{C}_{\epsilon }}\mathrm{M}_{\beta ,\mathfrak{t}}(%
\mathfrak{T}_{\epsilon }(\mathrm{B}_{\epsilon }+\mathrm{B}_{\mathrm{ext}%
}))\cdot \mathrm{\tilde{B}}_{\epsilon }\ \mathrm{d}^{3}\mathfrak{t}+\int_{%
\mathfrak{C}_{\epsilon }}\mathrm{M}_{\beta }(\mathfrak{T}_{\epsilon }\mathrm{%
\tilde{B}}_{\epsilon })\cdot \mathrm{\tilde{B}}_{\epsilon }\ \mathrm{d}^{3}%
\mathfrak{t}\ .  \label{euler B4}
\end{equation}%
By (\ref{magnetization0}) and the Cauchy--Schwarz inequality, we deduce from
(\ref{euler B4}) that
\begin{equation*}
\Vert \mathrm{\tilde{B}}_{\epsilon }\Vert _{2}\leq \vartheta |\mathfrak{C}%
\backslash \mathfrak{C}_{\epsilon }|^{1/2}+\Vert \mathrm{M}_{\beta }(%
\mathfrak{T}_{\epsilon }\mathrm{\tilde{B}}_{\epsilon })\Vert _{2}\ .
\end{equation*}%
(This bound is proven in the same way as (\ref{inequality de funes1+2}).)
\hfill $\Box $

Lemma \ref{lemma var prob3 copy(4)} directly yields the suppression of the
total magnetic induction within $\mathfrak{C}$ (a.e.) for sufficiently high
temperatures because in this case the map $\mathrm{B}\mapsto \mathrm{M}%
_{\beta }(\mathrm{B})$ from $L^{2}$ to $L^{2}(\mathfrak{C};\mathbb{R}^{3})$
satisfies $\Vert \mathrm{M}_{\beta }(\mathrm{B})\Vert _{2}\leq \mathbf{m}%
\Vert \mathrm{B}\Vert _{2}$ with $\mathbf{m}<1$:

\begin{lemma}[Variational problems $\mathfrak{B}_{\infty }^{(\protect%
\epsilon )}$ -- V]
\label{lemma var prob3 copy(7)}\mbox{ }\newline
Let $\epsilon \in \lbrack 0,\epsilon _{\xi })$ and $\mathrm{B}_{\mathrm{ext}%
}=\mathcal{S}_{0}(\mathrm{j}_{\mathrm{ext}})$ with $\mathrm{j}_{\mathrm{ext}%
}\in C_{0}^{\infty }\cap P^{\bot }\mathfrak{H}$. Assume that $\mathrm{B}%
_{\epsilon }$ is a minimizer of $\mathfrak{B}_{\infty }^{(\epsilon )}$. If $%
\beta <\vartheta ^{-1}$ then
\begin{equation*}
\Vert \mathrm{B}_{\epsilon }-\mathrm{B}_{\mathrm{int}}\Vert _{2}\leq \left(
\vartheta ^{-1}-\beta \right) ^{-1}|\mathfrak{C}\backslash \mathfrak{C}%
_{\epsilon }|^{1/2}
\end{equation*}%
with $\vartheta |\mathfrak{C}\backslash \mathfrak{C}_{\epsilon }|^{1/2}=%
\mathcal{O}(\sqrt{\epsilon })$, see (\ref{radius2}).
\end{lemma}

\noindent \textit{Proof. }For all $\mathrm{B}\in L^{2}(\mathfrak{C};\mathbb{R%
}^{3})$ and $\beta \in \mathbb{R}^{+}$, note that
\begin{equation*}
\Vert \mathrm{M}_{\beta }(\mathrm{B})\Vert _{2}^{2}\leq \int_{\mathfrak{C}%
}\vartheta ^{2}\beta ^{2}h_{\mathfrak{t}}^{2}\frac{\tanh ^{2}\left( \beta h_{%
\mathfrak{t}}\right) }{\beta ^{2}h_{\mathfrak{t}}^{2}}\ \mathrm{d}^{3}%
\mathfrak{t}\leq \vartheta ^{2}\beta ^{2}\Vert \mathrm{B}\Vert _{2}^{2}\ ,
\end{equation*}%
using (\ref{magnetization0}) and $\tanh \left( t\right) \leq t$ for all $%
t\in \mathbb{R}_{0}^{+}$. Since $\Vert \mathfrak{T}_{\epsilon }\Vert \leq 1$
for any $\epsilon \in \mathbb{R}_{0}^{+}$, we thus arrive at the assertion
by combining the last upper bound with $\vartheta \beta <1$ and Lemma \ref%
{lemma var prob3 copy(4)}.\hfill $\Box $

This last situation, i.e., the high temperature regime, is of course not the
main case of interest. Moreover, it is questionable from the physical point
of view, see discussions in Section \ref{Section Universality} (cf. \textbf{%
4.}). We are instead interested in showing the Mei{\ss }ner effect at large
enough inverse temperatures $\beta >>1$ and large BCS couplings $\gamma >>1$
to ensure the presence of a superconducting phase.

Now therefore, we pursue our analysis of the variational problem $\mathfrak{B%
}_{\infty }^{(\epsilon )}$ by using the asymptotics $\left\vert \mathrm{M}%
_{\beta ,\mathfrak{D}}\right\vert =\mathcal{O}(\mathrm{e}^{-\beta (\mathrm{h}%
_{\mathrm{c}}-\mathrm{h})})$ (cf. (\ref{magnetizationbis})) whenever (\ref%
{magnetizationbisbis}) is a.e. satisfied on some open subset $\mathfrak{D}%
\subseteq \mathfrak{C}$. We give below a sufficient (but \emph{not}
necessary) condition to prove in Theorem \ref{lemma var prob3 copy(5)} the
Mei{\ss }ner effect at low temperatures.

\begin{lemma}[Variational problems $\mathfrak{B}_{\infty }^{(\protect%
\epsilon )}$ -- VI]
\label{lemma var prob3 copy(8)}\mbox{ }\newline
Let $\epsilon \in \lbrack 0,\epsilon _{\xi })$ and $\mathrm{B}_{\mathrm{ext}%
}=\mathcal{S}_{0}(\mathrm{j}_{\mathrm{ext}})$ with $\mathrm{j}_{\mathrm{ext}%
}\in C_{0}^{\infty }\cap P^{\bot }\mathfrak{H}$. Assume that $\mathrm{B}%
_{\epsilon }$ is a minimizer of $\mathfrak{B}_{\infty }^{(\epsilon )}$ such
that%
\begin{equation*}
\mathfrak{z}:=\frac{\beta \cosh \left( \beta \mathrm{h}\right) }{\mathrm{e}%
^{-\beta \lambda }\cosh \left( \beta g_{\mathrm{r}_{\beta }(\mathfrak{T}%
_{\epsilon }(\mathrm{B}_{\epsilon }-\mathrm{B}_{\mathrm{int}}))}\right) }%
+\vartheta \mathrm{h}^{-1}<\vartheta ^{-1}
\end{equation*}%
for some $\mathrm{h}\in \mathbb{R}^{+}$. Here, $\mathrm{r}_{\beta }(%
\mathfrak{T}_{\epsilon }(\mathrm{B}_{\epsilon }-\mathrm{B}_{\mathrm{int}}))$
stands for any arbitrary solution of (\ref{var prob principal}) with $%
\mathrm{B}=\mathfrak{T}_{\epsilon }(\mathrm{B}_{\epsilon }-\mathrm{B}_{%
\mathrm{int}})$. Then,
\begin{equation*}
\left\Vert \mathrm{B}_{\epsilon }-\mathrm{B}_{\mathrm{int}}\right\Vert
_{2}\leq \left( \vartheta ^{-1}-\mathfrak{z}\right) ^{-1}|\mathfrak{C}%
\backslash \mathfrak{C}_{\epsilon }|^{1/2}
\end{equation*}%
with $\vartheta |\mathfrak{C}\backslash \mathfrak{C}_{\epsilon }|^{1/2}=%
\mathcal{O}(\sqrt{\epsilon })$, see (\ref{radius2}).
\end{lemma}

\noindent \textit{Proof. }Take any constant $\mathrm{h}\in \mathbb{R}^{+}$.
The magnetization density $\mathrm{M}_{\beta }\equiv \mathrm{M}_{\beta }(%
\mathrm{B})\in C_{0}^{\infty }$, defined by (\ref{magnetization0}) for all $%
\mathrm{B}\in L^{2}(\mathfrak{C};\mathbb{R}^{3})$, trivially satisfies
\begin{equation}
\left\Vert \mathrm{M}_{\beta }\right\Vert _{2}\leq \left\Vert \mathbf{1}[h_{%
\mathfrak{t}}\leq \mathrm{h}]\mathrm{M}_{\beta }\right\Vert _{2}+\left\Vert
\mathbf{1}[h_{\mathfrak{t}}\geq \mathrm{h}]\mathrm{M}_{\beta }\right\Vert
_{2}  \label{expo magn0}
\end{equation}%
with $h_{\mathfrak{t}}:=\vartheta \left\vert \mathrm{B}\left( \mathfrak{t}%
\right) \right\vert $ a.e. for $\mathfrak{t}\in \mathfrak{C}$. Using (\ref%
{magnetization0}) and the mean value theorem, one gets that
\begin{equation}
\left\Vert \mathbf{1}[h_{\mathfrak{t}}\leq \mathrm{h}]\mathrm{M}_{\beta
}\right\Vert _{2}\leq \frac{\beta \vartheta \cosh \left( \beta \mathrm{h}%
\right) }{\mathrm{e}^{-\beta \lambda }\cosh \left( \beta g_{\mathrm{r}%
_{\beta }}\right) }\left\Vert \mathrm{B}\right\Vert _{2}\ .
\label{expo magn0bis}
\end{equation}%
On the other hand,%
\begin{equation*}
\left\Vert \mathbf{1}[h_{\mathfrak{t}}\geq \mathrm{h}]\mathrm{M}_{\beta
}\right\Vert _{2}\leq \left\Vert \mathrm{M}_{\beta }\right\Vert
_{2}\left\Vert \mathbf{1}[h_{\mathfrak{t}}\geq \mathrm{h}]\right\Vert
_{2}\leq \vartheta ^{2}\mathrm{h}^{-1}\left\Vert \mathrm{B}\right\Vert _{2}\
.
\end{equation*}%
We combine this with (\ref{expo magn0})--(\ref{expo magn0bis}) to deduce that%
\begin{equation*}
\left\Vert \mathrm{M}_{\beta }\right\Vert _{2}\leq \left( \frac{\beta
\vartheta \cosh \left( \beta \mathrm{h}\right) }{\mathrm{e}^{-\beta \lambda
}\cosh \left( \beta g_{\mathrm{r}_{\beta }}\right) }+\vartheta ^{2}\mathrm{h}%
^{-1}\right) \left\Vert \mathrm{B}\right\Vert _{2}\ .
\end{equation*}%
Applying this inequality to $\mathrm{B}=\mathfrak{T}_{\epsilon }(\mathrm{B}%
_{\epsilon }-\mathrm{B}_{\mathrm{int}})$ and using Lemma \ref{lemma var
prob3 copy(4)} and $\Vert \mathfrak{T}_{\epsilon }\Vert \leq 1$, we finally
find that%
\begin{equation*}
\left\Vert \mathrm{B}_{\epsilon }-\mathrm{B}_{\mathrm{int}}\right\Vert
_{2}\leq \mathfrak{z}\vartheta \left\Vert \mathrm{B}_{\epsilon }-\mathrm{B}_{%
\mathrm{int}}\right\Vert _{2}+\vartheta |\mathfrak{C}\backslash \mathfrak{C}%
_{\epsilon }|^{1/2}
\end{equation*}%
from which we deduce the lemma \hfill $\Box $

Sufficient conditions to ensure that assumptions of the last lemma hold at
large $\beta >0$ are given by $\mu <-\vartheta ^{2}$ and $\gamma >|\mu
-\lambda |\Gamma _{0}$ with%
\begin{equation*}
\Gamma _{0}:=\frac{4}{1-\vartheta ^{2}|\mu |^{-1}}>4\ .
\end{equation*}%
See Lemmata \ref{lemma var prob3 copy(6)} and \ref{lemma var prob3 copy(4)}.
Indeed, the Mei{\ss }ner effect is directly related with the existence of a
superconducting phase, which is characterized by a strictly positive Cooper
pair condensate density for all minimizers $\omega _{\epsilon ,l}\in \mathit{%
\Omega }_{l}^{(\epsilon )}$ in the limit $\epsilon \rightarrow 0^{+}$, see (%
\ref{limit00}).

\begin{theorem}[Superconducting phase -- II]
\label{lemma var prob3 copy(5)}\mbox{ }\newline
Let $\mu <-\vartheta ^{2}$, $\gamma >|\mu -\lambda |\Gamma _{0}$ and $%
\mathrm{B}_{\mathrm{ext}}=\mathcal{S}_{0}(\mathrm{j}_{\mathrm{ext}})$ with $%
\mathrm{j}_{\mathrm{ext}}\in C_{0}^{\infty }\cap P^{\bot }\mathfrak{H}$.
Then, there is $\beta _{0}\in \mathbb{R}^{+}$ such that, for all $\beta
>\beta _{0}$: \newline
\emph{(i)} Any family $\{\mathrm{B}_{\epsilon }\}_{\epsilon \in \mathbb{R}%
^{+}}\subset \mathcal{B}$ of minimizers of $\mathfrak{B}_{\infty
}^{(\epsilon )}$ converges in norm to the unique minimizer $\mathrm{B}_{%
\mathrm{int}}$ of $\mathfrak{A}$ (cf. Lemma \ref{lemma var prob3 copy(2)}).
\newline
\emph{(ii)} $\mathrm{B}_{\mathrm{int}}$ is also the unique minimizer of $%
\mathfrak{B}_{\infty }^{(0)}$. \newline
\emph{(iii)} For any sequence of minimizers $\omega _{\epsilon ,l}\in
\mathit{\Omega }_{l}^{(\epsilon )}$,
\begin{equation*}
\underset{\epsilon \rightarrow 0^{+}}{\lim }\liminf_{l\rightarrow \infty
}\omega _{\epsilon ,l}\left( \frac{\mathfrak{c}_{0}^{\ast }\mathfrak{c}_{0}}{%
\left\vert \Lambda _{l}\right\vert }\right) =\underset{\epsilon \rightarrow
0^{+}}{\lim }\limsup_{l\rightarrow \infty }\omega _{\epsilon ,l}\left( \frac{%
\mathfrak{c}_{0}^{\ast }\mathfrak{c}_{0}}{\left\vert \Lambda _{l}\right\vert
}\right) =\mathrm{r}_{\beta }(0)
\end{equation*}%
with
\begin{equation*}
\mathrm{r}_{\beta }(0)\geq \Gamma _{0}^{-2}-\gamma ^{-2}\left( \mu -\lambda
\right) ^{2}>0
\end{equation*}%
being the unique solution of (\ref{var prob principal}) for $\mathrm{B}=0$.
\end{theorem}

\noindent \textit{Proof. }(i) For any $\mathrm{B}\in b_{\vartheta }(0)$, $%
\mu <-\vartheta ^{2}$ and $\gamma >|\mu -\lambda |\Gamma _{0}$, by Lemma \ref%
{lemma var prob3 copy(6)}, there is $\beta _{0}\in \mathbb{R}^{+}$ such
that, for all $\beta >\beta _{0}$,
\begin{equation*}
\mathrm{h}_{\mathrm{c}}\left( \mathrm{B}\right) :=g_{\mathrm{r}_{\beta
}}-\lambda >\mathrm{h}:=g_{\mathfrak{r}_{0}}-\lambda >\vartheta ^{2}
\end{equation*}%
with $\mathfrak{r}_{0}$ defined by (\ref{rminin}) for $\varepsilon =0$ and $%
R=\vartheta $. See also (\ref{magnetizationbisbis}). By Lemma \ref{lemma var
prob3 copy(4)}, it follows that the conditions of Lemma \ref{lemma var prob3
copy(8)} are satisfied for $\epsilon \in \mathbb{R}^{+}$. The latter yields
\begin{equation*}
\left\Vert \mathrm{B}_{\epsilon }-\mathrm{B}_{\mathrm{int}}\right\Vert _{2}=%
\mathcal{O}(\sqrt{\epsilon })\ .
\end{equation*}

(ii) We combine (i) and (\ref{de funes encore 00}) with Theorem \ref{thm
limit pressure} (ii) to check that $\mathrm{B}_{\mathrm{int}}$ is a
minimizer of $\mathfrak{B}_{\infty }^{(0)}$. On the other hand, recall that,
for any $\mathrm{B}\in b_{\vartheta }(0)$, the conditions of Lemma \ref%
{lemma var prob3 copy(8)} are also satisfied for $\epsilon =0$. Hence, $%
\mathrm{B}_{\mathrm{int}}$ is the unique minimizer of $\mathfrak{B}_{\infty
}^{(0)}$.

(iii) Since, by definition, $\omega _{\epsilon ,l}\in \mathit{\Omega }%
_{l}^{(\epsilon )}$ minimizes the magnetic free--energy density functional $%
\mathcal{F}_{l}^{(\epsilon )}$ (\ref{magnetic free energy}), every $\omega
_{\epsilon ,l}\in \mathit{\Omega }_{l}^{(\epsilon )}$ can be seen as a
tangent functional to the magnetic pressure $\mathcal{P}_{l}^{(\epsilon )}$ (%
\ref{magnetic pressure}). See \cite[Section 2.6]{BruPedra2} for further
details. In particular,
\begin{equation}
\underset{\delta \rightarrow 0^{+}}{\lim }\partial _{\gamma }\mathcal{P}%
_{l}^{(\epsilon )}\left( \gamma -\delta \right) \leq \omega _{\epsilon
,l}\left( \frac{\mathfrak{c}_{0}^{\ast }\mathfrak{c}_{0}}{\left\vert \Lambda
_{l}\right\vert }\right) \leq \underset{\delta \rightarrow 0^{+}}{\lim }%
\partial _{\gamma }\mathcal{P}_{l}^{(\epsilon )}\left( \gamma +\delta \right)
\label{bog ine}
\end{equation}%
for any $l\in \mathbb{N}$. Observe now that the finite volume magnetic
pressure $\mathcal{P}_{l}^{(\epsilon )}\equiv \mathcal{P}_{l}^{(\epsilon
)}\left( \gamma \right) $ is a continuous convex function of $\gamma \in
\mathbb{R}^{+}$. Indeed, $\mathcal{P}_{l}^{(\epsilon )}\left( \gamma \right)
$ is the supremum over a family of affine functions of $\gamma \in \mathbb{R}%
^{+}$. Therefore, by Griffiths arguments (see, e.g., \cite[Eq. (A.1)]%
{BruPedra1}) together with (\ref{bog ine}), the point--wise convergence of
functions $\mathcal{P}_{l}^{(\epsilon )}\equiv \mathcal{P}_{l}^{(\epsilon
)}\left( \gamma \right) $ towards the continuous convex function $\mathcal{P}%
_{\infty }^{(\epsilon )}\equiv \mathcal{P}_{\infty }^{(\epsilon )}\left(
\gamma \right) $ yields%
\begin{equation}
\underset{\delta \rightarrow 0^{+}}{\lim }\partial _{\gamma }\mathcal{P}%
_{\infty }^{(\epsilon )}\left( \gamma -\delta \right) \leq r_{\epsilon
}^{(-)}\leq r_{\epsilon }^{(+)}\leq \underset{\delta \rightarrow 0^{+}}{\lim
}\partial _{\gamma }\mathcal{P}_{\infty }^{(\epsilon )}\left( \gamma +\delta
\right)  \label{bog ine20}
\end{equation}%
for any $\epsilon \in \mathbb{R}^{+}$, where%
\begin{equation}
r_{\epsilon }^{(-)}:=\liminf_{l\rightarrow \infty }\ \omega _{\epsilon
,l}\left( \frac{\mathfrak{c}_{0}^{\ast }\mathfrak{c}_{0}}{\left\vert \Lambda
_{l}\right\vert }\right) \ ,\ r_{\epsilon }^{(+)}:=\limsup_{l\rightarrow
\infty }\ \omega _{\epsilon ,l}\left( \frac{\mathfrak{c}_{0}^{\ast }%
\mathfrak{c}_{0}}{\left\vert \Lambda _{l}\right\vert }\right) \ .
\label{bog ine22}
\end{equation}%
By applying once again \cite[Eq. (A.1)]{BruPedra1} to the family $\{\mathcal{%
P}_{\infty }^{(\epsilon )}(\gamma )\}_{\epsilon \in \mathbb{R}^{+}}$ of
continuous convex functions, we deduce from (\ref{perfect magnetic pressure}%
) and (\ref{bog ine20})--(\ref{bog ine22}) that the limits
\begin{equation}
r_{0}^{(-)}:=\liminf_{\epsilon \rightarrow 0^{+}}r_{\epsilon }^{(-)}\quad
\text{and}\quad r_{0}^{(+)}:=\limsup_{\epsilon \rightarrow 0^{+}}r_{\epsilon
}^{(+)}  \label{bog ine2-1}
\end{equation}%
must obey:%
\begin{equation}
\underset{\delta \rightarrow 0^{+}}{\lim }\partial _{\gamma }\mathcal{P}%
_{\infty }\left( \gamma -\delta \right) \leq r_{0}^{(-)}\leq r_{0}^{(+)}\leq
\underset{\delta \rightarrow 0^{+}}{\lim }\partial _{\gamma }\mathcal{P}%
_{\infty }\left( \gamma +\delta \right) \ .  \label{bog ine2}
\end{equation}%
Note that $\mathcal{P}_{\infty }\left( \gamma \right) $ is of course
well--defined for all $\gamma \in \mathbb{R}^{+}$.

In fact, we combine (ii) with Theorem \ref{thm limit pressure} (i),
Corollary \ref{corollary de funes} and Lemma \ref{lemma var prob3 copy(2)}
to obtain that, for any $\gamma >|\mu -\lambda |\Gamma _{0}$,
\begin{equation}
\mathcal{P}_{\infty }\equiv \mathcal{P}_{\infty }\left( \gamma \right) =-%
\frac{1}{2}\Vert \mathrm{\tilde{B}}_{\mathrm{int}}\Vert _{2}^{2}+\mathfrak{F}%
(\mathrm{r}_{\beta }(0),0)  \label{equality cool}
\end{equation}%
with $\mathrm{\tilde{B}}_{\mathrm{int}}:=\mathrm{B}_{\mathrm{int}}+\mathrm{B}%
_{\mathrm{ext}}$ and $\mathrm{r}_{\beta }(0)$ being a solution of (\ref{var
prob principal}) for $\mathrm{B}=0$. By Lemmata \ref{lemma var prob3 copy(6)}
together with \cite[Lemma 7.1]{BruPedra1}, the solution of (\ref{var prob
principal}) is unique. Using (\ref{equality cool}) while keeping in mind
that $\mathrm{B}_{\mathrm{int}}$ is the minimizer of $\mathfrak{B}_{\infty
}^{(0)}$ for all $\gamma \geq \gamma ^{\prime }>|\mu -\lambda |\Gamma _{0}$
(cf. (ii)) and $\beta >\beta _{0}=\beta _{0}\left( \gamma ^{\prime }\right) $
(cf. Lemma \ref{lemma var prob3 copy(6)}), we then conclude that
\begin{equation*}
\underset{\delta \rightarrow 0^{+}}{\lim }\partial _{\gamma }\mathcal{P}%
_{\infty }\left( \gamma -\delta \right) =\underset{\delta \rightarrow 0^{+}}{%
\lim }\partial _{\gamma }\mathcal{P}_{\infty }\left( \gamma +\delta \right) =%
\mathrm{r}_{\beta }(0)\ .
\end{equation*}%
Because of (\ref{bog ine2-1})--(\ref{bog ine2}), the latter yields
(iii).\hfill $\Box $

\subsection{Sketch of the Proof of Theorem \protect\ref{Theorem inv trans}
\label{Section inv transl}}

As compared to the model without hopping term, i.e., $\mathfrak{e}=0$ in (%
\ref{model kinetic}), there are two main new technical difficulties to be
managed:

\begin{itemize}
\item The model $H_{l,\mathfrak{e}}$ with $\mathfrak{e}\neq 0$ is not
anymore permutation invariant, but only translation invariant. This implies
that one cannot express important quantities in the one--point CAR $C^{\ast }
$--algebra $\mathcal{U}_{\left\{ 0\right\} }$ generated by the identity $%
\mathbf{1}$ and $\{a_{0,\mathrm{s}}\}_{\mathrm{s}\in \{\uparrow ,\downarrow
\}}$. Instead, the objects to be analyzed will be defined with respect to
the full CAR $C^{\ast }$--algebra generated by the identity $\mathbf{1}$ and
$\{a_{x,\mathrm{s}}\}_{x\in \mathbb{Z}^{3},\mathrm{s}\in \{\uparrow
,\downarrow \}}$.

\item The functional $\mathfrak{F}$ defined by (\ref{expliciti function2}),
which is an approximating free--energy density, is not anymore explicitly
given. It can only be represented as an absolutely converging series and $%
\left\Vert \mathfrak{e}\right\Vert _{1}$ and $|\lambda |$ have to be small
as compared to $\gamma $. This is achieved by combining Grassmann
integration and Brydges--Kennedy tree expansion methods together with
determinant bounds \cite[Definition 1.2, Theorem 1.3]{Pedra-Salmhofer}.
\end{itemize}

Therefore, we focus on these technical aspects and the corresponding changes
implied by them. We separate this sketch in ten points permitting the reader
to compare the detailed proofs for $\mathfrak{e}=0$ given above with the
more general case for which $\mathfrak{e}$ is only a summable real function.

For technical simplicity, Theorem \ref{Theorem inv trans} deals with Mei{\ss
}ner effect at large but finite inverse temperature $\beta \in \mathbb{R}%
^{+} $. We additionally show after Points 1--10 some additional results
(Theorem \ref{lemma var prob3 copy(13)} and Corollary \ref{corollary de
funes copy(1)}) showing how to manage the zero temperature case ($\beta
=\infty $). A complete analysis of the zero temperature regime will be the
subject of a companion paper. See also Remark \ref{dielectric copy(1)}.
\smallskip

\noindent \textbf{1.} \textit{Cf. Lemma \ref{lemma var pro}.} For each $%
\mathrm{B}\in L^{2}(\mathfrak{C};\mathbb{R}^{3})$, the functional $\mathfrak{%
F}_{\mathfrak{e}}$ is defined on $\mathbb{R}_{0}^{+}$ by%
\begin{equation}
\mathfrak{F}_{\mathfrak{e}}(r)\equiv \mathfrak{F}_{\mathfrak{e}}\left( r,%
\mathrm{B}\right) :=\mu +\beta ^{-1}\ln 2-\gamma r+\beta ^{-1}\int_{%
\mathfrak{C}}\mathrm{\tilde{p}}_{\mathfrak{e}}\left( r,\mathfrak{t}\right)
\mathrm{d}^{3}\mathfrak{t}\ .  \label{new functional f}
\end{equation}%
Here, the real function
\begin{equation*}
\mathrm{\tilde{p}}_{\mathfrak{e}}\left( r,\mathfrak{t}\right) \equiv \mathrm{%
\tilde{p}}_{\mathfrak{e}}\left( r,\mathfrak{t},\mathrm{B}\right) :=\underset{%
l\rightarrow \infty }{\lim }\left\vert \Lambda _{l}\right\vert ^{-1}\ln
\mathrm{Trace}_{\wedge \mathcal{H}_{\Lambda _{l}}}\left( \mathrm{e}^{-\beta
H_{l,\mathfrak{e}}\left( r,\mathrm{B}\left( \mathfrak{t}\right) \right)
}\right)
\end{equation*}%
defined at $r\in \mathbb{R}_{0}^{+}$, $\mathfrak{t}\in \mathfrak{C}$ (a.e.),
and $\mathrm{B}\in L^{2}(\mathfrak{C};\mathbb{R}^{3})$ is (up to a factor $%
\beta ^{-1}$) the pressure associated with the approximating Hamiltonian
defined, for any $r\in \mathbb{R}_{0}^{+}$ and $\mathrm{B}\in \mathbb{R}^{3}$%
, by%
\begin{eqnarray}
H_{l,\mathfrak{e}}\left( r,\mathrm{B}\right)  &:=&\sum\limits_{x\in \Lambda
_{l}}\left\{ -\mu (n_{x,\uparrow }+n_{x,\downarrow })+2\lambda n_{x,\uparrow
}n_{x,\downarrow }-\gamma \sqrt{r}(a_{x,\uparrow }^{\ast }a_{x,\downarrow
}^{\ast }+a_{x,\downarrow }a_{x,\uparrow })-\mathrm{B}\cdot \mathrm{M}%
^{x}\right\}   \notag \\
&&+\sum\limits_{x,y\in \Lambda _{l}}\mathfrak{e}(x-y)\left( a_{x,\downarrow
}^{\ast }a_{y,\downarrow }+a_{x,\uparrow }^{\ast }a_{y,\uparrow }\right) \ .
\label{approximating hamiltoian}
\end{eqnarray}%
Recall again that $\mathrm{M}^{x}:=(\mathrm{m}_{1}^{x},\mathrm{m}_{2}^{x},%
\mathrm{m}_{3}^{x})$ is defined via (\ref{magne1}). Compare with Definition (%
\ref{one site hamil}) of $u\left( r,\mathfrak{t}\right) $ and note that $%
\mathfrak{F}_{0}=\mathfrak{F}$, see (\ref{expliciti function1}). Then, for
any $\mathrm{B}\in C^{0}(\mathfrak{C};\mathbb{R}^{3})$,
\begin{equation*}
p_{\infty }\left( \mathrm{B}\right) :=\underset{l\rightarrow \infty }{\lim }%
p_{l}\left( \mathrm{B}\right) =\underset{r\geq 0}{\sup }\ \mathfrak{F}_{%
\mathfrak{e}}(r,\mathrm{B})<\infty \ .
\end{equation*}%
The proof uses essentially the same arguments as the one done in \cite[%
Theorem 4.1]{BruPedra-homog} by approximating the pressure of (piece--wise)
translation invariant models by the pressure of permutation invariant model.
See also \cite[Lemma 6.7]{BruPedra2} for more details of such an argument.
We aim to show this kind of result for a much more general class of models
in a separated paper. \smallskip

\noindent \textbf{2.} \textit{Cf. Theorem \ref{thm limit pressure}. }To
prove Theorem \ref{thm limit pressure} in the case $\mathfrak{e}\neq 0$, we
need to replace Inequality (\ref{perssure1bis}) for any $\mathrm{B},\mathrm{C%
}\in L^{2}(\mathfrak{C};\mathbb{R}^{3})$ and $\mathfrak{t}\in \mathfrak{C}$
(a.e.) by
\begin{equation*}
\left\vert \mathrm{\tilde{p}}_{\mathfrak{e}}\left( r,\mathfrak{t},\mathrm{B}%
\right) -\mathrm{\tilde{p}}_{\mathfrak{e}}\left( r,\mathfrak{t},\mathrm{C}%
\right) \right\vert \leq 2\sqrt{3}\vartheta \beta \left\vert \mathrm{B}%
\left( \mathfrak{t}\right) -\mathrm{C}\left( \mathfrak{t}\right) \right\vert
\ ,
\end{equation*}%
which is also a consequence of \cite[Eq. (3.11)]{BruPedra2}.\smallskip

\noindent \textbf{3.} \textit{Cf. Lemma \ref{lemma var prob3 copy(6)}. }Let $%
R\in \mathbb{R}^{+}$. For every $\mu <-R\vartheta $, we choose $\varepsilon
\in \mathbb{R}^{+}$\ such that $\gamma >\left\vert \mu -\lambda \right\vert
\Gamma _{\varepsilon }$, with $\Gamma _{\varepsilon }$ being defined by (\ref%
{GammaGamma}). Then, there is a decreasing function $\beta _{0}\equiv \beta
_{0}(\gamma )\in \mathbb{R}^{+}$ of $\gamma $ which does not depend on
sufficiently small $\varepsilon \in \mathbb{R}^{+}$ and with $\beta
_{0}(\gamma )\rightarrow 0$, as $\gamma \rightarrow \infty $, such that, for
all $\beta >\beta _{0}$, $\mathrm{B}\in b_{R}(0)$ and $r\in \lbrack 0,%
\mathfrak{r}_{\varepsilon }]$ (see (\ref{rminin})),%
\begin{equation*}
\beta ^{-1}\int_{\mathfrak{C}}\left( \mathrm{\tilde{p}}_{0}\left( r,%
\mathfrak{t}\right) -\mathrm{\tilde{p}}_{0}\left( 0,\mathfrak{t}\right)
\right) \mathrm{d}^{3}\mathfrak{t}\geq \frac{\varepsilon \gamma r}{4}\ .
\end{equation*}%
This assertion follows from explicit computations, see (\ref{debile1}).
Thus, if the hopping amplitude satisfies the inequality
\begin{equation}
\left\Vert \mathfrak{e}\right\Vert _{1}<\frac{\varepsilon \gamma \mathfrak{r}%
_{\varepsilon }}{24}\ ,  \label{important toto}
\end{equation}%
then, by using (\ref{new functional f}) and \cite[Eq. (3.11)]{BruPedra2},
\begin{equation*}
\int_{\mathfrak{C}}\mathrm{\tilde{p}}_{\mathfrak{e}}\left( r,\mathfrak{t}%
\right) \mathrm{d}^{3}\mathfrak{t}<\int_{\mathfrak{C}}\mathrm{\tilde{p}}_{%
\mathfrak{e}}\left( \tilde{r},\mathfrak{t}\right) \mathrm{d}^{3}\mathfrak{t}%
\ ,\qquad r\in \left[ 0,\mathfrak{r}_{\varepsilon }/3\right] \ ,\ \tilde{r}%
\in \left[ 2\mathfrak{r}_{\varepsilon }/3,\mathfrak{r}_{\varepsilon }\right]
\ ,
\end{equation*}%
for $\beta >\beta _{0}$, $\mathrm{B}\in b_{R}(0)$ and all sufficiently small
$\varepsilon \in \mathbb{R}^{+}$ such that $\gamma >\left\vert \mu -\lambda
\right\vert \Gamma _{\varepsilon }$. As a consequence, if (\ref{important
toto}) holds true at $\varepsilon =0$ then there is $\beta _{0}\in \mathbb{R}%
^{+}$ such that, for all $\beta >\beta _{0}$, any maximizer $\mathrm{r}%
_{\beta }\equiv \mathrm{r}_{\beta }\left( \mathrm{B}\right) \in \mathbb{R}%
_{0}^{+}$ of
\begin{equation}
\underset{r\geq 0}{\sup }\ \mathfrak{F}_{\mathfrak{e}}(r,\mathrm{B})=%
\mathfrak{F}_{\mathfrak{e}}(\mathrm{r}_{\beta },\mathrm{B})\ ,\quad \mathrm{B%
}\in L^{2}(\mathfrak{C};\mathbb{R}^{3})\ ,  \label{var blilbi}
\end{equation}%
satisfies the inequality%
\begin{equation}
\underset{\mathrm{B}\in b_{R}(0)}{\inf }\mathrm{r}_{\beta }(\mathrm{B})\geq
\frac{\mathfrak{r}_{0}}{3}>0\ .  \label{var blilbi2}
\end{equation}%
\smallskip

\noindent \textbf{4.} \textit{Approximating minimizers of the free--energy.}
Instead of the elementary boxes (\ref{elementary box}), use the here more
convenient definition
\begin{equation*}
\mathfrak{\tilde{G}}_{l,k}:=\left\{ [-\ell ^{\eta },\ell ^{\eta })\times
\lbrack -\ell ^{\eta ^{\bot }},\ell ^{\eta ^{\bot }})^{2}+(2k_{1}\ell ^{\eta
},2k_{2}\ell ^{\eta ^{\bot }},2k_{3}\ell ^{\eta ^{\bot }})\right\} \cap
\Lambda _{l}
\end{equation*}%
with $\ell :=l-1$ for $l>1$, $0<\eta ^{\bot }<\eta <1$, and%
\begin{equation*}
k\in \mathcal{\tilde{R}}_{l}:=\{(k_{1},k_{2},k_{3})\in \mathbb{Z}%
^{3}:|k_{1}|<(\ell ^{1-\eta }+1)\ ,\ |k_{2,3}|<(\ell ^{1-\eta ^{\bot
}}+1)\}\ .
\end{equation*}%
Note that, for any $l>1$ and $k,q\in \mathcal{\tilde{R}}_{l}$,
\begin{equation}
\mathfrak{\tilde{G}}_{l,k}\cap \mathfrak{\tilde{G}}_{l,q}=\emptyset \qquad
\text{and}\qquad \Lambda _{l}=\underset{k\in \mathcal{\tilde{R}}_{l}}{%
\bigcup }\mathfrak{\tilde{G}}_{l,k}\ .  \label{toto box}
\end{equation}%
For each $l>1$, $k\in \mathcal{\tilde{R}}_{l}$ and $\mathrm{B}\in L^{2}(%
\mathfrak{C};\mathbb{R}^{3})$, define $\mathrm{\bar{B}}_{l,k}:=(\mathrm{\bar{%
b}}_{1,l,k},\mathrm{\bar{b}}_{2,l,k},\mathrm{\bar{b}}_{3,l,k})$ with
\begin{equation}
\mathrm{\bar{b}}_{i,l,k}:=\frac{1}{8\ell ^{\eta }\ell ^{2\eta ^{\bot }}}%
\int_{[-\ell ^{\eta },\ell ^{\eta })\times \lbrack -\ell ^{\eta ^{\bot
}},\ell ^{\eta ^{\bot }})^{2}}\mathrm{b}_{1}\left( \frac{y+(2k_{1}\ell
^{\eta },2k_{2}\ell ^{\eta ^{\bot }},2k_{3}\ell ^{\eta ^{\bot }})}{2l}%
\right) \mathrm{d}^{3}y\ ,\quad i\in \{1,2,3\}\ ,  \label{b bar}
\end{equation}
as well as the magnetization observable%
\begin{eqnarray}
\mathcal{\tilde{M}}_{l,k} &:=&-\vartheta \sum\limits_{x\in \mathfrak{\tilde{G%
}}_{l,k}}\left( a_{x,\uparrow }^{\ast }a_{x,\downarrow }+a_{x,\downarrow
}^{\ast }a_{x,\uparrow }\right) \mathrm{\bar{b}}_{1,l,k}+i\vartheta
\sum\limits_{x\in \mathfrak{\tilde{G}}_{l,k}}\left( a_{x,\uparrow }^{\ast
}a_{x,\downarrow }-a_{x,\downarrow }^{\ast }a_{x,\uparrow }\right) \mathrm{%
\bar{b}}_{2,l,k}  \notag \\
&&-\vartheta \sum\limits_{x\in \mathfrak{\tilde{G}}_{l,k}}\left(
n_{x,\uparrow }-n_{x,\downarrow }\right) \mathrm{\bar{b}}_{3,l,k}\ .
\label{magnetizasion cellule}
\end{eqnarray}%
Consider the approximating Hamiltonian with periodic boundary condition in
the elementary boxes $\mathfrak{\tilde{G}}_{l,k}$ defined by%
\begin{eqnarray*}
\tilde{H}_{l,\mathfrak{e}}\left( r,k\right)  &:=&\mathcal{\tilde{M}}%
_{l,k}+\sum\limits_{x\in \mathfrak{\tilde{G}}_{l,k}}\left\{ -\mu
(n_{x,\uparrow }+n_{x,\downarrow })+2\lambda n_{x,\uparrow }n_{x,\downarrow
}-\gamma \sqrt{r}(a_{x,\uparrow }^{\ast }a_{x,\downarrow }^{\ast
}+a_{x,\downarrow }a_{x,\uparrow })\right\}  \\
&&+\sum\limits_{x,y\in \mathfrak{\tilde{G}}_{l,k}}\sum\limits_{\substack{ %
z_{1}\in |\mathbb{Z}\cap \{2k_{1}\ell ^{\eta }+[-\ell ^{\eta },\ell ^{\eta
})\}|\mathbb{Z} \\ z_{2}\in |\mathbb{Z}\cap \{2k_{2}\ell ^{\eta ^{\bot
}}+[-\ell ^{\eta ^{\bot }},\ell ^{\eta ^{\bot }})\}|\mathbb{Z} \\ z_{3}\in |%
\mathbb{Z}\cap \{2k_{3}\ell ^{\eta ^{\bot }}+[-\ell ^{\eta ^{\bot }},\ell
^{\eta ^{\bot }})\}|\mathbb{Z}}}\mathfrak{e}\left( x-y+\left(
z_{1},z_{2},z_{3}\right) \right) \left( a_{x,\downarrow }^{\ast
}a_{y,\downarrow }+a_{x,\uparrow }^{\ast }a_{y,\uparrow }\right)
\end{eqnarray*}%
for each $l>1$, $r\in \mathbb{R}_{0}^{+}$ and $k\in \mathcal{\tilde{R}}_{l}$%
. The corresponding Gibbs state $\tilde{\omega}_{l,k}\equiv \tilde{\omega}%
_{l,k,r,\mathrm{B}}\in E_{\mathfrak{\tilde{G}}_{l,k}}$ is then defined by
\begin{equation*}
\tilde{\omega}_{l,k}\left( A\right) :=\mathrm{Trace}_{\wedge \mathcal{H}_{%
\mathfrak{\tilde{G}}_{l,k}}}\left( A\frac{\mathrm{e}^{-\beta \tilde{H}_{l,%
\mathfrak{e}}\left( r,k\right) }}{\mathrm{Trace}_{\wedge \mathcal{H}_{%
\mathfrak{\tilde{G}}_{l,k}}}(\mathrm{e}^{-\beta \tilde{H}_{l,\mathfrak{e}%
}\left( r,k\right) })}\right) \ ,\quad A\in \mathcal{U}_{\mathfrak{\tilde{G}}%
_{l,k}}\ ,
\end{equation*}%
for any inverse temperature $\beta \in \mathbb{R}^{+}$, $r\in \mathbb{R}%
_{0}^{+}$, $l>1$ and $k\in \mathcal{\tilde{R}}_{l}$. Note that the
expectation value of the current observable $\mathrm{I}_{l}^{x,y}$, which is
defined by (\ref{curents observable}) for any $x,y\in \Lambda _{l}$,
vanishes. Indeed, for any $l>1$, $k\in \mathcal{\tilde{R}}_{l}$ and $x,y\in
\mathfrak{\tilde{G}}_{l,k}$,
\begin{equation}
\tilde{\omega}_{l,k}\left( a_{x,\uparrow }^{\ast }a_{x,\downarrow }^{\ast
}a_{y,\downarrow }a_{y,\uparrow }\right) =\tilde{\omega}_{l,k}\left(
a_{y,\uparrow }^{\ast }a_{y,\downarrow }^{\ast }a_{x,\downarrow
}a_{x,\uparrow }\right) \ ,  \label{real quantity}
\end{equation}%
by translation and reflection invariance of $\tilde{H}_{l,\mathfrak{e}%
}\left( r,k\right) $ within $\mathfrak{\tilde{G}}_{l,k}$ seen as the torus.
In particular, by hermicity of states, (\ref{real quantity}) is a real
quantity and, as a consequence,
\begin{equation}
\tilde{\omega}_{l,k}\left( \mathrm{I}_{l}^{x,y}\right) =0\ ,\qquad x,y\in
\mathfrak{\tilde{G}}_{l,k}\ ,\ k\in \mathcal{\tilde{R}}_{l}\ ,\ l>1\ .
\label{current zero pair}
\end{equation}%
Furthermore, by combining \cite[Eq. (A.6)]{BruPedra1} with translation
invariance of $\tilde{\omega}_{l,k}$ in the torus, one obtains that
\begin{equation}
\tilde{\omega}_{l,k}\left( a_{x,\downarrow }a_{x,\uparrow }\right) \in
\mathbb{R}\ ,\qquad x\in \mathfrak{\tilde{G}}_{l,k}\ ,\ k\in \mathcal{\tilde{%
R}}_{l}\ ,\ l>1\ .  \label{real idiot0}
\end{equation}%
In particular, for any $\mathrm{B}\in L^{2}(\mathfrak{C};\mathbb{R}^{3})$, $%
\gamma \in \mathbb{R}^{+}$, $r\in \mathbb{R}_{0}^{+}$, $l>1$, $k\in \mathcal{%
\tilde{R}}_{l}$, and $x\in \mathfrak{\tilde{G}}_{l,k}$,
\begin{equation}
\tilde{\omega}_{l,k}\left( a_{x,\downarrow }a_{x,\uparrow }\right) =\gamma
^{-1}\sqrt{r}\partial _{r}\mathrm{\hat{p}}_{l,\mathfrak{e}}\left( \mathrm{B}%
,r,k\right) \ ,  \label{real idiot}
\end{equation}%
where%
\begin{equation*}
\mathrm{\hat{p}}_{l,\mathfrak{e}}\left( \mathrm{B},r,k\right) :=(\beta |%
\mathfrak{\tilde{G}}_{l,k}|)^{-1}\ln \mathrm{Trace}_{\wedge \mathcal{H}_{%
\mathfrak{\tilde{G}}_{l,k}}}(\mathrm{e}^{-\beta \tilde{H}_{l,\mathfrak{e}%
}\left( r,k\right) })\ .
\end{equation*}

Similar to (\ref{gibbs box0}), we next define the approximating states by
product states of the form
\begin{equation*}
\mathfrak{\tilde{g}}_{l,r}\equiv \mathfrak{\tilde{g}}_{l,r,\mathrm{B}}:=%
\underset{k\in \mathcal{\tilde{R}}_{l}}{\bigotimes }\tilde{\omega}_{l,k}\in
E_{\Lambda _{l}}
\end{equation*}%
for all $l>1$, $r\in \mathbb{R}_{0}^{+}$ and $\mathrm{B}\in L^{2}(\mathfrak{C%
};\mathbb{R}^{3})$, see (\ref{toto box}). Observe that $\tilde{\omega}_{l,k}$
are even states and the above product is thus well--defined, see \cite[%
Theorem 11.2.]{Araki-Moriya}. Moreover, from (\ref{current zero pair})--(\ref%
{real idiot0}),
\begin{equation*}
\mathfrak{\tilde{g}}_{l,r}\left( \mathrm{I}_{l}^{x,y}\right) =0\ ,\qquad
x,y\in \Lambda _{l}\ ,\ l>1\ ,\ r\in \mathbb{R}_{0}^{+}\ .
\end{equation*}%
By (\ref{curents observable}), observe that $\mathrm{I}_{l}^{x,y}$ is the
Cooper pair component of the current. The hopping term also yields an
electronic component of the current defined by
\begin{equation*}
\mathrm{i}^{x,y}:=\mathop{\rm Im}\left\{ \mathfrak{e}\left( x-y\right)
\left( a_{x,\uparrow }^{\ast }a_{y,\uparrow }+a_{x,\downarrow }^{\ast
}a_{y,\downarrow }\right) \right\} \ ,\qquad x,y\in \mathbb{Z}^{3}\ .
\end{equation*}%
With this definition, the following conservation equation holds:
\begin{equation*}
\frac{\mathrm{d}}{\mathrm{d}t}\left\{ \mathrm{e}^{itH_{l,\mathfrak{e}%
}}\left( n_{x,\uparrow }+n_{x,\downarrow }\right) \mathrm{e}^{-itH_{l,%
\mathfrak{e}}}\right\} \Big |_{t=0}=\sum_{y\in \Lambda _{l}}\left( \mathrm{I}%
_{l}^{x,y}+\mathrm{i}^{x,y}\right) \ ,\qquad x\in \Lambda _{l}\ ,\ l>1\ .
\end{equation*}%
By using again the translation and reflection invariance of $\tilde{H}_{l,%
\mathfrak{e}}\left( r,k\right) $ within $\mathfrak{\tilde{G}}_{l,k}$ seen as
the torus as well as the fact that, for any $l>1$ and $k\in \mathcal{\tilde{R%
}}_{l}$, $\tilde{\omega}_{l,k}$ is an even state, we arrive at
\begin{equation*}
\mathfrak{\tilde{g}}_{l,r}\left( \mathrm{i}^{x,y}\right) =0\ ,\qquad x,y\in
\Lambda _{l}\ ,\ l>1\ ,\ r\in \mathbb{R}_{0}^{+}\ .
\end{equation*}%
\smallskip

\noindent \textbf{5.} \textit{Cf. Lemma \ref{lemma free energy copy(2)}.}
Assume that $\mathrm{B}\in C^{0}(\mathfrak{C};\mathbb{R}^{3})$. Define the
Bogoliubov (unitary) transformation $\mathrm{U}\in \mathcal{U}_{\mathfrak{%
\tilde{G}}_{l,k}}$ by
\begin{equation*}
\mathrm{U}a_{x,\uparrow }\mathrm{U}^{\ast }:=\frac{1}{\sqrt{2}}\left(
a_{x,\uparrow }+a_{x,\downarrow }^{\ast }\right) \ ,\qquad \mathrm{U}%
a_{x,\downarrow }\mathrm{U}^{\ast }:=\frac{1}{\sqrt{2}}\left(
a_{x,\downarrow }-a_{x,\uparrow }^{\ast }\right) \ ,
\end{equation*}%
for any $x\in \mathbb{Z}^{3}$. Let
\begin{equation*}
h_{l,k}:=\vartheta \left\vert \mathrm{\bar{B}}_{l,k}\right\vert \ ,\qquad
k\in \mathcal{\tilde{R}}_{l}\ ,\ l>1\ ,
\end{equation*}%
see (\ref{b bar}). Then, assuming without loss of generality that $\mathrm{B}
$ is oriented along the $z$--axis, one gets that, for any $k\in \mathcal{%
\tilde{R}}_{l}$, $r\in \mathbb{R}_{0}^{+}$, and $l>1$,
\begin{equation*}
\mathrm{U}^{\ast }\tilde{H}_{l,\mathfrak{e}}\left( r,k\right) \mathrm{U}%
=\sum\limits_{x\in \mathfrak{\tilde{G}}_{l,k}}\left\{ \left( \sqrt{\mu
^{2}+\gamma ^{2}r}-h_{l,k}\right) a_{x,\uparrow }^{\ast }a_{x,\uparrow
}+\left( \sqrt{\mu ^{2}+\gamma ^{2}r}+h_{l,k}\right) a_{x,\downarrow }^{\ast
}a_{x,\downarrow }\right\} +\Phi _{\mathfrak{e},\lambda }+C\ ,
\end{equation*}%
with $C\in \mathbb{R}$ being some real constant and where $\Phi _{\mathfrak{e%
},\lambda }$ is a term of the form
\begin{multline*}
\Phi _{\mathfrak{e},\lambda }\left( \varphi \right) =\sum\limits_{\substack{ %
\nu _{1},\nu _{2}\in \left\{ \ast ,-\right\} \  \\ \mathrm{s}_{1},\mathrm{s}%
_{2}\in \left\{ \uparrow ,\downarrow \right\}  \\ x_{1},x_{2}\in \mathfrak{%
\tilde{G}}_{l,k}}}\varphi _{2}\left( \left( \nu _{1},\mathrm{s}%
_{1},x_{1}\right) ,\left( \nu _{2},\mathrm{s}_{2},x_{2}\right) \right)
:a_{x_{1},\mathrm{s}_{1}}^{\nu _{1}}a_{x_{2},\mathrm{s}_{2}}^{\nu _{2}}: \\
+\sum\limits_{\substack{ \nu _{1},\nu _{2},\nu _{3},\nu _{4}\in \left\{ \ast
,-\right\}  \\ \mathrm{s}_{1},\mathrm{s}_{2},\mathrm{s}_{3},\mathrm{s}%
_{4}\in \left\{ \uparrow ,\downarrow \right\}  \\ x_{1},x_{2},x_{3},x_{4}\in
\mathfrak{\tilde{G}}_{l,k}}}\varphi _{4}\left( \left( \nu _{1},\mathrm{s}%
_{1},x_{1}\right) ,\ldots ,\left( \nu _{4},\mathrm{s}_{4},x_{4}\right)
\right) :a_{x_{1},\mathrm{s}_{1}}^{\nu _{1}}a_{x_{2},\mathrm{s}_{2}}^{\nu
_{2}}a_{x_{3},\mathrm{s}_{3}}^{\nu _{3}}a_{x_{4},\mathrm{s}_{4}}^{\nu
_{4}}:\ .
\end{multline*}%
Here, the notation
\begin{equation}
:a_{x_{1},\mathrm{s}_{1}}^{\nu _{1}}\dots a_{x_{n},\mathrm{s}_{n}}^{\nu
_{n}}:\quad :=(-1)^{\varsigma }a_{x_{\varsigma (1)},\mathrm{s}_{\varsigma
(1)}}^{\nu _{\varsigma (1)}}\dots a_{x_{\varsigma (n)},\mathrm{s}_{\varsigma
(n)}}^{\nu _{\varsigma (n)}}  \label{normal}
\end{equation}%
stands for the normal ordered product defined via any permutation $\varsigma
$ of the set $\{1,\dots ,n\}$ moving all creation operators in the product $%
a_{x_{\varsigma (1)},\mathrm{s}_{\varsigma (1)}}^{\nu _{\varsigma (1)}}\dots
a_{x_{\varsigma (n)},\mathrm{s}_{\varsigma (n)}}^{\nu _{\varsigma (n)}}$ to
the left of all annihilation operators. $\varphi _{2}$ and $\varphi _{4}$
are anti--symmetric complex functions satisfying
\begin{eqnarray*}
\varphi _{2}(\left( \nu _{1},\mathrm{s}_{1},x_{1}\right) ,\left( \nu _{2},%
\mathrm{s}_{2},x_{2}\right) ) &=&\overline{\varphi _{2}(\left( \overline{\nu
_{2}},\mathrm{s}_{2},x_{2}\right) ,\left( \overline{\nu _{1}},\mathrm{s}%
_{1},x_{1}\right) )} \\
\varphi _{4}(\left( \nu _{1},\mathrm{s}_{1},x_{1}\right) ,\ldots ,\left( \nu
_{4},\mathrm{s}_{4},x_{4}\right) ) &=&\overline{\varphi _{4}(\left(
\overline{\nu _{4}},\mathrm{s}_{4},x_{4}\right) ,\ldots ,\left( \overline{%
\nu _{1}},\mathrm{s}_{1},x_{1}\right) )}
\end{eqnarray*}%
with $\overline{\ast }:=-$ and $\overline{-}:=\ast $. One computes that%
\begin{eqnarray*}
\underset{\nu _{1}\in \left\{ \ast ,-\right\} ,\mathrm{s}_{1}\in \left\{
\uparrow ,\downarrow \right\} ,x_{1}\in \mathfrak{\tilde{G}}_{l,k}}{\max }%
\sum\limits_{\substack{ \nu _{2}\in \left\{ \ast ,-\right\}  \\ \mathrm{s}%
_{2}\in \left\{ \uparrow ,\downarrow \right\}  \\ x_{2}\in \mathfrak{\tilde{G%
}}_{l,k}}}\left\vert \varphi _{2}\left( \left( \nu _{1},\mathrm{s}%
_{1},x_{1}\right) ,\left( \nu _{2},\mathrm{s}_{2},x_{2}\right) \right)
\right\vert  &=&\mathcal{O}\left( \Vert \mathfrak{e}\Vert _{1}\right) \ , \\
\underset{\nu _{1}\in \left\{ \ast ,-\right\} ,\mathrm{s}_{1}\in \left\{
\uparrow ,\downarrow \right\} ,x_{1}\in \mathfrak{\tilde{G}}_{l,k}}{\max }%
\sum\limits_{\substack{ \nu _{2},\nu _{3},\nu _{4}\in \left\{ \ast
,-\right\}  \\ \mathrm{s}_{2},\mathrm{s}_{3},\mathrm{s}_{4}\in \left\{
\uparrow ,\downarrow \right\}  \\ x_{2},x_{3},x_{4}\in \mathfrak{\tilde{G}}%
_{l,k}}}\left\vert \varphi _{4}\left( \left( \nu _{1},\mathrm{s}%
_{1},x_{1}\right) ,\ldots ,\left( \nu _{4},\mathrm{s}_{4},x_{4}\right)
\right) \right\vert  &=&\mathcal{O}\left( \left\vert \lambda \right\vert
\right) \ ,
\end{eqnarray*}%
uniformly with respect to $k\in \mathcal{\tilde{R}}_{l}$\ and $l>1$. For all
$x_{1},x_{2}\in \mathbb{Z}^{3}$, $\mathrm{s}_{1},\mathrm{s}_{2}\in \left\{
\uparrow ,\downarrow \right\} $, and $\tau _{1},\tau _{2}\in \lbrack -\beta
,\beta )$, we define the fermionic imaginary time covariance by%
\begin{equation}
\mathcal{C}\left( \left( x_{1},\mathrm{s}_{1},\tau _{1}\right) ,\left( x_{2},%
\mathrm{s}_{2},\tau _{2}\right) \right) =\frac{\mathrm{e}^{-\left( \tau
_{1}-\tau _{2}\right) \left( \sqrt{\mu ^{2}+\gamma ^{2}r}-h_{l,k}\right) }}{%
1+\mathrm{e}^{-\beta \left( \sqrt{\mu ^{2}+\gamma ^{2}r}-h_{l,k}\right) }}%
\delta _{x_{1},x_{2}}\delta _{\mathrm{s}_{1},\uparrow }\delta _{\mathrm{s}%
_{2},\uparrow }+\frac{\mathrm{e}^{-\left( \tau _{1}-\tau _{2}\right) \left(
\sqrt{\mu ^{2}+\gamma ^{2}r}+h_{l,k}\right) }}{1+\mathrm{e}^{-\beta \left(
\sqrt{\mu ^{2}+\gamma ^{2}r}+h_{l,k}\right) }}\delta _{x_{1},x_{2}}\delta _{%
\mathrm{s}_{1},\downarrow }\delta _{\mathrm{s}_{2},\downarrow }
\label{covariance derivative}
\end{equation}%
when $\tau _{1}\geq \tau _{2}$, $|\tau _{1}-\tau _{2}|<\beta $, and by
\begin{equation*}
\mathcal{C}\left( \left( x_{1},\mathrm{s}_{1},\tau _{1}\right) ,\left( x_{2},%
\mathrm{s}_{2},\tau _{2}\right) \right) =-\mathcal{C}\left( \left( x_{1},%
\mathrm{s}_{1},\tau _{1}-\tau _{2}+\beta \right) ,\left( x_{2},\mathrm{s}%
_{2},0\right) \right)
\end{equation*}%
when $\tau _{2}>\tau _{1}$, $|\tau _{1}-\tau _{2}|<\beta $. For $|\tau
_{1}-\tau _{2}|\geq \beta $, we impose the $2\beta $--periodicity:
\begin{equation*}
\mathcal{C}\left( \left( x_{1},\mathrm{s}_{1},\tau _{1}\pm 2\beta \right)
,\left( x_{2},\mathrm{s}_{2},\tau _{2}\right) \right) =\mathcal{C}\left(
\left( x_{1},\mathrm{s}_{1},\tau _{1}\right) ,\left( x_{2},\mathrm{s}%
_{2},\tau _{2}\right) \right) =\mathcal{C}\left( \left( x_{1},\mathrm{s}%
_{1},\tau _{1}\right) ,\left( x_{2},\mathrm{s}_{2},\tau _{2}\pm 2\beta
\right) \right) \ .
\end{equation*}%
By \cite[Theorem 1.3]{Pedra-Salmhofer}, this covariance obeys the following
determinant bound:
\begin{equation}
\underset{\left\{ m_{i,j}\right\} _{i,j=1}^{N}\geq 0\ ,\ |m_{i,j}|\leq 1}{%
\sup }\left\vert \det \left\{ m_{i,j}\ \mathcal{C}\left( \left( x_{i},%
\mathrm{s}_{i},\tau _{i}\right) ,\left( x_{j},\mathrm{s}_{j},\tau
_{j}\right) \right) \right\} _{i,j=1}^{N}\right\vert \leq 4^{N}
\label{moreover}
\end{equation}%
for all $x_{i},x_{j}\in \mathbb{Z}^{3}$, $\mathrm{s}_{i},\mathrm{s}_{j}\in
\left\{ \uparrow ,\downarrow \right\} $, $\tau _{i},\tau _{j}\in \lbrack
-\beta ,\beta )$ with $N\in \mathbb{N}$ and $i,j\in \{1,\ldots ,N\}$.
Moreover,
\begin{equation}
\underset{x_{1}\in \mathbb{Z}^{3},\mathrm{s}_{1}\in \left\{ \uparrow
,\downarrow \right\} ,\tau _{1}\in \lbrack -\beta ,\beta )}{\max }\sum\limits
_{\substack{ x_{2}\in \mathbb{Z}^{3} \\ \mathrm{s}_{2}\in \left\{ \uparrow
,\downarrow \right\} }}\int_{[-\beta ,\beta )}\left\vert \mathcal{C}(\left(
x_{1},\mathrm{s}_{1},\tau _{1}\right) ,\left( x_{2},\mathrm{s}_{2},\tau
_{2}\right) )\right\vert \mathrm{d}\tau _{2}=\mathcal{O}\left( \beta \right)
\ .  \label{moreoverbis}
\end{equation}%
Equations (\ref{moreover})--(\ref{moreoverbis}) imply that, for any inverse
temperature $\beta \in \mathbb{R}^{+}$ such that $\beta \left( |\lambda
|+\Vert \mathfrak{e}\Vert _{1}\right) $ is sufficiently small, all $%
\mathfrak{t}\in \mathfrak{C}$, and any sequence $\{k_{l,\mathfrak{t}%
}\}_{l=2}^{\infty }$ with $k_{l,\mathfrak{t}}=(k_{1,l,\mathfrak{t}},k_{2,l,%
\mathfrak{t}},k_{3,l,\mathfrak{t}})\in \mathcal{\tilde{R}}_{l}$ and
\begin{equation*}
\underset{l\rightarrow \infty }{\lim }\left\vert \mathfrak{t}-(\ell ^{\eta
-1}k_{1,l,\mathfrak{t}},\ell ^{\eta ^{\bot }-1}k_{2,l,\mathfrak{t}},\ell
^{\eta ^{\bot }-1}k_{3,l,\mathfrak{t}})\right\vert =0\ ,
\end{equation*}%
one has
\begin{equation}
\underset{l\rightarrow \infty }{\lim }\mathrm{\hat{p}}_{l,\mathfrak{e}%
}\left( \mathrm{B},r,k_{l,\mathfrak{t}}\right) =\mathrm{\tilde{p}}_{%
\mathfrak{e}}\left( \mathrm{B},r,\mathfrak{t}\right) \equiv \mathrm{\tilde{p}%
}_{\mathfrak{e}}\left( r,\mathfrak{t}\right) \ .  \label{triviality}
\end{equation}%
Moreover, for all $\mathfrak{t}\in \mathfrak{C}$, $r\mapsto \mathrm{\tilde{p}%
}_{\mathfrak{e}}\left( r,\mathfrak{t}\right) $ is a differentiable function
of $r\in \mathbb{R}_{0}^{+}$ and the derivative $\partial _{r}\mathrm{\tilde{%
p}}_{\mathfrak{e}}$ is a continuous function of parameters $\mathrm{B}(%
\mathfrak{t})\in \mathbb{R}^{3}$ and $r\in \mathbb{R}_{0}^{+}$ at $\mathfrak{%
t}\in \mathfrak{C}$. The proof of these facts uses Grassmann integration and
Brydges--Kennedy tree expansions together with (\ref{moreover})--(\ref%
{moreoverbis}). For more details, see \cite{Pedra-Salmhofer} and the
references therein. Analogously, under the same condition, the function
\begin{equation*}
r\mapsto \int_{\mathfrak{C}}\mathrm{\tilde{p}}_{\mathfrak{e}}\left( r,%
\mathfrak{t}\right) \mathrm{d}^{3}\mathfrak{t}
\end{equation*}%
is differentiable and
\begin{equation*}
\partial _{r}\left( \int_{\mathfrak{C}}\mathrm{\tilde{p}}_{\mathfrak{e}%
}\left( r,\mathfrak{t}\right) \mathrm{d}^{3}\mathfrak{t}\right) =\int_{%
\mathfrak{C}}\partial _{r}\mathrm{\tilde{p}}_{\mathfrak{e}}\left( r,%
\mathfrak{t}\right) \mathrm{d}^{3}\mathfrak{t}\ .
\end{equation*}%
It follows that any non--zero solution $\mathrm{r}_{\beta }$ of the
variational problem of (\ref{var blilbi}) has to be solution of the gap
equation (or Euler--Lagrange equation):%
\begin{equation}
\beta ^{-1}\int_{\mathfrak{C}}\partial _{r}\mathrm{\tilde{p}}_{\mathfrak{e}%
}\left( r,\mathfrak{t}\right) \mathrm{d}^{3}\mathfrak{t}=\gamma \ .
\label{gap equation new}
\end{equation}%
Furthermore, by using Griffiths arguments \cite[Appendix]{BruPedra1}, one
deduces from (\ref{real idiot}) and (\ref{triviality}) that%
\begin{equation}
\underset{l\rightarrow \infty }{\lim }\left\{ \underset{k\in \mathcal{\tilde{%
R}}_{l}}{\sum }\frac{|\mathfrak{\tilde{G}}_{l,k}|}{\left\vert \Lambda
_{l}\right\vert }\tilde{\omega}_{l,k}\left( a_{x,\downarrow }a_{x,\uparrow
}\right) \right\} =\gamma ^{-1}\sqrt{r}\int_{\mathfrak{C}}\partial _{r}%
\mathrm{\tilde{p}}_{\mathfrak{e}}\left( r,\mathfrak{t}\right) \mathrm{d}^{3}%
\mathfrak{t}\ .  \label{gap equation 3-2 new}
\end{equation}%
Compare with (\ref{gap equation 3-2}).

Therefore, by (\ref{triviality})--(\ref{gap equation 3-2 new}), for all $%
\beta \in \mathbb{R}^{+}$ such that $\beta \left( |\lambda |+\Vert \mathfrak{%
e}\Vert _{1}\right) $ is sufficiently small, Lemma \ref{lemma free energy
copy(2)} holds true in the case $\mathfrak{e}\neq 0$ with $\mathfrak{\tilde{g%
}}_{l,\mathrm{r}_{\beta }}$ replacing the state $\mathfrak{g}_{l,\mathrm{r}%
_{\beta }}$.\smallskip

\noindent \textbf{6.} \textit{Cf. Theorem \ref{lemma free energy copy(1)}.}
We want to prove the norm equicontinuity of the collection
\begin{equation*}
\{\mathrm{B}\mapsto f_{l}(\mathrm{B},\mathfrak{\tilde{g}}_{l,r,\mathrm{B}%
})\}_{l\in \mathbb{N}}
\end{equation*}%
of maps from $L^{2}(\mathfrak{C};\mathbb{R}^{3})$ to $\mathbb{R}$. As in the
proof of Theorem \ref{lemma free energy copy(1)}, we prove separately the
equicontinuity of the families
\begin{equation}
\text{(i) }\{\mathrm{B}\mapsto f_{l}(0,\mathfrak{\tilde{g}}_{l,r,\mathrm{B}%
})\}_{l\in \mathbb{N}}\qquad \text{and}\qquad \text{(ii) }\left\{ \mathrm{B}%
\mapsto \langle \mathrm{B},\mathfrak{m}_{l}\left( \mathrm{B}\right) \rangle
_{2}\right\} _{l\in \mathbb{N}}  \label{family00}
\end{equation}%
of maps from $L^{2}(\mathfrak{C};\mathbb{R}^{3})$ to $\mathbb{R}$, see (\ref%
{map a la condebi+}). Starting with (i), we observe that, for any $\mathrm{B}%
\in L^{2}(\mathfrak{C};\mathbb{R}^{3})$, $l>1$ and $r\in \mathbb{R}_{0}^{+}$,%
\begin{eqnarray}
f_{l}(0,\mathfrak{\tilde{g}}_{l,r,\mathrm{B}}) &=&-\sum\limits_{k\in
\mathcal{\tilde{R}}_{l}}\frac{|\mathfrak{\tilde{G}}_{l,k}|}{\left\vert
\Lambda _{l}\right\vert }\left\{ \mathrm{\hat{p}}_{l,\mathfrak{e}}\left(
\mathrm{B},r,k\right) +\frac{1}{|\mathfrak{\tilde{G}}_{l,k}|}\tilde{\omega}%
_{l,k}\left( \mathcal{\tilde{M}}_{l,k}\right) \right\}   \notag \\
&&-\frac{1}{\left\vert \Lambda _{l}\right\vert }\sum\limits_{x\in \Lambda
_{l}}\gamma \sqrt{r}\mathfrak{\tilde{g}}_{l,r,\mathrm{B}}(a_{x,\uparrow
}^{\ast }a_{x,\downarrow }^{\ast }+a_{x,\downarrow }a_{x,\uparrow })  \notag
\\
&&-\frac{\gamma }{\left\vert \Lambda _{l}\right\vert ^{2}}\sum_{x,y\in
\Lambda _{l}}\mathfrak{\tilde{g}}_{l,r,\mathrm{B}}\left( a_{x,\uparrow
}^{\ast }a_{x,\downarrow }^{\ast }a_{y,\downarrow }a_{y,\uparrow }\right) \ .
\label{family0}
\end{eqnarray}%
The collection
\begin{equation}
\{\mathrm{\bar{B}}_{l,k}\mapsto \mathrm{\hat{p}}_{l,\mathfrak{e}}\left(
\mathrm{\bar{B}}_{l,k},r,k\right) \}_{k\in \mathcal{\tilde{R}}_{l},l\in
\mathbb{N}}  \label{family1}
\end{equation}%
of maps from $\mathbb{R}^{3}$ to $\mathbb{R}$ is norm equicontinuous, by
\cite[Eq. (3.11)]{BruPedra2}. Moreover, using again determinant bounds,
Grassmann integration and Brydges--Kennedy tree expansion for the Gibbs
states $\{\tilde{\omega}_{l,k}\}_{l>1,k\in \mathcal{\tilde{R}}_{l}}$ we
obtain that, for all $\beta \in \mathbb{R}^{+}$ such that $\beta \left(
|\lambda |+\Vert \mathfrak{e}\Vert _{1}\right) $ is sufficiently small, the
families
\begin{eqnarray}
&&\left\{ \mathrm{\bar{B}}_{l,k}\mapsto |\mathfrak{\tilde{G}}_{l,k}|^{-1}%
\tilde{\omega}_{l,k}\left( \mathcal{\tilde{M}}_{l,k}\right) \right\} _{k\in
\mathcal{\tilde{R}}_{l},l\in \mathbb{N}}  \label{family2} \\
&&\left\{ \mathrm{\bar{B}}_{l,k}\mapsto |\mathfrak{\tilde{G}}%
_{l,k}|^{-1}\sum\limits_{x\in \mathfrak{\tilde{G}}_{l,k}}\tilde{\omega}%
_{l,k}(a_{x,\downarrow }a_{x,\uparrow })\right\} _{k\in \mathcal{\tilde{R}}%
_{l},l\in \mathbb{N}}  \label{family3} \\
&&\left\{ \mathrm{\bar{B}}_{l,k}\mapsto |\mathfrak{\tilde{G}}%
_{l,k}|^{-2}\sum\limits_{x,y\in \mathfrak{\tilde{G}}_{l,k}}\tilde{\omega}%
_{l,k}\left( a_{x,\uparrow }^{\ast }a_{x,\downarrow }^{\ast }a_{y,\downarrow
}a_{y,\uparrow }\right) \right\} _{k\in \mathcal{\tilde{R}}_{l},l\in \mathbb{%
N}}  \label{family4}
\end{eqnarray}%
of maps from $\mathbb{R}^{3}$ to $\mathbb{R}$ are also uniformly Lipschitz
equicontinuous. By using the Cauchy--Schwarz inequality and Jensen's
inequality together with (\ref{family1})--(\ref{family4}) we then deduce
from (\ref{family0}) that the first collection (i) of maps from $L^{2}(%
\mathfrak{C};\mathbb{R}^{3})$ to $\mathbb{R}$ in (\ref{family00}) is norm
equicontinuous. By similar arguments, the second family (ii) in (\ref%
{family00}) is also norm equicontinuous. Theorem \ref{lemma free energy
copy(1)} follows in the case $\mathfrak{e}\neq 0$ with $\mathfrak{\tilde{g}}%
_{l,\mathrm{r}_{\beta }}$ replacing the state $\mathfrak{g}_{l,\mathrm{r}%
_{\beta }}$, provided $\beta \left( |\lambda |+\Vert \mathfrak{e}\Vert
_{1}\right) $ is sufficiently small. \smallskip

\noindent \textbf{7.} \textit{Cf. Lemmata \ref{lemma free energy}--\ref%
{propositioncool}}. We define from $\{\mathfrak{g}_{l,\mathrm{r}_{\beta
}}\}_{l\in \mathbb{N}}$ states manifesting some current in subregions of the
box $\Lambda _{l}$ with very small volumes with respect to the total volume $%
|\Lambda _{l}|=(2l+1)^{3}$. This is done exactly as explained after Theorem %
\ref{lemma free energy copy(1)} with $\mathfrak{\tilde{g}}_{l,\mathrm{r}%
_{\beta }}$ replacing the state $\mathfrak{g}_{l,\mathrm{r}_{\beta }}$.
Observe indeed that the state $\varpi \in E_{\left\{ 0\right\} }$ is even.
In particular,%
\begin{equation*}
\left( \underset{z\in \mathfrak{G}_{l,k}}{\otimes }\varpi _{z}\right) \left(
\mathrm{i}_{l}^{x,y}\right) =0\ ,\qquad x,y\in \mathfrak{G}_{l,k}\ ,\ l>1\
,\ k\in \mathcal{\tilde{R}}_{l}\ .
\end{equation*}%
Then, the total current in the new state $\rho _{l}$ equals its Cooper pair
component. As a consequence, Lemmata \ref{lemma free energy}--\ref%
{propositioncool} hold true when $\mathfrak{e}\neq 0$, provided $\beta
\left( |\lambda |+\Vert \mathfrak{e}\Vert _{1}\right) $ is sufficiently
small.\smallskip

\noindent \textbf{8.} \textit{Cf. Lemma \ref{lemma var prob3 copy(10)}}.
This lemma also holds true when $\mathfrak{e}\neq 0$, provided $\beta \left(
|\lambda |+\Vert \mathfrak{e}\Vert _{1}\right) $ is sufficiently small.
Indeed, observe that the map (\ref{magnetization0bis}) is replaced by the
map
\begin{equation*}
\mathrm{B}\mapsto \underset{l\rightarrow \infty }{\lim }\left\vert \Lambda
_{l}\right\vert ^{-1}\ln \mathrm{Trace}_{\wedge \mathcal{H}_{\Lambda
_{l}}}\left( \mathrm{e}^{-\beta H_{l,\mathfrak{e}}\left( r,\mathrm{B}\right)
}\right)
\end{equation*}%
from $\mathbb{R}^{3}$ to $\mathbb{R}^{+}$, with $H_{l,\mathfrak{e}}\left( r,%
\mathrm{B}\right) $ defined by (\ref{approximating hamiltoian}) for any $%
r\in \mathbb{R}_{0}^{+}$ and $\mathrm{B}\in \mathbb{R}^{3}$. It is clearly a
continuous convex function at any fixed $(\beta ,\mu ,\lambda ,\gamma
,\vartheta ,r)$. \smallskip

\noindent \textbf{9.} \textit{Cf. Lemmata \ref{lemma var prob3 copy(9)}--\ref%
{lemma var prob3 copy(4)}}. For any $r\in \mathbb{R}_{0}^{+}$ and $\mathrm{B}%
\in \mathbb{R}^{3}$, we define the magnetization by
\begin{equation*}
\mathbf{M}_{\beta }\left( \mathrm{B}\right) :=\partial _{\mathrm{B}}\left(
\underset{l\rightarrow \infty }{\lim }\left\vert \Lambda _{l}\right\vert
^{-1}\ln \mathrm{Trace}_{\wedge \mathcal{H}_{\Lambda _{l}}}\left( \mathrm{e}%
^{-\beta H_{l,\mathfrak{e}}\left( r,\mathrm{B}\right) }\right) \right) \in
\mathbb{R}^{3}\ ,
\end{equation*}%
where $\partial _{\mathrm{B}}f\left( \mathrm{b}_{1},\mathrm{b}_{2},\mathrm{b}%
_{3}\right) :=\left( \partial _{\mathrm{b}_{1}}f,\partial _{\mathrm{b}%
_{2}}f,\partial _{\mathrm{b}_{3}}f\right) $. This quantity is well--defined
if $\beta \left( |\lambda |+\Vert \mathfrak{e}\Vert _{1}\right) $ is
sufficiently small, again by determinant bounds, Grassmann integration and
Brydges--Kennedy tree expansions. Let the magnetization density $\mathrm{M}%
_{\beta }\equiv \mathrm{M}_{\beta }(\mathrm{B})\in L^{2}$ be defined a.e. on
$\mathbb{R}^{3}$, for any $\mathrm{B}\in L^{2}$, by
\begin{equation}
\mathrm{M}_{\beta ,\mathfrak{t}}\left( \mathrm{B}\right) :=\mathbf{1}[%
\mathfrak{t}\in \mathfrak{C}]\mathbf{M}_{\beta }\left( \mathrm{B}\left(
\mathfrak{t}\right) \right) \ .  \label{magnetization000}
\end{equation}%
In fact, Grassmann integration and Brydges--Kennedy tree expansion methods
together with determinant bounds yield Euler--Lagrange equations stated in
Lemma \ref{lemma var prob3 copy(9)} with the magnetization density $\mathrm{M%
}_{\beta }\equiv \mathrm{M}_{\beta }(\mathrm{B})$ defined on $\mathbb{R}^{3}$
by (\ref{magnetization000}) instead of (\ref{magnetization0}) for all $%
\mathrm{B}\in L^{2}$, provided $\beta \left( |\lambda |+\Vert \mathfrak{e}%
\Vert _{1}\right) $ is sufficiently small. Since, by using Griffiths
arguments \cite[Appendix]{BruPedra1}, $\left\Vert \mathrm{M}_{\beta }\left(
\mathrm{B}\right) \right\Vert _{2}\leq \vartheta $ for any $\mathrm{B}\in
\mathbb{R}^{3}$, Lemma \ref{lemma var prob3 copy(4)} follows, when $\beta
\left( |\lambda |+\Vert \mathfrak{e}\Vert _{1}\right) $ is sufficiently
small. \smallskip

\noindent \textbf{10.} \textit{Cf. Lemma \ref{lemma var prob3 copy(8)} and
Theorem \ref{lemma var prob3 copy(5)}}. Fix $\mu <-\vartheta ^{2}$ and $%
\beta _{0}\in \mathbb{R}^{+}$. There is $\gamma _{0}>\left\vert \mu
\right\vert \Gamma _{0}$ such that, for all $\beta \geq \beta _{0}$,
Equation (\ref{var blilbi2}) holds true. For all $\gamma \in \lbrack \gamma
_{0},\infty )$, $\beta \in \lbrack \beta _{0},2\beta _{0}]$, $\lambda \in
\mathbb{R}$ and any hopping amplitude $\mathfrak{e}$ such that $\lambda $
and $\Vert \mathfrak{e}\Vert _{1}$ are sufficiently small, $\mathrm{M}%
_{\beta }$ exists and%
\begin{equation*}
\left\Vert \mathbf{1}[h_{\mathfrak{t}}\leq \mathrm{h}]\mathrm{M}_{\beta
}\right\Vert _{2}\leq \left( \frac{\beta \vartheta \cosh \left( \beta
\mathrm{h}\right) }{\mathrm{e}^{-\beta \lambda }\cosh \left( \beta g_{%
\mathrm{r}_{\beta }}\right) }+C\left( \Vert \mathfrak{e}\Vert
_{1}+\left\vert \lambda \right\vert \right) \right) \left\Vert \mathrm{B}%
\right\Vert _{2}\ ,
\end{equation*}%
for some constant $C\in \mathbb{R}^{+}$ not depending on $\mathrm{h}\in
\mathbb{R}_{0}^{+}$, $\beta \in \lbrack \beta _{0},2\beta _{0}]$, $\gamma
\in \lbrack \gamma _{0},\infty )$ and $\mathfrak{e}$. This inequality
follows again from Grassmann integration, Brydges--Kennedy tree expansion
and determinant bounds. Recall that $h_{\mathfrak{t}}:=\vartheta \left\vert
\mathrm{B}\left( \mathfrak{t}\right) \right\vert $ a.e. in the unit box $%
\mathfrak{C}$. From this, we deduce Lemma \ref{lemma var prob3 copy(8)},
provided $\Vert \mathfrak{e}\Vert _{1}$ is sufficiently small to
additionally ensure that%
\begin{equation*}
\frac{\beta \cosh \left( \beta \mathrm{h}\right) }{\mathrm{e}^{-\beta
\lambda }\cosh \left( \beta g_{\mathrm{r}_{\beta }(\mathfrak{T}_{\epsilon }(%
\mathrm{B}_{\epsilon }-\mathrm{B}_{\mathrm{int}}))}\right) }+C\vartheta
^{-1}\left( \Vert \mathfrak{e}\Vert _{1}+\left\vert \lambda \right\vert
\right) +\vartheta \mathrm{h}^{-1}<\vartheta ^{-1}\ .
\end{equation*}%
Theorem \ref{lemma var prob3 copy(5)} in the case $\mathfrak{e}\neq 0$ is
then a direct consequence of previous assertions. This concludes the sketch
of the proof of Theorem \ref{Theorem inv trans}. \smallskip

We conclude now this section by an additional result which shows, among
other things, how to manage the zero temperature case.

\begin{theorem}[Asymptotics of $\mathfrak{F}_{\mathfrak{e}}$ arround $\left(
\mathrm{r}_{\protect\beta }\left( \mathrm{B}_{\mathrm{int}}+\mathrm{B}_{%
\mathrm{ext}}\right) ,\mathrm{B}_{\mathrm{int}}+\mathrm{B}_{\mathrm{ext}%
}\right) $]
\label{lemma var prob3 copy(13)}\mbox{ }\newline
Fix $\mu \in \mathbb{R}$, $\lambda \in \mathbb{R}$ and $\mathrm{B}_{\mathrm{%
ext}}=\mathcal{S}_{0}(\mathrm{j}_{\mathrm{ext}})$ with $\mathrm{j}_{\mathrm{%
ext}}\in C_{0}^{\infty }\cap P^{\bot }\mathfrak{H}$. Then, there are
constants $\gamma _{0},\beta _{0},r_{0},\kappa _{0},C_{1},C_{2}\in \mathbb{R}%
^{+}$ such that $r_{0}<\mathrm{r}_{\beta }\left( \mathrm{B}_{\mathrm{int}}+%
\mathrm{B}_{\mathrm{ext}}\right) $ and, for all $\gamma \in \lbrack \gamma
_{0},\infty )$, $\beta \in \lbrack \beta _{0},\infty )$, $r\in \lbrack
r_{0},\infty )$, hopping amplitude $\mathfrak{e}$ satisfying $\Vert
\mathfrak{e}\Vert _{1}<\kappa _{0}$,
\begin{equation*}
\left\vert \mathfrak{F}_{\mathfrak{e}}\left( \mathrm{r}_{\beta }\left(
\mathrm{B}_{\mathrm{int}}+\mathrm{B}_{\mathrm{ext}}\right) ,\mathrm{B}_{%
\mathrm{int}}+\mathrm{B}_{\mathrm{ext}}\right) -\mathfrak{F}_{\mathfrak{e}%
}\left( r,\mathrm{B}+\mathrm{B}_{\mathrm{ext}}\right) \right\vert \leq
C_{1}\left( \Vert \mathrm{B}-\mathrm{B}_{\mathrm{int}}\Vert
_{2}^{2}+\left\vert r-\mathrm{r}_{\beta }\left( \mathrm{B}_{\mathrm{int}}+%
\mathrm{B}_{\mathrm{ext}}\right) \right\vert ^{2}\right) \ ,
\end{equation*}%
while%
\begin{equation*}
\left\vert \mathrm{r}_{\beta }\left( \mathrm{B}_{\mathrm{int}}+\mathrm{B}_{%
\mathrm{ext}}\right) -\mathrm{r}_{\beta }\left( \mathrm{B}+\mathrm{B}_{%
\mathrm{ext}}\right) \right\vert \leq C_{2}\Vert \mathrm{B}-\mathrm{B}_{%
\mathrm{int}}\Vert _{2}\ .
\end{equation*}
\end{theorem}

\noindent \textit{Proof.} Choose $\gamma _{0}\in \mathbb{R}^{+}$
sufficiently large such that (\ref{var blilbi2}) holds true. Take $r_{0}=%
\mathfrak{r}_{0}/3$. Let $\mathfrak{D}_{\mathrm{B}}$ be any measurable
subset of $\mathfrak{C}$ such that $\left\vert \mathrm{B}+\mathrm{B}_{%
\mathrm{ext}}\right\vert \leq \sqrt{r_{0}}\gamma _{0}/2$ a.e. in $\mathfrak{D%
}_{\mathrm{B}}$. If the parameter $h_{l,k}$ in Definition (\ref{covariance
derivative}) has absolute value less than $\sqrt{r_{0}}\gamma _{0}/2$ then
\begin{equation*}
\underset{x_{1}\in \mathbb{Z}^{3},\mathrm{s}_{1}\in \left\{ \uparrow
,\downarrow \right\} ,\tau _{1}\in \lbrack -\beta ,\beta )}{\max }\sum\limits
_{\substack{ \mathrm{s}_{2}\in \left\{ \uparrow ,\downarrow \right\}  \\ %
x_{2}\in \mathbb{Z}^{3}}}\int_{[-\beta ,\beta )}\left\vert \mathcal{C}%
(\left( x_{1},\mathrm{s}_{1},\tau _{1}\right) ,\left( x_{2},\mathrm{s}%
_{2},\tau _{2}\right) )\right\vert \mathrm{d}\tau _{2}=\mathcal{O}\left(
r_{0}^{-1/2}\gamma _{0}^{-1}\right) \ ,
\end{equation*}%
uniformly in the parameters $\gamma \in \lbrack \gamma _{0},\infty )$, $%
\beta \in \mathbb{R}^{+}$ and $r\in \lbrack r_{0},\infty )$. The above
covariance also obeys the determinant bound (\ref{moreover}). As a
consequence, by using Grassmann integration and Brydges--Kennedy tree
expansions, we deduce that the map
\begin{equation*}
\left( r,\mathrm{B}\right) \mapsto \underset{l\rightarrow \infty }{\lim }%
\left\vert \Lambda _{l}\right\vert ^{-1}\ln \mathrm{Trace}_{\wedge \mathcal{H%
}_{\Lambda _{l}}}\left( \mathrm{e}^{-\beta H_{l,\mathfrak{e}}\left( r,%
\mathrm{B}\right) }\right)
\end{equation*}%
from $\mathbb{R}_{0}^{+}\times \mathbb{R}^{3}$\ to $\mathbb{R}$ satisfies
the bounds%
\begin{equation*}
\partial _{r}^{i_{0}}\partial _{\mathrm{b}_{1}}^{i_{1}}\partial _{\mathrm{b}%
_{2}}^{i_{2}}\partial _{\mathrm{b}_{3}}^{i_{3}}\left( \underset{l\rightarrow
\infty }{\lim }\left\vert \Lambda _{l}\right\vert ^{-1}\ln \mathrm{Trace}%
_{\wedge \mathcal{H}_{\Lambda _{l}}}\left( \mathrm{e}^{-\beta H_{l,\mathfrak{%
e}}\left( r,\mathrm{B}\right) }\right) \right) \leq
C^{1+i_{0}+i_{1}+i_{2}+i_{3}}\left( i_{0}!\right) \left( i_{1}!\right)
\left( i_{2}!\right) \left( i_{3}!\right)
\end{equation*}%
for all $r\in \lbrack r_{0},\infty )$, $\left\vert \mathrm{B}\right\vert
\leq \sqrt{r_{0}}\gamma _{0}/2$, $i_{0},i_{1},i_{2},i_{3}\in \mathbb{N}_{0}$
and hopping amplitude $\mathfrak{e}$ satisfying $\Vert \mathfrak{e}\Vert
_{1}<1$. Here, $C\in \mathbb{R}^{+}$ does not depend on $\gamma \in \lbrack
\gamma _{0},\infty )$, $\beta \in \mathbb{R}^{+}$, $r\in \lbrack
r_{0},\infty )$ and $\mathfrak{e}$. It follows that%
\begin{equation}
\left\vert \int_{\mathfrak{C}\backslash \mathfrak{D}_{\mathrm{B}}}\left(
\mathrm{\tilde{p}}_{\mathfrak{e}}\left( r,\mathfrak{t},\mathrm{B}+\mathrm{B}%
_{\mathrm{ext}}\right) -\mathrm{\tilde{p}}_{\mathfrak{e}}\left( r,\mathfrak{t%
},\mathrm{B}_{\mathrm{int}}+\mathrm{B}_{\mathrm{ext}}\right) \right) \mathrm{%
d}^{3}\mathfrak{t}\right\vert \leq C_{0}\Vert \mathrm{B}-\mathrm{B}_{\mathrm{%
int}}\Vert _{2}^{2}  \label{toto1}
\end{equation}%
for some constant $C_{0}\in \mathbb{R}^{+}$ not depending on $\gamma \in
\lbrack \gamma _{0},\infty )$, $\beta \in \mathbb{R}^{+}$, $r\in \lbrack
r_{0},\infty )$ and $\mathfrak{e}$. Indeed, note that
\begin{equation*}
\partial _{\mathrm{b}_{1}}^{i_{1}}\partial _{\mathrm{b}_{2}}^{i_{2}}\partial
_{\mathrm{b}_{3}}^{i_{3}}\left( \underset{l\rightarrow \infty }{\lim }%
\left\vert \Lambda _{l}\right\vert ^{-1}\ln \mathrm{Trace}_{\wedge \mathcal{H%
}_{\Lambda _{l}}}\left( \mathrm{e}^{-\beta H_{l,\mathfrak{e}}\left( r,%
\mathrm{B}\right) }\right) \right) =0
\end{equation*}%
for all $i_{1},i_{2},i_{3}\in \mathbb{N}_{0}$, $i_{1}+i_{2}+i_{3}=1$, as
\begin{equation*}
\underset{l\rightarrow \infty }{\lim }\left\vert \Lambda _{l}\right\vert
^{-1}\ln \mathrm{Trace}_{\wedge \mathcal{H}_{\Lambda _{l}}}\left( \mathrm{e}%
^{-\beta H_{l,\mathfrak{e}}\left( r,\mathrm{B}\right) }\right) =\underset{%
l\rightarrow \infty }{\lim }\left\vert \Lambda _{l}\right\vert ^{-1}\ln
\mathrm{Trace}_{\wedge \mathcal{H}_{\Lambda _{l}}}\left( \mathrm{e}^{-\beta
H_{l,\mathfrak{e}}\left( r,-\mathrm{B}\right) }\right) \ .
\end{equation*}%
On the other hand, by \cite[Eq. (3.11)]{BruPedra2} and the Cauchy--Schwarz
inequality,
\begin{equation}
\left\vert \int_{\mathfrak{D}_{\mathrm{B}}}\left( \mathrm{\tilde{p}}_{%
\mathfrak{e}}\left( r,\mathfrak{t},\mathrm{B}+\mathrm{B}_{\mathrm{ext}%
}\right) -\mathrm{\tilde{p}}_{\mathfrak{e}}\left( r,\mathfrak{t},\mathrm{B}_{%
\mathrm{int}}+\mathrm{B}_{\mathrm{ext}}\right) \right) \mathrm{d}^{3}%
\mathfrak{t}\right\vert \leq \left\vert \mathfrak{D}_{\mathrm{B}}\right\vert
\vartheta \leq \frac{4\vartheta }{r_{0}\gamma _{0}^{2}}\Vert \mathrm{B}-%
\mathrm{B}_{\mathrm{int}}\Vert _{2}^{2}\ .  \label{toto2}
\end{equation}%
From (\ref{toto1})--(\ref{toto2}) it follows that%
\begin{equation}
\left\vert \int_{\mathfrak{C}}\left( \mathrm{\tilde{p}}_{\mathfrak{e}}\left(
r,\mathfrak{t},\mathrm{B}+\mathrm{B}_{\mathrm{ext}}\right) -\mathrm{\tilde{p}%
}_{\mathfrak{e}}\left( r,\mathfrak{t},\mathrm{B}_{\mathrm{int}}+\mathrm{B}_{%
\mathrm{ext}}\right) \right) \mathrm{d}^{3}\mathfrak{t}\right\vert \leq
C_{1}\Vert \mathrm{B}-\mathrm{B}_{\mathrm{int}}\Vert _{2}^{2}  \label{toto3}
\end{equation}%
for some constant $C_{1}\in \mathbb{R}^{+}$ not depending on $\gamma \in
\lbrack \gamma _{0},\infty )$, $\beta \in \mathbb{R}^{+}$, $r\in \lbrack
r_{0},\infty )$ and hopping amplitude $\mathfrak{e}$ with $\Vert \mathfrak{e}%
\Vert _{1}<1$. Now, using Lemma \ref{lemma var prob3 copy(2)}, Grassmann
integration and Brydges--Kennedy tree expansion together with determinant
bounds, one shows that $\mathrm{r}_{\beta }\left( \mathrm{B}_{\mathrm{int}}+%
\mathrm{B}_{\mathrm{ext}}\right) $ solves Equation (\ref{gap equation new})
and therefore,
\begin{equation}
\left\vert \int_{\mathfrak{C}}\left( \mathrm{\tilde{p}}_{\mathfrak{e}}\left(
r,\mathfrak{t},\mathrm{B}_{\mathrm{int}}+\mathrm{B}_{\mathrm{ext}}\right) -%
\mathrm{\tilde{p}}_{\mathfrak{e}}\left( \mathrm{r}_{\beta }\left( \mathrm{B}%
_{\mathrm{int}}+\mathrm{B}_{\mathrm{ext}}\right) ,\mathfrak{t},\mathrm{B}_{%
\mathrm{int}}+\mathrm{B}_{\mathrm{ext}}\right) \right) \mathrm{d}^{3}%
\mathfrak{t}\right\vert \leq C_{2}\left( \left\vert r-\mathrm{r}_{\beta
}\left( \mathrm{B}_{\mathrm{int}}+\mathrm{B}_{\mathrm{ext}}\right)
\right\vert ^{2}\right)   \label{toto4}
\end{equation}%
for some constant $C_{2}\in \mathbb{R}^{+}$ not depending on $\gamma \in
\lbrack \gamma _{0},\infty )$, $\beta \in \mathbb{R}^{+}$, $r\in \lbrack
r_{0},\infty )$ and hopping amplitude $\mathfrak{e}$ with $\Vert \mathfrak{e}%
\Vert _{1}<1$. From (\ref{toto3})--(\ref{toto4}) we deduce the first upper
bound of the theorem.

The second assertion is proven in a similar way. Indeed, by (\ref{toto3}),
\begin{equation}
\int_{\mathfrak{C}}\left( \mathrm{\tilde{p}}_{\mathfrak{e}}\left( r,%
\mathfrak{t},\mathrm{B}+\mathrm{B}_{\mathrm{ext}}\right) -\mathrm{\tilde{p}}%
_{\mathfrak{e}}\left( r,\mathfrak{t},\mathrm{B}_{\mathrm{int}}+\mathrm{B}_{%
\mathrm{ext}}\right) \right) \mathrm{d}^{3}\mathfrak{t}\geq -C_{1}\Vert
\mathrm{B}-\mathrm{B}_{\mathrm{int}}\Vert _{2}^{2}\ \ .  \label{toto5}
\end{equation}%
On the other hand, we can combine Grassmann integration and Brydges--Kennedy
tree expansion method with determinant bounds, Lemma \ref{lemma var prob3
copy(2)}, explicit computations for the case $\mathfrak{e}=0$, and the fact
that $\mathrm{r}_{\beta }\left( \mathrm{B}_{\mathrm{int}}+\mathrm{B}_{%
\mathrm{ext}}\right) $ solves Equation (\ref{gap equation new}), to show
that, at fixed $\beta _{0},r_{0}\in \mathbb{R}^{+}$ and sufficiently large $%
\gamma _{0}$,
\begin{equation}
\int_{\mathfrak{C}}\left( \mathrm{\tilde{p}}_{\mathfrak{e}}\left( r,%
\mathfrak{t},\mathrm{B}_{\mathrm{int}}+\mathrm{B}_{\mathrm{ext}}\right) -%
\mathrm{\tilde{p}}_{\mathfrak{e}}\left( \mathrm{r}_{\beta }\left( \mathrm{B}%
_{\mathrm{int}}+\mathrm{B}_{\mathrm{ext}}\right) ,\mathfrak{t},\mathrm{B}_{%
\mathrm{int}}+\mathrm{B}_{\mathrm{ext}}\right) \right) \mathrm{d}^{3}%
\mathfrak{t}\geq \left\vert r-\mathrm{r}_{\beta }\left( \mathrm{B}_{\mathrm{%
int}}+\mathrm{B}_{\mathrm{ext}}\right) \right\vert ^{2}  \label{toto6}
\end{equation}%
for all $\gamma \in \lbrack \gamma _{0},\infty )$, $\beta \in \mathbb{R}^{+}$%
, $r\in \lbrack r_{0},\infty )$ and any hopping amplitude $\mathfrak{e}$
with $\Vert \mathfrak{e}\Vert _{1}<1$. We then infer from (\ref{toto5})--(%
\ref{toto6}) that%
\begin{multline*}
-C_{1}\Vert \mathrm{B}-\mathrm{B}_{\mathrm{int}}\Vert _{2}^{2}+\left\vert r-%
\mathrm{r}_{\beta }\left( \mathrm{B}_{\mathrm{int}}+\mathrm{B}_{\mathrm{ext}%
}\right) \right\vert ^{2}\leq \mathfrak{F}_{\mathfrak{e}}\left( r,\mathrm{B}+%
\mathrm{B}_{\mathrm{ext}}\right) -\mathfrak{F}_{\mathfrak{e}}\left( \mathrm{r%
}_{\beta }\left( \mathrm{B}_{\mathrm{int}}+\mathrm{B}_{\mathrm{ext}}\right) ,%
\mathrm{B}_{\mathrm{int}}+\mathrm{B}_{\mathrm{ext}}\right) \\
\leq C_{1}\left( \Vert \mathrm{B}-\mathrm{B}_{\mathrm{int}}\Vert
_{2}^{2}+\left\vert r-\mathrm{r}_{\beta }\left( \mathrm{B}_{\mathrm{int}}+%
\mathrm{B}_{\mathrm{ext}}\right) \right\vert ^{2}\right)
\end{multline*}%
for all $\gamma \in \lbrack \gamma _{0},\infty )$, $\beta \in \mathbb{R}^{+}$%
, $r\in \lbrack r_{0},\infty )$ and any hopping amplitude $\mathfrak{e}$
with $\Vert \mathfrak{e}\Vert _{1}<1$. It follows that%
\begin{equation*}
\left\vert \mathrm{r}_{\beta }\left( \mathrm{B}_{\mathrm{int}}+\mathrm{B}_{%
\mathrm{ext}}\right) -\mathrm{r}_{\beta }\left( \mathrm{B}+\mathrm{B}_{%
\mathrm{ext}}\right) \right\vert \leq 2C_{1}\Vert \mathrm{B}-\mathrm{B}_{%
\mathrm{int}}\Vert _{2}\ .
\end{equation*}%
\hfill $\Box $

\begin{corollary}[$\mathrm{B}_{\mathrm{int}}$ as a critical point at large $%
\protect\gamma $]
\label{corollary de funes copy(1)}\mbox{ }\newline
Fix $\mu \in \mathbb{R}$, $\lambda \in \mathbb{R}$ and $\mathrm{B}_{\mathrm{%
ext}}=\mathcal{S}_{0}(\mathrm{j}_{\mathrm{ext}})$ with $\mathrm{j}_{\mathrm{%
ext}}\in C_{0}^{\infty }\cap P^{\bot }\mathfrak{H}$. Then, there are
constants $\gamma _{0},\beta _{0},\kappa _{0}\in \mathbb{R}^{+}$ such that,
for all $\gamma \in \lbrack \gamma _{0},\infty )$, $\beta \in \lbrack \beta
_{0},\infty )$ and any hopping amplitude $\mathfrak{e}$ satisfying $\Vert
\mathfrak{e}\Vert _{1}<\kappa _{0}$, $\mathrm{B}_{\mathrm{int}}$ is a
critical point of the map $\mathcal{G}$ (\ref{map+sup}) from $\mathcal{B}$\
to $\mathbb{R}$, i.e., $\mathcal{G}$ is Fr\'{e}chet differentiable at $%
\mathrm{B}_{\mathrm{int}}$ with vanishing Fr\'{e}chet derivative at this
point.
\end{corollary}

\noindent \textit{Acknowledgments:} This work has been supported by the
grant MTM2010-16843 of the Spanish \textquotedblleft Ministerio de Ciencia e
Innovaci{\'{o}}n\textquotedblright {}and a grant of the \textquotedblleft
Inneruniversit{\"{a}}re Forschungsf{\"{o}rderung}\textquotedblright {}of the
Johannes Gutenberg University in Mainz.

\end{document}